\documentclass[aps,prd,superscriptaddress,nofootinbib,11pt]{revtex4}

\usepackage{array}
\usepackage{color}
\usepackage{rotating}

\usepackage{amssymb}
\usepackage{graphicx}
\usepackage{color}
\usepackage{amscd}
\usepackage{amsmath}

\catcode`\@=11
\def\marginnote#1{}
\def\numberbysection{\@addtoreset{equation}{section}
      \def\theequation{\thesection.\arabic{equation}}}
\numberbysection

\newcommand\tablenum[1]{%
 \def\thetable{#1}%
 \let\@currentlabel\thetable
 \addtocounter{table}{\m@ne}%
}%

\usepackage{float}

\long\def\symbolfootnote[#1]#2{\begingroup%
\def\thefootnote{\fnsymbol{footnote}}\footnote[#1]{#2}\endgroup}

\begin{document}




\title{Signal Propagation in \\Nonlinear Stochastic Gene
Regulatory Networks}

\date{May 18, 2005}




\author{\large Sever Achimescu}
\email{sachimescu@mcg.edu}


\author{\large Ovidiu Lipan}
\email{ olipan@mcg.edu}
\affiliation{Center for Biotechnology and Genomic Medicine\\
Medical College of Georgia, \\1120m 15th St., Ca-4124, Augusta, GA
30912 }

\begin{abstract}
\bf{The structure of a stochastic nonlinear gene regulatory
network is uncovered by studying its response to input signal
generators. Four applications are studied in detail: a nonlinear
connection of two linear systems, the design of a logic pulse, a
molecular amplifier and the interference of three signal
generators in E2F1 regulatory element. The gene interactions are
presented using molecular diagrams that have a precise
mathematical structure and retain the biological meaning of the
processes.}
\end{abstract}
\maketitle

Excerpts from this manuscript were presented  at the 3rd
International Conference on Pathways, Networks, and Systems:
Theory and Experiments, October 2-7, Rhodes Greece 2005,
\cite{Rhodes}.

\parskip 1pt

\symbolfootnote[0]{To whom correspondence should be addressed:
olipan@mcg.edu}

\section{Introduction}

A living organism is a complex interconnection of many control
units that form a gene regulatory network. In developmental
biology, clusters of DNA sequence elements (cis-regulatory module)
are target sites for transcription factors.
 One cis-regulatory module controls a set of gene expression both in space (location)
and time \cite{Davidson2}. One transcription factor can interact
with many modules, and one module is controlled by many
transcription factors. Thus, the time and space variation of a
gene expression is a consequence of an interconnected network of
interactions.

With the advent of high throughput technologies (microarrays,
proteomics tools) the need for quantitative models of gene
networks becomes a reality \cite{Palsson,Hood,Elf}. In recent
years we have witnessed a growing interest in experiments within
the field of systems biology that require mathematical models to
describe the experimental results
\cite{Alex3,Weiss,Alon1,Gardner,Jeff,Dohlman,Becskei}. A
mathematical model for gene regulatory networks is also closely
related with synthetic biology, the engineering counterpart of
systems biology \cite{Brent,ArkinSynthetic,Endy}. Similar with the
development of the field of electronics, where complex equipment
is built on interconnected simple devices, the field of synthetic
biology aims to build simple molecular devices for later use in
more complex molecular machines \cite{Alon,Leibler2}. To build a
robust, reliable, and simple device, the molecular engineer needs
to have a mathematical description of the system in order to
evaluate the number, range and meaning of a group of parameters
that are critical for the device functionality.
  As with any mathematical model of a natural system, the models of a gene regulatory network must fulfill
  certain constraints. At present, the community of researchers agrees that
a gene regulation model must be nonlinear and stochastic.
Nonlinearity is a widespread phenomena in life science
\cite{MathBook,May}. When two molecules dimerize to form a
complex, the concentration of the complex is proportional to the
product of the concentrations of the molecules involved in the
process. Multimerization will require polynomial functions in
molecular concentrations \cite{VanKampen}; furthermore, rational
functions are used to model the most simple gene autoregulatory
system\cite{Alex}. The model should be also stochastic, its
molecules being subjected to the thermodynamics laws of
fluctuation. Two cell lines with identical genetic background can
show variation in phenotype due only to a probabilistic
distribution of the molecule number inside the cell
\cite{ArkinMcAdams}. The stochastic process that describe
molecular interactions are
 fundamentally discrete; the molecule number can change by an
 integer amount only. Fluctuations in biological systems created
 by these stochastic processes, are described by a Master Equation
 \cite{VanKampen}. Approximations to the Master Equation, like the
 Fokker-Planck, Langevin and $\Omega$ expansion, are often use
 \cite{AlexPedraza,ArkinRao,ElfOmega,Paulsson}.
 However, many biological regulatory systems function with
 molecules present in low numbers \cite{Berg4}. For such systems,
 the Master Equation should not be approximated
 \cite{Mukund,Siggia,Osc,Othmer}.
   Another requirement for the model is to explain the flow of a
 signal as it passes through the genetic network. To reveal the structure of a gene regulatory
interactions, input signals (growth factors, heat shocks, drugs)
are inserted into different positions within a gene regulatory
network. Then, output signals are measured at some other
positions. A coherent model of the network must explain the
measured input-output relations and must provide predictions that
can be experimentally tested. It is desirable to know how measured
data is related with the input signal \cite{Hood,Osc,Arkin}. In
engineering sciences \cite{Hassan}, control theory explains in
mathematical terms the
 relation between the input and output signals (measured data). The same
 theory provides ways to design stable systems using feedback
 signals. Ideas from Control Theory translated into molecular
 biology will help the design process in synthetic biology.

  In \cite{Osc} it is proposed that signal generators controlled by light,
 \cite{Quail}, should be incorporated into the gene regulatory
 network. With the help of these light-controlled signal
 generators, different types of signal perturbations can be imposed
 on the gene regulatory network. In \cite{Osc} the Master Equation
 was solved for systems that are linear in the transition
 probabilities. The network's response to signal generators was
 expressed in terms of a transfer matrix for the first and second
 order moments of the stochastic process.
  Our goal in this article is to construct a mathematical description
of a gene regulatory network that is nonlinear, stochastic and
explains the input-output relations as in control theory. The
nonlinearity means that the transition probabilities are rational
functions in the molecule numbers. We do not use Fokker-Planck,
Langevin or $\Omega$ expansion to approximate the Master Equation.
For polynomial transition probabilities, the time evolution
equation for the factorial cumulants will have the following form:
\begin{equation}\label{eclinieInt}
\dot {X}_{\alpha_1\alpha_2...\alpha_n}=\left \{
R_{\alpha_1\alpha_2...\alpha_k
}^m(t)\sum_{Y,\sigma}X_{\alpha_{k+1}\mid...\mid\alpha_n\mid
Y[m^\sigma]}\right \}_{\alpha}
\end{equation}
 If the transition probabilities are rational functions, the equations are (\ref{EqRational}).
 The signal generators are part of the coefficients $R_{\alpha_1\alpha_2...\alpha_k
}^m(t)$.
 Before the theory is presented, we will solve four genetic
regulatory networks
 to explain all the notations and meaning of the variables in (\ref{eclinieInt}) and (\ref{EqRational}). The examples are also meant to provide practical
 applications of the underlying theory.

\section{Results}

\subsection{Two Genes Coupled by a Nonlinear Interaction }

 The aim of this subsection is twofold: to introduce the
 notations that will later be generalized, and to
 explain how the classical control theory model changes due to stochastic effects.
   Each genetic regulatory network is built on a set of interacting molecular
 species. In the present case the entire network is composed of System 1, with mRNA1 and
 protein1 as molecular species, coupled with System 2 with mRNA2
 and protein2 as its molecular species.
\begin{figure}
\centering
\includegraphics[width=12cm]{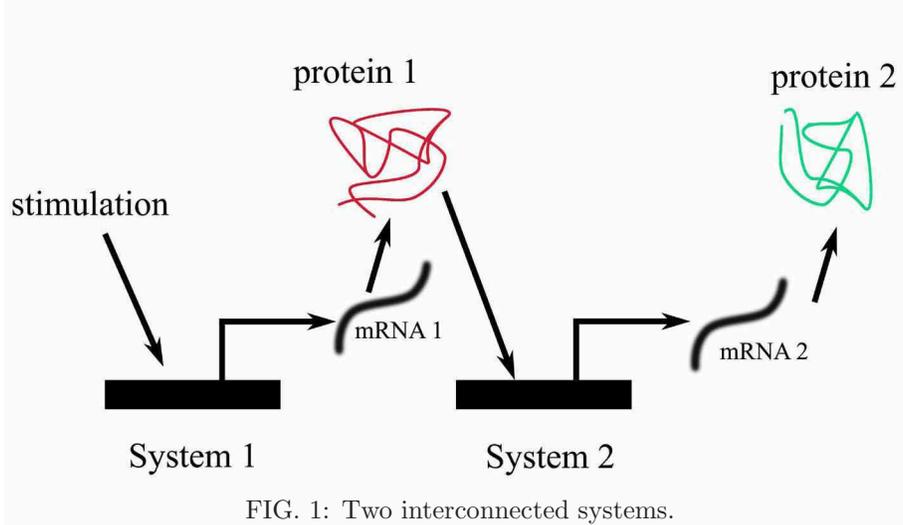}
\vspace{-25pt}\caption{Two interconnected systems.}
\end{figure}
  The number of mRNA1 is
 denoted by $r_1$ and similar notations for the other three
 components. At a given time, the $\it state$ of the network will be a
 denoted by
 \begin{eqnarray}\label{state}
  q\equiv(q_1,q_2,q_3,q_4)=(r_1,p_1,r_2,p_2).
\end{eqnarray}
   The mRNA1 is controlled by an input generator and thus the state $q$
 will evolve in time. Such a generator
 can be practically constructed using an yeast two-hybrid system,
 as is described in \cite{Quail}.
 The light-switch is based on phytochrome that is synthesized in darkness in the Q1
form, Fig. 2. A red light photon of wavelength 664 nm shined on
the Q1 form of the protein transforms it in the form Q2. Fig. 2
presents the state of the switch after the effect of the
corresponding wavelength took place. When Q2 absorbs a far red
light of wavelength 748 nm, the molecule Q goes back to its
original form, Q1. These transitions take milliseconds. The
protein P interacts only with the Q2 form, recruiting thus the
activation domain to the target promoter. In this position, the
promoter is open and the gene is transcribed. After the desired
time elapsed, the gene can be turned off by a photon from a far
red light source. Using a sequence of red and far red light pulses
the molecular switch can be opened and closed.

 The time evolution of the $\it state $ depends on
 all possible $\it transitions$ that can appear in the system. For
 example, from the state $(r_1,p_1,r_2,p_2)$ at time $t$, the
 system can move to the state with one more protein1 molecule
 $(r_1,p_1+1,r_2,p_2)$ because mRNA1 is translated. This transition is
 described by a vector $\epsilon_2=(0,1,0,0)$ that shows the change
 in the state: $(r_1,p_1+1,r_2,p_2)=(r_1,p_1,r_2,p_2)+\epsilon_2$.
 The list of all possible transitions is described in the first column of Table
 1. Which transition will actually take place is governed by a
 stochastic process \cite{VanKampen}. A third element in the model (beside the
 $\it state$ and $\it transitions $) is the set of all transition probabilities
 $T_{\epsilon}$. If the state at time $t$ is $q$, then the
 probability of the system to jump in the state $q+\epsilon$ at the
 time $t+dt$ is $T_{\epsilon}(q,t)dt$. The presence of the time
 $t$ in the argument of the transition function show that, in general, the
 transition depends not only on the number of molecules in the system, but also on the moment in time when it is recorded. $T_{\epsilon_1}$
 in Table 1 is a time dependent transition probability. It
 contains the signal generator $G(t)$ that modulates the mRNA1 transcription.

\begin{figure}[h]
\centering
\includegraphics[width=14cm]{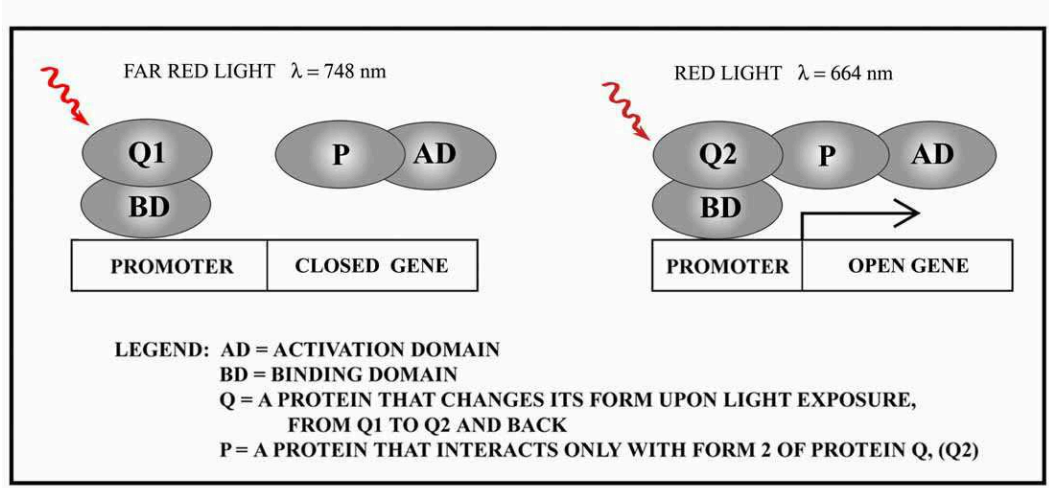}
\vspace{-25pt}\caption{Signal generator. Adapted from ref.
\cite{Quail}.}
\end{figure}

 All transition probabilities are given in the second column in
 Table 1. The coefficients that multiply the molecule number in
 a transition probability are the parameters of the model. Higher values of the coefficient will make that transition to
 appear more often per unit of time. The last column of the table
 will be explained later in the text. The the dynamic of the genetic network is given by the time evolution
 of the probability of the network to be in the state $q$ at time
 $t$: $P(q,t)$. The equation for the time evolution of the state
 probability is know as the  Master Equation  \cite{VanKampen}.

\vspace{15pt}

 \begin{table}[h]
 \tablenum{\large \bf {1}}
 \caption{${\mbox {\large \bf
Nonlinear Coupling of Two Linear Systems }}$}\vspace{-10pt}
\begin{center}
\resizebox{!}{3.10cm}{\begin{tabular}{c|c|c|c|c|}
  \cline{2-5}
   & Transitions & $\phantom{\frac{\sigma_D^{\frac{R}{T}}}{\sigma_D^{\frac{R}{T}}}}\begin{array}{c}\hbox{Transition}\\[0pt] \hbox{probabilities}\end{array}\phantom{\frac{\sigma_D^{\frac{R}{T}}}{\sigma_D^{\frac{R}{T}}}}$ & Coefficients & Polynomial basis \\
\hline
 Signal Generator & $\phantom{\frac{\sigma_D^{Y}}{\sigma_D^{Y}}}\epsilon_1=(1,0,0,0)\phantom{\frac{\sigma_D^{Y}}{\sigma_D^{Y}}}$ & $\phantom{\frac{\sigma_D^{Y}}{\sigma_D^{Y}}}T_{\epsilon_1}(q)=G(t)\phantom{\frac{\sigma_D^{Y}}{\sigma_D^{Y}}}$ & $\phantom{\frac{\sigma_D^{Y}}{\sigma_D^{Y}}}M_{\epsilon_1}(t)=G(t)\phantom{\frac{\sigma_D^{Y}}{\sigma_D^{Y}}}$ & ${\bf e}_{(0,0,0,0)}=1$ \\
\hline
   &$\phantom{\frac{\sigma_D^{Y}}{\sigma_D^{Y}}}\epsilon_{-1}=(-1,0,0,0)\phantom{\frac{\sigma_D^{Y}}{\sigma_D^{Y}}}$ & $T_{\epsilon_{-1}}(q)=\gamma_{r_1}r_1$ & $M_{\epsilon_{-1}}^1=\gamma_{r_1}$ & ${\bf e}_{(1,0,0,0)}=r_1$ \\
\cline{2-5}
   Linear  System 1 &$\phantom{\frac{\sigma_D^{Y}}{\sigma_D^{Y}}}\epsilon_2=(0,1,0,0)\phantom{\frac{\sigma_D^{Y}}{\sigma_D^{Y}}}$ & $T_{\epsilon_2}(q)=k_1r_1$ & $M_{\epsilon_2}^1=k_1$ & ${\bf e}_{(1,0,0,0)}=r_1$ \\
\cline{2-5}
   &$\phantom{\frac{\sigma_D^{Y}}{\sigma_D^{Y}}}\epsilon_{-2}=(0,-1,0,0)\phantom{\frac{\sigma_D^{Y}}{\sigma_D^{Y}}}$ & $T_{\epsilon_{-2}}(q)=\gamma_{p_1}p_1$ & $M_{\epsilon_{-2}}^2=\gamma_{p_1}$ & ${\bf e}_{(0,1,0,0)}=p_1$ \\
\hline
  Nonlinear  Coupling &$\epsilon_3=(0,0,1,0)$ & $\phantom{\frac{\sigma_D^{\frac{R^{\sigma}}{T}}}{\sigma_D^{\frac{R^{\sigma}}{T}}}}\begin{array}{c}\hspace{-6pt}T_{\epsilon_3}(q)=hp_1^2= \\[2pt]=hp_1(p_1-1)+hp_1\end{array}\phantom{\frac{\sigma_D^{\frac{R^{\sigma}}{T}}}{\sigma_D^{\frac{R^{\sigma}}{T}}}}$ & $\begin{array}{c}M_{\epsilon_3}^2=h \\[2pt]M_{\epsilon_3}^{22}=h\end{array}$ & $\begin{array}{c}{\bf e}_{(0,1,0,0)}=p_1 \\[2pt]{\bf e}_{(0,2,0,0)}=p_1\,(p_1-1)\end{array}$ \\
\hline
   &$\phantom{\frac{\sigma_D^{Y}}{\sigma_D^{Y}}}\epsilon_{-3}=(0,0,-1,0)\phantom{\frac{\sigma_D^{Y}}{\sigma_D^{Y}}}$ & $T_{\epsilon_{-3}}(q)=\gamma_{r_2}r_2$ & $M_{\epsilon_{-3}}^3=\gamma_{r_2}$ & ${\bf e}_{(0,0,1,0)}=r_2$ \\
\cline{2-5}
  Linear System 2 &$\phantom{\frac{\sigma_D^{Y}}{\sigma_D^{Y}}}\epsilon_4=(0,0,0,1)\phantom{\frac{\sigma_D^{Y}}{\sigma_D^{Y}}}$ & $T_{\epsilon_4}(q)=k_2r_2$ & $M_{\epsilon_4}^3=k_2$ & ${\bf e}_{(0,0,1,0)}=r_2$ \\
\cline{2-5}
    &$\phantom{\frac{\sigma_D^{Y}}{\sigma_D^{Y}}}\epsilon_{-4}=(0,0,0,-1)\phantom{\frac{\sigma_D^{Y}}{\sigma_D^{Y}}}$ & $T_{\epsilon_{-4}}(q)=\gamma_{p_2}p_2$ & $M_{\epsilon_{-4}}^4=\gamma_{p_2}$ & ${\bf e}_{(0,0,0,1)}=p_2$ \\
\hline
\end{tabular}}
\end{center}
\end{table}

\vspace{-30pt}
\begin{align}\label{MasterProb}
 &P (r_1,p_1,r_2,p_2,t+dt) =  \\\nonumber
  & P(r_1-1,p_1,r_2,p_2,t)\;T_{\epsilon_1}(r_1-1,p_1,r_2,p_2,t)dt+P(r_1+1,p_1,r_2,p_2,t)\;T_{\epsilon_{-1}}(r_1+1,p_1,r_2,p_2,t)dt\;\; +
  \\\nonumber
  &  P(r_1,p_1-1,r_2,p_2,t)\;T_{\epsilon_2}(r_1,p_1-1,r_2,p_2,t)dt+P(r_1,p_1+1,r_2,p_2,t)\;T_{\epsilon_{-2}}(r_1,p_1+1,r_2,p_2,t)dt\;\; +\\\nonumber
   &P(r_1,p_1,r_2-1,p_2,t)\;T_{\epsilon_3}(r_1,p_1,r_2-1,p_2,t)dt +P(r_1,p_1,r_2+1,p_2,t)\;T_{\epsilon_{-3}}(r_1,p_1,r_2+1,p_2,t)dt\;\; +\\\nonumber
  & P(r_1,p_1,r_2,p_2-1,t)\;T_{\epsilon_4}(r_1,p_1,r_2,p_2-1,t)dt+P(r_1,p_1,r_2,p_2+1,t)\;T_{\epsilon_{-4}}(r_1,p_1,r_2,p_2+1,t)dt\; +\\\nonumber
   &P(r_1,p_1,r_2,p_2,t)\left(\vphantom{\sum}1-T_{\epsilon_1}(r_1,p_1,r_2,p_2,t)dt-T_{\epsilon_{-1}}(r_1,p_1,r_2,p_2,t)dt-T_{\epsilon_2}(r_1,p_1,r_2,p_2,t)dt\;- \right.\\\nonumber
  &\hspace{115pt} T_{\epsilon_{-2}}(r_1,p_1,r_2,p_2,t)dt-T_{\epsilon_3}(r_1,p_1,r_2,p_2,t)dt-T_{\epsilon_{-3}}(r_1,p_1,r_2,p_2,t)dt-\\\nonumber
  &
  \hspace{108pt}\left.\;T_{\epsilon_4}(r_1,p_1,r_2,p_2,t)dt-T_{\epsilon_{-4}}(r_1,p_1,r_2,p_2,t)dt\vphantom{\sum}\right)\;.
\end{align}

 Given that the system was at time $t$ in the state
$q=(r_1,p_1,r_2,p_2)$, the first 8 terms in (\ref{MasterProb})
represent the probability that the system will be in a new state
at the time $t+dt$. The new state depends on which transition
actually took place. The rest of the terms in (\ref{MasterProb})
express the probability that no transition will take place in
$(t,t+dt)$. Dividing by $dt$ and taking the limit $dt\rightarrow
0$, the above relation between probabilities takes the form of a
partial differential equation:

\begin{eqnarray}\label{MasterEquation}
  {\frac {\partial}{\partial t}}P \left(q, t \right)=\sum_\epsilon T_\epsilon (q-\epsilon,t)P(q-\epsilon,t)-\sum_\epsilon
  T_\epsilon(q,t)P(q,t)\;.
\end{eqnarray}

This equation is known as the Master Equation for the jump Markov
processes with discrete states \cite{GillespieBook,VanKampen}. The
summation is over all possible transitions.
 It is hard to solve this
equation, even in very simple examples. However, we are interested
in the mean number of molecules of different species and in their
standard deviation. Or more generally, we want to know the
correlation between different molecular species. The quantities of
interest are thus means of products of state components:

\begin{eqnarray}
 \langle q ^{\overline m} \rangle  &=& \sum_{q}q^{\overline
 m}P(q,t)\;,
\end{eqnarray}
where the sum goes over all possible states. The notation
$\overline m$ stands for a vector $\overline m=(m_1,m_2,m_3,m_4)$
of integer numbers. The number of components of $\overline m$ is
the same as the number of components in the state $q$. The power
of $q$ to the $\overline m$ is defined as $q^{\overline
m}=q_1^{m_1}q_2^{m_2}q_3^{m_3}q_4^{m_4}$. We need a line on top of
$m$ to distinguish it from a simple $m$ that will be used heavily
in what follows.

 The time evolution for $\langle q ^{\overline m} \rangle$ can be
 obtained from the Master Equation (\ref{MasterEquation}) using the
 $z$-transform of the state probability $P(q,t)$:

 \begin{eqnarray}\label{Ztransform}
   F(z,t) &=& {\cal Z}(P(q,t))\equiv \sum_{q}z^q P(q,t)
 \end{eqnarray}

where $z=(z_1,z_2,z_3,z_4)$ are variables in the complex plane and
the power $z^q$ was defined above. Quantities of interest are
means of polynomials in the components of the state variable $q$,
like $ \langle q ^{\overline m} \rangle$. A natural way to obtain
means of polynomials in $q$ is by taking derivatives with respect
to $z$ in (\ref{Ztransform}) and then put $z_i=1,\,i=1\dots 4$.
For example:
\begin{eqnarray}
  \langle r_1 \rangle &=& \frac{\partial F(z,t)}{\partial z_1}\mid_{z=1}\,, \\
  \langle p_2(p_2-1) \rangle  &=&\frac{\partial F(z,t)}{\partial
  z_4^2}\mid_{z=1}\,.
\end{eqnarray}

We notice that the derivatives of $F(z,t)$ bring us to a set of
polynomials that are known as decreasing factorials:

\begin{eqnarray}\label{DecreasingFactorial}
{ e}_{m_k}(q_k)&=&q_k(q_k-1)...(q_k-m_k+1)\, ,\\
{\bf e}_{\overline
m}(q)&=&e_{m_1}(q_1)e_{m_2}(q_2)e_{m_3}(q_3)e_{m_4}(q_4)\,.
\end{eqnarray}

We will use the polynomials $e_{\overline m}(q)$ as a base, to
express all the transition probabilities, as is explained in the
last column of Table 1. In this base, the results are easy to
express in terms of the derivatives of $F(z,t)$. A decreasing
factorial has a physical interpretation, \cite{VanKampen}. In a
system with $q_k$ molecules of specie $k$, the probability for a
collision involving $m_k$ such molecules is proportional with ${
e}_{m_k}(q_k)$. In other words, the probability for multimer
formation is described by a decreasing factorial.

 Every polynomial
can be expressed as a linear combination of the basic polynomials
$e_{\overline m}(q)$. To make a distinction between the moments
$\langle q ^{\overline m} \rangle$ and $\langle e_{\overline
m}(q)\rangle$, the later one is known as a factorial moment. The
first order moments (which are actually the means of the state
variable) and the first order factorial moments are equal.

 The variables that describe the  system are thus
 \begin{eqnarray} \label{DynamicalF}
   \langle {\bf e}_{\overline m}(q) \rangle =
   \partial_{m}F(z,t)\mid_{ z=1}\;,
 \end{eqnarray}
which also displays the tensor index $m$. From a vector index
\begin{eqnarray}
 \overline{m}=(m_1,m_2,m_3,m_4)\;,
\end{eqnarray}
we can construct a tensor index
\begin{eqnarray}
  m=\underbrace{11...1}_{m_1}\underbrace{22...2}_{m_2}\underbrace{33...3}_{m_3}\underbrace{44...4}_{m_4}\;,
\end{eqnarray}
and vice versa. The tensor index $m$ is useful for ordering the
 variables as they come from partial derivatives.
\begin{eqnarray}\nonumber
  \partial_{m} &=& \partial{\underbrace {z_1\dots z_1}_{m_1},\underbrace{z_2\dots z_2}_{m_2},\underbrace{z_3\dots z_3}_{m_3},\underbrace{z_4\dots
  z_4}_{m_4}}\;.
\end{eqnarray}
For example
\begin{eqnarray}\nonumber
  \langle r_1r_2p_2(p_2-1)(p_2-2)\rangle &=&
  \partial_{13444}F\mid_{z=1}\;,
\end{eqnarray}
because $r_1$ is on the first position in the state
$q=(r_1,p_1,r_2,p_2)$, and only one derivative with respect to
$z_1$ is required. The protein $p_2$ is on the 4th position and it
takes 3 derivatives to obtain $p_2(p_2-1)(p_2-2)$.

Instead of always showing that after a partial derivative we have
to insert $z=1$ in the expression, we will use the following
notation

\begin{eqnarray}
  \partial_m F\mid_{z=1} &=& F_{m}\,.
\end{eqnarray}

For special examples like the one we work with, we can use
suggestive notations for the indices
$z=(z_{r_1},z_{p_1},z_{r_2},z_{p_2})$. We list the first $F_m$
 variables in order as they appear in the Taylor
expansion of the function $F(z,t)$ about $z=1$.

\begin{eqnarray}\nonumber
  F_{r_1},F_{r_2},F_{p_1},F_{p_2},F_{r_1r_1},F_{r_1p_1},F_{r_1r_2},F_{r_1p_2},\dots,F_{p_2p_2},F_{r_1r_1r_1},F_{r_1r_1r_2}\dots
  \;.
\end{eqnarray}

These variables change in time as the generator $G(t)$ drives the
system. From these variables we can read the mean values of the
molecules like $\langle r_2 \rangle=F_{r_2}$ and also their
standard deviation from the mean, $\langle (r_2-\langle r_2
\rangle)^2\rangle=F_{r_2r_2}+F_{r_2}-F_{r_2}^2$. The time
evolution of the variables $F_m$ is a consequence of the Master
Equation in $F(z,t)$:

\begin{eqnarray}\label{Fzt}
   \partial_tF(z,t)&=& G(t)(z_{r_1}-1)F(z,t)+ \\
  \nonumber & &  \gamma_{r_1}(1-z_{r_1})\partial_{z_{r_1}}F(z,t)+k_1(z_{p_1}-1)z_{r_1}\partial_{z_{r_1}}F(z,t)+\gamma_{p_1}(1-z_{p_1})\partial_{z_{p_1}}F(z,t)+\\
  \nonumber & &  h(z_{r_2}-1)\left (z_{p_1}^2\partial_{{z_{p_1}}{z_{p_1}}}F(z,t)+z_{p_1}\partial_{
p_1}F(z,t)\right )+\\
  \nonumber & &
  \gamma_{r_2}(1-z_{r_2})\partial_{z_{r_2}}F(z,t)+k_2(z_{p_2}-1)z_{r_2}\partial_{z_{r_2}}F(z,t)+\gamma_{p_2}(1-z_{p_2})\partial_{z_{p_2}}F(z,t)\;.
\end{eqnarray}

The right side of this equation is composed of three pieces. The
first piece contains variables only from the System 1 (first 4
terms). Then comes a term proportional with the coupling constant
$h$ that show how the two systems are connected. The last three
terms are specific to System 2. The coupling terms contain the
second derivative with respect the protein 1, which is the input
signal into the second system. Derivatives of order more than 1
are a sign of nonlinearity. Here the coupling transition
probability $T_{\epsilon_3}$ is a quadratic function in the
protein number of System 1.

The time evolution equations of the $F_m$ variables  can be
obtained from (\ref{Fzt}) by taking partial derivatives with
respect to different combinations of the components of $z$ and
then inserting $z=1$. Through this procedure, the partial
differential equation for $F(z,t)$ transforms into an infinite
system of ordinary differential equations for the $F_m$ variables
(the tensor index $m$ takes all the possible values). We present
the first few:

\begin{eqnarray}\nonumber
    \dot{F}_{r_1}&=&G(t)-\gamma_{r_1}F_{r_1}\\\nonumber
    \dot{F}_{p_1}&=&k_1F_{r_1}-\gamma_{p_1}F_{p_1}\\\nonumber
    \dot{F}_{r_2}&=&h(F_{p_1}+F_{p_1p_1})-\gamma_{r_2}F_{r_2}\\\nonumber
    \dot{F}_{p_2}&=&k_2F_{r_2}-\gamma_{p_2}F_{p_2}\\\nonumber
    \dot{F}_{r_1p_2}&=&G(t)F_{p_2}+k_2F_{r_1r_2}-(\gamma_{r_1}+\gamma_{p_2})F_{r_1p_2}\\\nonumber
    \dot{F}_{r_2r_2}&=&2h(F_{p_1r_2}+F_{p_1p_1r_2})-2\gamma_{r_2}F_{r_2r_2}\;.
\end{eqnarray}

If the system is linear, i.e. all the transition probabilities are
linear in the state variables, then the infinite system can be
closed to a finite one. Namely, if we collect all variables up to
the modulus  $\left|\overline m \right|=Max$, ($\left|\overline m
\right|=m_1+\dots +m_n$), than we create a system of equations
that do not depend on a
 \clearpage
\begin{figure}
\centering
\includegraphics[height=21cm,angle=0]{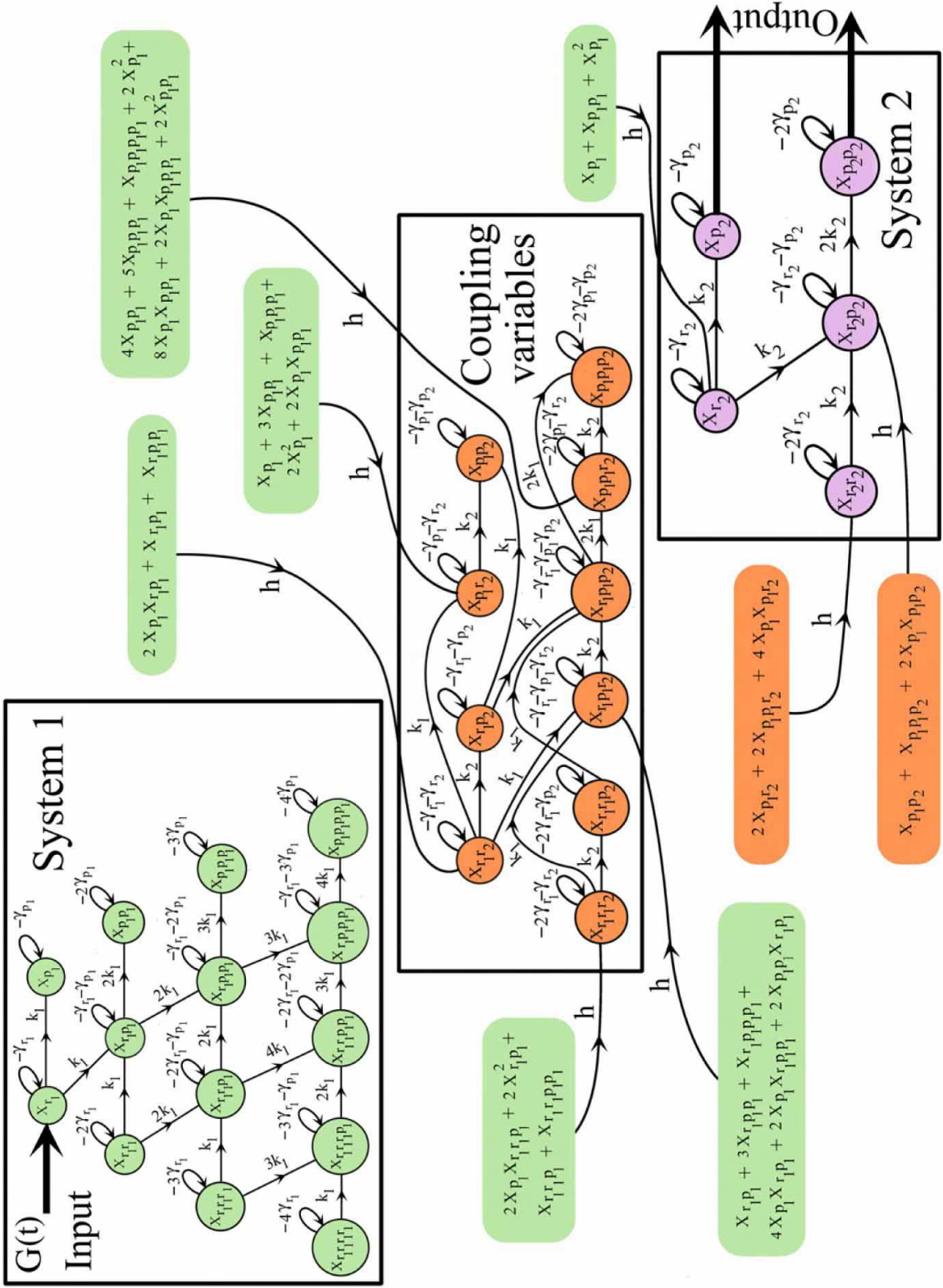}
\caption{Graph of the equations for the coupled systems.}
\end{figure}
\clearpage
 $\overline m$ that has a modulus greater than $Max$.
The system, being finite, is completely solvable \cite{Osc}. For
the nonlinear transition probabilities, the system is usually not
finite. The variable $F_m$ will depend on some other $F_{m^{'}}$
with $\left| m'\right|
> \left| m \right|$.

 Thus we need to cut the infinite system to obtain an ordinary system of
 equations: discard all $F_m$s with $\left| m\right|$ greater then sum cutoff value.
  The problem that we run
 into is that the solution for variables $F_m$ with small $\left| m \right|$ depends on the cut
 even if the cut is taken at high values of $\left| m
 \right|$.
 Moreover, from a stable infinite system, by cutting we obtain an
 unstable system. The cause of this behavior is that as $\left|
 m\right|$ increases, the values of $F_m$ increases because it
 represents a mean of a high power polynomial in the state
 variables. Discarding such high values from the equations causes
 the aforementioned instability. However, a finite system
 can be obtained if instead of the variables $F_m$ we change to a
 new set of variables that have small values that can be neglected
 for higher values of $\left| m\right| $. These variables represent for the factorial moments what cumulants represent for
 the classical moments \cite{McCullagh}. The new function, call it $X(z,t)$ is
 given by
\begin{eqnarray}\label{Xcumulants}
   F(z,t) &=& e^{X(z,t)}\,.
 \end{eqnarray}

 The equation for the time evolution of $X(z,t)$ is
\begin{eqnarray}
  \nonumber \partial_t X(z,t)&=& G(t)(z_{r_1}-1)+ \\
  \nonumber & &  \gamma_{r_1}(1-z_{r_1})\partial_{z_{r_1}}X(z,t)+k_1(z_{p_1}-1)z_{r_1}\partial_{z_{r_1}}X(z,t)+\gamma_{p_1}(1-z_{p_1})\partial_{z_{p_1}}X(z,t)+\\
  \nonumber & &  h (z_{r_2}-1)\left (z_{p_1}^2\partial_{z_{p_1}z_{p_1}}X(z,t)+z_{p_1}^2\left ( \partial_{z_{p_1}}X(z,t)\right )^2+z_{p_1}\partial_{z_{p_1}}X(z,t)\right )+\\
  \nonumber & &
  \gamma_{r_2}(1-z_{r_2})\partial_{z_{r_2}}X(z,t)+k_2(z_{p_2}-1)z_{r_2}\partial_{z_{r_2}}X(z,t)+\gamma_{p_2}(1-z_{p_2})\partial_{z_{p_2}}X(z,t)\;.
\end{eqnarray}
 We notice that  the term proportional with $h$ couples
the two linear  systems. Taking partial derivatives with respect
to $z$ and inserting $z=1$, we obtain the variables $X_m$, indexed
by the tensorial index $m$. The system of equations for these
variables is represented as a graph in Fig. 3.
 Each node represents the time derivative of the
variable written inside the node. A line entering the node
corresponds to one term on the right side of the equation for that
node; the term is the product of the variable written inside the
start node and the coefficient above the line. For example, the
equation for the variable $X_{r_2p_2}$ is
\begin{equation}
    \dot
{X}_{r_2p_2}=k_2X_{r_2}+k_2X_{r_2r_2}-(\gamma_{r_2}+\gamma_{p_2}){X}_{r_2p_2}+h(X_{p_1p_2}+X_{p_1p_1p_2}+2X_{p_1}X_{p_1p_2}).
\end{equation}
The indices $r_2p_2$ of ${X}_{r_2p_2}$ belong to System 2 as well
as three other terms that have indices from that same system. The
term $k_2X_{r_2}$ is represented by the line starting on $X_{r_2}$
and ending on the ${X}_{r_2p_2}$ node. The coupling coefficient
$h$ multiplies the polynomial combination
$X_{p_1p_2}+X_{p_1p_1p_2}+2X_{p_1}X_{p_1p_2}$. These polynomial
combinations stem from the nonlinear coupling and their role is to
connect System 1 with the System 2. If we are interested only in
the mean value of the protein $p_2$, that is $X_{p_2}$, then we
only need to solve the equations for the System 1, compute the
coupling factor $X_{p_1}+X_{p_1p_1}+X_{p_1}$  and solve the
equations for System 2. However, if we ask for the standard
deviation of the number of protein molecules $p_2$, then we need
$X_{p_2p_2}$ which requires solving for the coupling variables
$X_{r_1r_2},\,X_{r_1p_2},\dots X_{p_1p_1p_2}$, see Fig.3. When it
comes to solving for the coupling variables, we find that
variables up to fourth order in System 1 are necessary (like
$X_{r_1r_1r_1r_1}$). In general, suppose we need to solve System 2
in order n, that is  we need $X_m$ with $\left|m\right|=n$ and all
components of the index $m$ contain $r_2$ or $p_2$. Then, we need
an order $n+1$ in the coupling variables and $n+2$ in System 1.
 Another observation is that even if the coupling is nonlinear,
 the entire system of equations is finite.
 This property is a
 reminiscent of the fact that System 1 and 2 are linear. A linear
 system, whose equations are depicted in Fig. 3 in the upper left
 corner, has the property that the order $n$ depends only on
 orders that are smaller or equal to $n$.
 Thus, the equations for a linear system have a hierarchical
 structure, Fig 3, System 1. In \cite{Osc} the equations for the first two
 orders for a linear system were solved. The second order variables
 like $X_{r_1r_1},X_{p_1p_1}$ were called non-Poisson components in
 \cite{Osc}. This name came from the fact that the standard
 deviation can be expressed as $\sigma_{r}^2=X_{r}+X_{rr}$ and
 for a Poisson process $\sigma_{r}^2=X_{r}$.
 We conclude this section by noting that two stochastic
 systems (System 1 and System 2) are not coupled by simply taking
 output variables from System 1 and input them into System 2. The
 stochastic coupling requires that the output of the System 1
 should first pass through an intermediate system and then enter
 into the System 2.

\subsection{Molecular Diagrams}

A graphical representation of gene regulatory networks is
essential in order to capture information about gene interaction.
 A notational system should satisfy four
important criteria \cite{Kitano}:

(1) Expressiveness: the diagram should describe every possible
relationship among molecules.

(2) Semantically unambiguous: different symbols should be assigned
to different semantics.

(3) Extension capability: the notation system should be flexible,
so that new symbols can be added in a consistent manner.

(4) Mathematical translation: each diagram should be able to be
converted into a mathematical formalism for use in quantitative
computations.
\vspace{-10pt}
\begin{figure}[H]
\centering
\includegraphics[width=17cm]{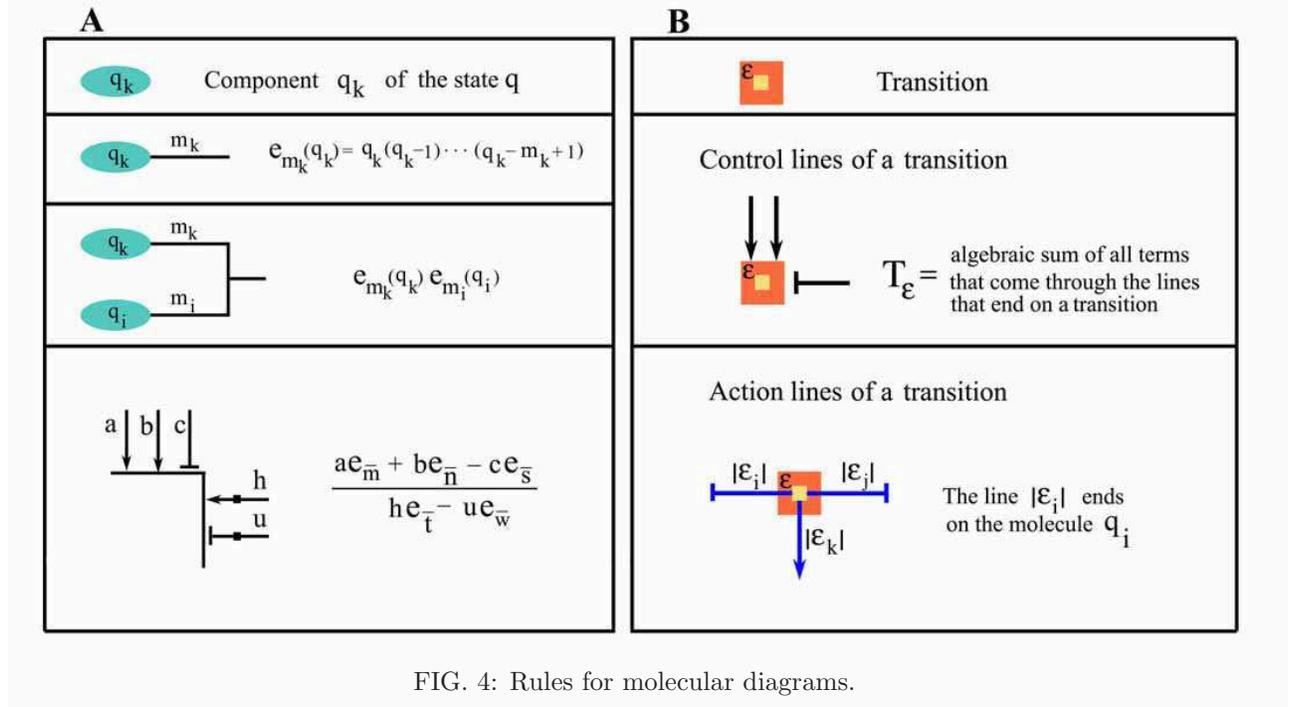}\vspace{-30pt}
\caption{Rules for molecular diagrams.}
\end{figure}

Based on the mathematical model for the stochastic genetic
networks, we can assemble a set of rules to construct diagrams
that obey the above criteria. The building blocks of the model
are: the state $q$, the transitions $\epsilon$ and the transition
probabilities $T_\epsilon$. Each of these building blocks will be
represented in a diagram, whose graphical notations are depicted
in Fig. 4.
  The component $q_k$ is represented by an oval, Fig. 4A,
first row. If the molecules from a specific biological context are
classified in families (antibodies, cytokines, etc.) then, instead
of using an oval for each molecule, a specific geometric shape can
be associated with each family. For example, to distinguish
between an mRNA and a protein, we used a quadrilateral symbol for
mRNA and an oval for the protein. Each transition will be
represented by a square, Fig. 4B first row. If necessary,  the
transitions can be grouped in phosphorylation, transport,
transcription, etc. For each class of transition a different
geometric symbol can be used.
 The transition probabilities are built upon decreasing factorials
 (\ref{DecreasingFactorial}). A decreasing factorial ${\bf e}_{m_k}(q_k)$ will be
 represented by a line starting from the component $q_k$, Fig. 4A second row.
If the coefficient in front of a decreasing factorial is positive,
the line will carry an arrow; otherwise the line will
 end in a bar. We will not write $m_k$ on top of its corresponding line if $m_k=1$.
 A product of two decreasing factorials is represented by joining the
lines of each of the term in the product, Fig. 4A
 third row. Graphical representation of a rational function is
 depicted in Fig. 4A fourth row. The lines representing the terms from the denominator of
 the rational function are marked by a filled square. Two types of lines are associated with a transition
 $\epsilon $. One type ends on the border of the square representing the transition, Fig. 4B second row.
 These lines originate on different components $q_k$ on which the
 transition probability $T_{\epsilon}(q)$ depends. The components
 $q_k$ control the transition probability $T_{\epsilon}(q)$ so the
 lines that end on the boundary of the transition symbol are
 called {\it control lines}. Each transition will act on some
 molecules to change their number. This action is described by the
 components of the transition $\epsilon$. Each nonzero
 component of an $\epsilon$ will be associated with a line that
 starts from the center of the transition symbol, Fig. 4B third row. These are
 called {\it action lines}. Each action line ends on a component
 $q_k$ that is changed by the corresponding transition. If the
 component of a transition $\epsilon $ is positive, the line is
 marked by an arrow and by a bar if the component is negative. In
 Fig. 4B third row, $\epsilon _i<0$, $\epsilon_j<0$ and
 $\epsilon_k >0$. If $\left|\epsilon_i\right|=1$ we will not write
 it on its corresponding line. Finally, terms that correspond to lines that
 end on a transition $\epsilon$ must be
 summed to form the transition probability, Fig. 4B second row.
The molecular diagram for the system  under study, Fig. 5, follows
from Table 1 and the rules from Fig. 4.
\begin{figure}
\centering
\includegraphics[width=15cm]{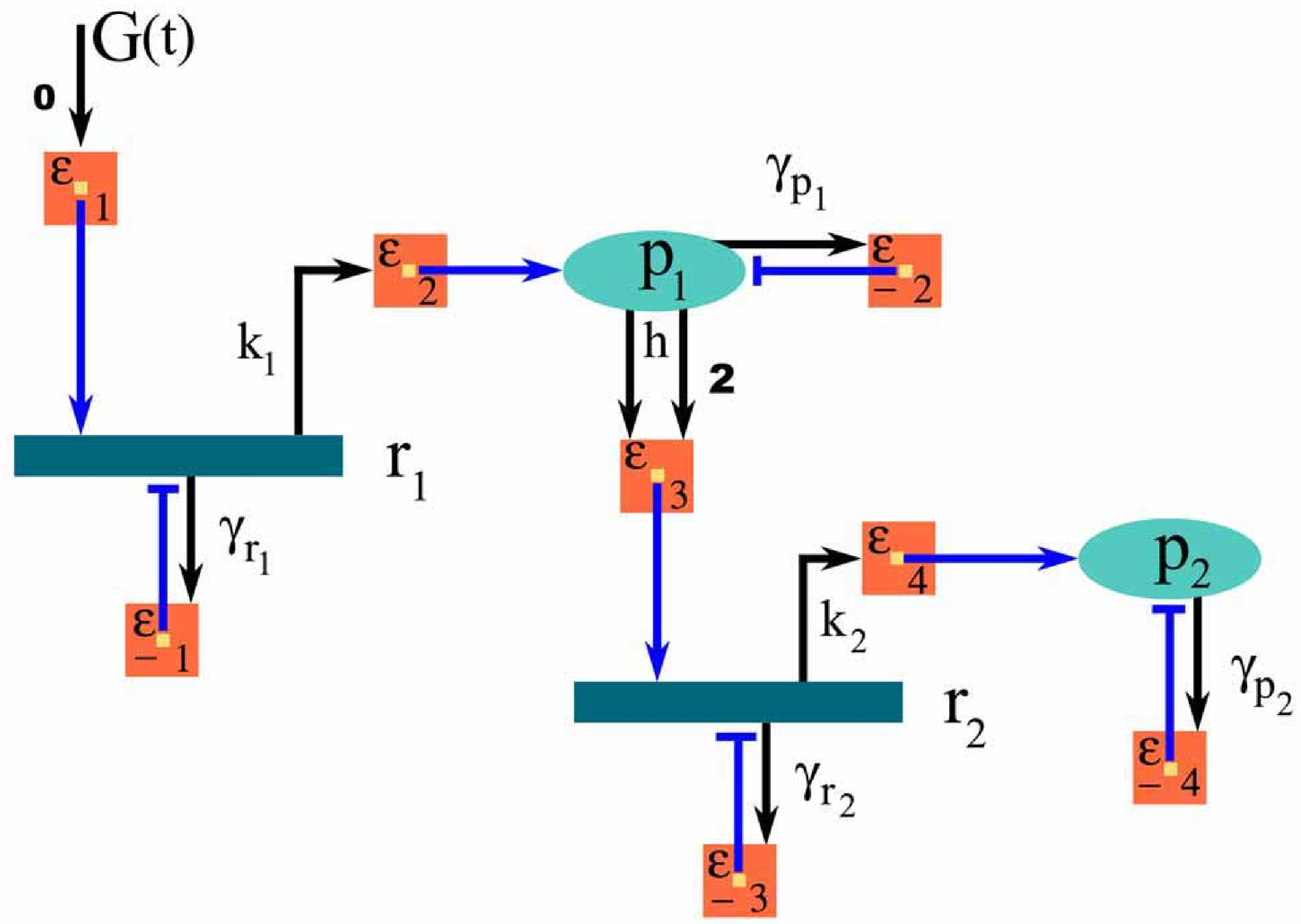}
\caption{Molecular diagram for two coupled systems}
\end{figure}

   The
generator line has $m=0$ written on it, which corresponds to ${\bf
e}_{(0,0,0,0)}=1$, from Table 1. This line does not reach the
center of the transition $\epsilon_1$, indicating that it is a
control line. The transition is controlled by the generator and
its transition probability is readable from the diagram:
$T_{\epsilon_1}=G(t){\bf e}_{(0,0,0,0)}=G(t)$. From the center of
the transition $\epsilon_1$ starts a line that points to the mRNA
symbol $r_1$. This line, because it starts from the center,
corresponds to the effect of the transition and tells that only
one component of $\epsilon_1$ is not zero, $\epsilon_1=(1,0,0,0)$.
The line has an arrow that indicates that this component is
positive, so the transition will cause an increase of mRNA by one
molecule. All other lines can be read in the same manner. From the
protein $p_1$ symbol there are two lines that control the
transition $\epsilon_3$. These lines represent the nonlinear
coupling $T_{\epsilon_3}=hp_1^2=hp_1+hp_1(p_1-1)$ decomposed in
the factorial bases. In Fig. 5, the line marked with the number 2
corresponds to the term $hp_1(p_1-1)$ whereas the other line
indicates the term $hp_1$.
  We will use molecular
diagrams to present the interactions for each example that
follows.

\subsection{Units of Measurement}

The transition probabilities $T_{\epsilon}(q,t)$ are measured in
$[{\mbox {seconds}}]^{-1}$ and thus the coefficient
$M_{\epsilon}^m(t)$ from
$T_{\epsilon}(q,t)=\sum_{m}M_{\epsilon}^m(t){\bf e}_m(q)$ is
measured in $[{\mbox {moles}}]^{-\left|m\right|}[{\mbox
{seconds}}]^{-1}$. At present, the numerical values for the
molecular constants $M_{\epsilon}^m(t)$ that govern the mechanisms
inside the cell can not be determined precisely through laboratory
experiments. In what follows a numerical coefficients will be
written without a unit of measurement, to show that it is not
experimentally determined. We use numerical values to compare the
analytical theory with Monte Carlo simulations and to show how the
general formulas can be use in practical applications.

\subsection{Hill Feedback Control}

One of the basic elements of a gene regulatory network is a gene
that controls its own transcription \cite{Young}. The protein acts
on mRNA production through a term of the form
\begin{eqnarray}
  \frac{a}{b+p^2}\;.
\end{eqnarray}

When the number of protein molecules increases, the rate of mRNA
production will decrease, stabilizing the system's transcription
and translation. This kind of feedback control is employed in the
description of many biological systems. In \cite{Alex} it is used
to explain the appearance of multistability in the lactose
utilization network of Escherichia coli. In \cite{Leibler} it is
used to describe a stable oscillator constructed from three genes
that repress themselves in a closed loop.

 Our study focuses on the case where a signal generator accompanies the feedback. When the generator
 is turned off, the gene is driven slowly by the nonlinear
 feedback. This special case will be addressed in the following subparagraph. From a mathematical
 point of view, this system is interesting because the transition
 probability for repression is a rational function. The Table 2
 and Fig.6 presents the structure of the system
\vspace{5pt}

\begin{table}[h]
 \tablenum{{\large \bf 2}}
 \caption{${\mbox {\large \bf Hill Feedback
 Control}}$}
\begin{center}
\resizebox{!}{1cm}
{\begin{tabular}{|p{2.4in}|p{1.1in}|p{1in}|p{1.1in}|}
  \hline
  $\phantom{\frac{\sigma_n^{H^{\sigma ^N}}}{\sigma_n^H}}\hspace{60pt}{\epsilon_1 = (1, 0)}$ & $\hspace{10pt}\epsilon_{-1} = (-1, 0)$ & $\hspace{10pt}\epsilon_2=(0,1)$ & $\hspace{10pt}\epsilon_{-2} = (0, -1)$ \\\hline
  $\phantom{\frac{\sigma_n^{H^{\sigma ^N}}}{\sigma_n^H}}\vspace{1.9pt}T_{\epsilon_1}=G(t)+\displaystyle {\frac{a_1+a_2p}{b_1+b_2p+b_3p(p-1)}}\vspace{1.9pt} $& $\hspace{10pt}T_{\epsilon_{-1}}=\gamma_rr$ & $\hspace{10pt}T_{\epsilon_2}=K r$ & $\hspace{10pt}T_{\epsilon_{-2}}=\gamma_pp$ \\
  \hline
\end{tabular}}
\end{center}
\end{table}
\begin{figure}
\centering
  \includegraphics[width=14cm]{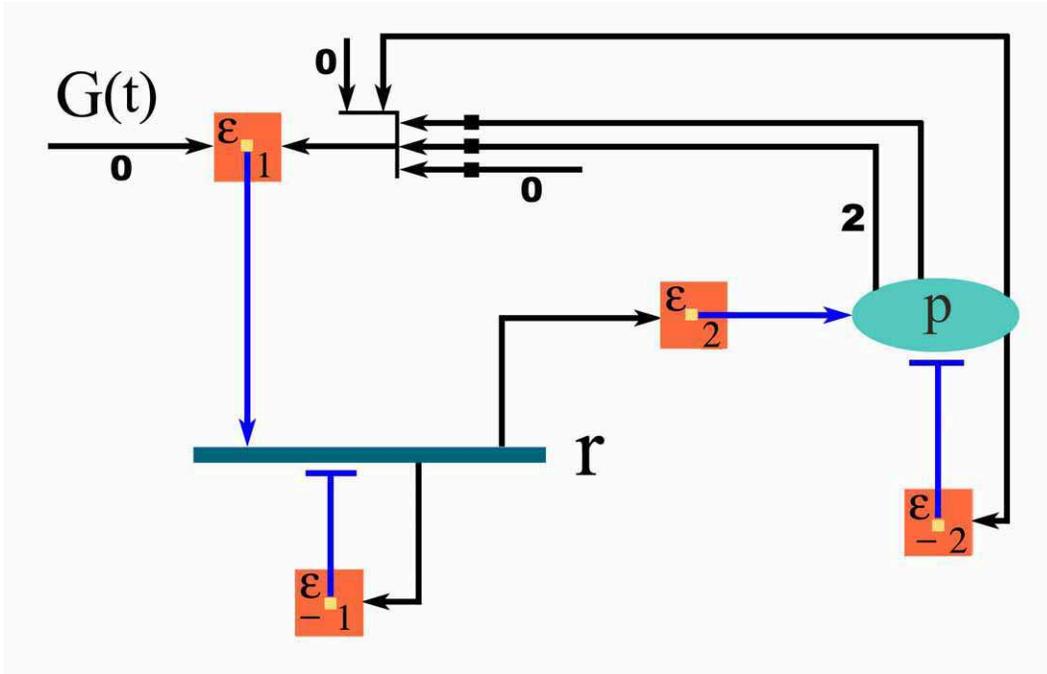}
  \caption{Autoregulatory gene. The feedback has a Hill coefficient of 2.}
\end{figure}
The lines that do not start on any molecule represent the
coefficients $a_1$ and $b_1$. The coefficient $b_1$, being a term
in the denominator of the feedback transition probability, has a
small square superimposed upon it.

 The Master Equation for the state probability is transformed by
 multiplying the whole equation (\ref{MasterEquation}) with $ b_1+b_2\,p+b_3\,p\,(p-1)$.

\begin{eqnarray}
  \nonumber & &\left( b_1+b_2\,p+b_3\,p\,(p-1)\right )\frac{\partial}{\partial t} {P}(r,p,t)=\left (a_1+a_2\, p\, \right ) \left (
  P(r-1,p,t)-P(r,p,t)\right )+\\ \nonumber
  & &\left (  b_1+b_2\,p+b_3\,p\,(p-1)  \right ) [\,G\left (t \right )\left (
  P(r-1,p,t)-P(r,p,t)\right ) +\gamma_r\, (r+1)\,P(r+1,p,t)-\gamma_r\,
  r\,
  P(r,p,t)+\\ \nonumber
  & &K\, r\, P(r,p-1,t)-K\, r\, P(r,p,t) +
  \gamma_p \,(p+1)\,P(r,p+1,t)-\gamma_p\, p\, P(r,p,t)\, ]\;.
\end{eqnarray}

It is now possible to take the $z$-transform (\ref{Ztransform})and
then change
 the variable from $F(z,t)$ to $X(z,t)$, (\ref{Xcumulants}). The dynamical equation
 for $X(z,t)$ contains both the effects of the nonlinear feedback
 as well as the input signal generator $G(t)$:
\begin{eqnarray}
  \nonumber
 b_{{1}} \partial_t{X}  +
  b_{{2}}z_{{p}} \left( \partial_p \partial_t{X} +
   {\partial_t X}{\partial_p X}  \right) +
  b_{{3}}{z_{{p}}}^{2} \left(
          \partial_{pp}\partial_t X +
            \partial_t X \partial_{pp} X   +
                2\,\partial_p X  \partial_t\partial_p X +
                 \left(\partial_p X   \right) ^{2}\partial_t X
                \right)&=&
                 \\ \nonumber
 G \left( t \right)  \left( z_{{r}}-1 \right)  \left( b_{{1}}+b_{{2}}z_
      {{p}}\partial_p X+b_{{3}}{z_{{p}}}^{2}
      ( \partial_{pp} X + \left(
        \partial_p X \right) ^{2}
        )  \right) +
\left( z_{{r}}-1 \right)  \left( a_{{1}}+a_{{2}}z_ {{p}}
     \partial_p X \right) +\\ \nonumber
\gamma_{{r}} \left( 1-z_{{r}} \right)
  \left(\phantom {\frac{\sigma}{\sigma}}\hspace{-12pt}\right . b_{{1}}
  \partial_r X +b_{{2}}z_{{p}} \left(
   \partial_{rp}X +
  \partial_r X
      \partial_p X\right) +
    \\ \nonumber
 b_{{3}}{z_{{p}}}^{2} (
   \partial_{rpp}X+
     \partial_{pp}X
    \partial_r X+2\,
    \partial_p X
    \partial_{rp}X+ \left(
    \partial_p X\right) ^{2}
    \partial_r X )
    \left .\phantom {\frac{\sigma}{\sigma}}\hspace{-12pt}\right )
    +\\ \nonumber
Kz_{{r} }
  \left(\phantom {\frac{\sigma}{\sigma}}\hspace{-12pt}\right .
   b_{{1}} \left( z_{{p}}-1 \right)
  \partial_r X+b_{{2 }}z_{{p}}
  \partial_r X+b_{{2}}z_{{p}} \left( z_{{p}}-1
    \right)  \left(  \partial_{rp}X+ \partial_r X
    \partial_p X
    \right) +
    \\ \nonumber
2\,b_{{3}}{z_{{p} }}^{2} \left(
    \partial_{rp}X+ \left(
   \partial_r X \right)
    \partial_p X
    \right) +
    \\ \nonumber
b_{{3}}{z_{{p}}}^{2}
   \left( z_{{p}}-1 \right)  ( \partial_{rpp}X
    +  \partial_{pp}X
  \partial_r X+2\,
    \partial_p X
    \partial_{rp}X+ \left(
    \partial_p X \right) ^{2}
    \partial_r X )
   \left .\phantom {\frac{\sigma}{\sigma}}\hspace{-12pt}\right )
    +\\ \nonumber
\gamma_{{p}}
    \left( \phantom {\frac{\sigma}{\sigma}}\hspace{-12pt}\right .
     b_{{1}} \left( 1-z_{{p }} \right)
    \partial_p X-b_{{2}}z_{{p}}
    \partial_p X +b_{{2}}z_{{p}} \left( 1-z_{{p}}
    \right)  ( \partial_{pp}X+
    \left( \partial_p X \right) ^{2} )-\\ \nonumber
2\,b_ {{3}}{z_{{p}}}^{2} (
    \partial_{pp}X+ \left(
    \partial_p X\right) ^{2} )+\\ \nonumber
b_{{3}}{z_{{p}}}^{2}
    \left( 1-z_{{p}} \right)  ( \partial_{ppp}X+3\,
    \partial_{pp}X \partial_p X
    + \left( \partial_p X\right) ^{3}
    )
    \left .\phantom {\frac{\sigma}{\sigma}}\hspace{-12pt}\right )\;.
\end{eqnarray}

The time dependent variables are the factorial cumulants and can
be obtained from the Taylor expansion of $X(z,t)$ about $z=1$.
Thus, we obtain an infinite number of equations; the first three
of these are:
\begin{eqnarray}\label{CumulantsForFeedback}
  & &b_{{2}}\dot X_{{p}}   +b_{{3}} \left( \dot X_{{{\it pp}}}   +2\,X_{{p}}   \dot X_{{p}}
  \right)=K
  \left( \phantom {\frac{\sigma}{\sigma}}\hspace{-12pt}\right .
  b_{{2}}X_{{r}} +2\,b_{{3}} \left( X_{{{\it rp}}}   +X_{{r}}   X_{{p}}  \right)
  \left .\phantom {\frac{\sigma}{\sigma}}\hspace{-12pt}\right ) \\\nonumber
  & &-\gamma_{{p}} \left( b_{{2}}X_{{p}}
   +2\,b_{{3}} ( X_{{{\it pp}}}   +
 \left( X_{{p}}    \right) ^{2} )  \right)\;,\\\nonumber
 \\
  \nonumber
 & &b_{{1}}\dot X_{{r}}  +b_{{2}} \left( X_{{p}}
  \dot X_{{r}}  +\dot X_{{{\it rp}}}   \right) +b_{{3}} \left(
2\,X_{{p} }  \dot X_{{{\it rp}}}  +X_{ {{\it pp}}}  \dot X_{{r}}
 + 2\,X_{{{\it rp}}}  \dot X_{{p}}
 + \left( X_{{p}}   \right) ^{2}\dot X_{{r}}  +\dot X_{{{\it rpp}}}
 \right)=\\\nonumber
 & &G\left( t \right)  \left( b_{{1}}+b_{{2}}X_{{p}}  +b_
{{3}} ( X_{{{\it pp}}}  + \left( X_{{p}}   \right) ^{2} ) \right)
+a_{{1}}+a_{{2}}X_{{p}}\\ \nonumber
 & & -\gamma_{{r}} \left( b_{{1}}X_{{r}}  +b_{{2}} \left( X_{{{\it rp}}}  +X_{{r}}
  X_{{p}}   \right) +b_{{3}} ( X_
{{{\it rpp}}}  +X_{{{\it pp}}}  X_{{r} } +2\,X_{{p}}  X_{{{\it
rp}}}  + \left( X_{{p}}   \right) ^{2}X_{{r}}
   )  \right) \\ \nonumber
 & & + K \left( \phantom {\frac{\sigma}{\sigma}}\hspace{-12pt}\right .
 b_{{2}}(X_{{r}} +X_{{{\it rr}}})  +2\,b_{{3}} \left( X_{{{\it rp}}}  +X_{{r}}
  X_{{p}} + X_{{ {\it rrp}}}
+X_{{{\it rr}}}  X_{{p}}
  +X_{{r}}  X_{{{\it rp}}}   \right)
  \left .\phantom {\frac{\sigma}{\sigma}}\hspace{-12pt}\right ) \\ \nonumber
 & &-\gamma_{{p}}
 \left( \phantom {\frac{\sigma}{\sigma}}\hspace{-12pt}\right .
 b_{{2}}X_{{{\it rp}}}
  +2\,b_{{3}} \left( X_{{{\it rpp}}}
+2\,X_{{p}}  X_{{{\it rp}}}  \right) \left .\phantom
{\frac{\sigma}{\sigma}}\hspace{-12pt}\right )
 \;,\\ \nonumber
 \\ \nonumber
 \end{eqnarray}
 \begin{eqnarray}
 & &b_{{1}}\dot X_{{p}}  +b_{{2}}\dot
X_{{p}}  +b_{{2}} \left( \dot X_{{{\it pp}}}
  +X_{{p}}  \dot X_{{p}}
   \right) +2\,b_{{3}} \left( \dot X_{{{\it
pp}}}  +2\,X_{{p}}  \dot X_{ {p}}  \right) \\\nonumber & &+b_{{3}}
\left( \left( X_{{p}}   \right) ^{2}\dot X_{{p}}  +2\,X_{ {p}}
\dot X_{{{\it pp}}}  + \dot X_{{{\it ppp}}}  +3\,X_{{{\it pp}}}
  \dot X_{{p}}   \right)=\\ \nonumber
 & & K
\left( \phantom {\frac{\sigma}{\sigma}}\hspace{-12pt}\right .
 b_1X_r+b_2(X_r+2X_{rp}+X_rX_p)+b_3(4X_{rp}+4X_rX_p+4X_pX_{rp}+3X_{pp}X_r+3X_{rpp}+(X_p)^2X_r)
\left.\phantom {\frac{\sigma}{\sigma}}\hspace{-12pt}\right
)\\\nonumber & &
 -\gamma_{{p}}
\left( \phantom {\frac{\sigma}{\sigma}}\hspace{-12pt}\right .
b_1X_p+b_2(X_p+2X_{pp}+(X_p)^2)+
b_3(4X_{pp}+4(X_p)^2+3X_{ppp}+7X_{pp}X_p+(X_p)^3)
 \left .\phantom{\frac{\sigma}{\sigma}}\hspace{-12pt}\right )
 \;.
\end{eqnarray}
 If $a_1=0$ and $a_2=0$, then the nonlinear feedback
disappears and the system becomes linear. The equations then
factorize into the simple equations discussed before for System 1,
Fig. 3. When the feedback is nonzero, that is $a_1\neq 0$ and
$a_2\neq 0$, the equations does not factorize. The left side of
each equation in (\ref{CumulantsForFeedback}) is polynomial in the
time derivative ${\dot X}_m$ and $X_m$. This is characteristic for
the rational transition probabilities, as will be proven in
section 4.
 To assure that the transition probabilities are positive, we will
 work with a positive signal generator $G(t)>0$.
We will split the generator
 $G(t)$ into a constant component $G$ and a time variable component
 $g(t)$
  \begin{eqnarray}\label{Generator}
    G(t) &=& G+g(t)\,.
  \end{eqnarray}

 For the constant component we take $G=(G_{max}-G_{min})/2$
with $G_{max}$ and $G_{min}$ being the maximum and respectively
the minimum value of $G(t)$. With this choice for $G$ we have
$\left|g(t)\right|<G$, so solutions to the equations
(\ref{CumulantsForFeedback}) can be found by the method of
expansion with respect to a small parameter. Namely, insert a
parameter $\eta$ in
\begin{eqnarray}\label{GeneratorWithEta}
    G(t) &=& G+\eta g(t)\,,
  \end{eqnarray}
and generate approximations, $X_{m,k}$, by collecting the like
powers in $\eta$ :
\begin{eqnarray}\label{SolXVariational}
  X_m(t) &=& X_{m,0}+\eta X_{m,1}(t)+\eta^2 X_{m,2}(t)+\dots\;;
\end{eqnarray}

then eliminate $\eta$ by setting it to $1$.
 The constant term $G$ fixes a stationary level $X_{m,0}$ that obeys a system of
equations obtained from (\ref{CumulantsForFeedback}) by
eliminating any time derivative and putting $G$ instead of $g(t)$.
The solution to the stationary case is interesting from a
practical point of view and will explored it in the next
paragraph; afterwards, we will return to study $X_{m,1}(t)$.

\subsubsection{Designing the Shape of a Logic Pulse}

In electrical engineering systems, properly connecting equipment
along a signal path requires strict compliance with various
standards. The logic 1's and 0's must be designed in such a way
that they will be detected correctly  after passing through chains
of devices. A TTL device is guaranteed to interpret any input
above 2 volts as a logic 1 or true and any input below 0.8 volts
as a logic 0 or false; thus, there is a 1.2 volts protection
against noise.
  Translating these ideas to a molecular device, we want to use
  the autoregulatory system to generate a logic pulse in protein
  numbers. Then let $G_1$ be the input signal for a protein level
  that represents a logic 0 and $G_2$ for a logic 1.

\begin{figure}[H]
\centering
  \includegraphics[width=9cm]{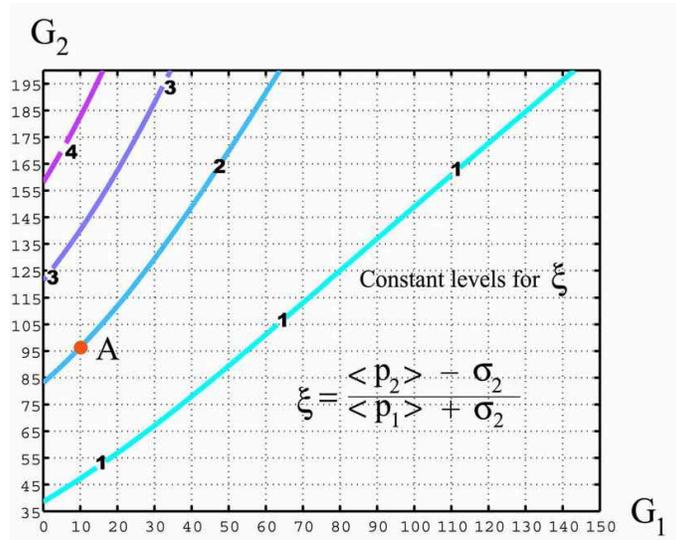}
  \caption{Constant Levels for $\xi$. }
\end{figure}
Because the
  system is intrinsically stochastic the logic levels refer to the
  mean values of the protein, $\langle p_1\rangle$ and $\langle p_2 \rangle$ respectively.
  The protein values will fluctuate around these mean values. A
  measure of this fluctuation is the standard deviation of the
  protein numbers $\sigma_1=\sqrt(\langle (p_1-\langle p_1\rangle)^2
\rangle)$ and similar for $\sigma_2$. To separate the logical
  levels we will ask that the following ratio:
  \begin{eqnarray}
    \xi &=& \frac{\langle p_2\rangle -\sigma _2}{\langle p_1\rangle +\sigma
    _1}
  \end{eqnarray}
be high enough. The constant contour plot of the ratio $\xi $ is
presented in Fig. 7 for the following set of numerical parameters
\begin{eqnarray}\label{NumericalParametersFeedback}
\left\{ \gamma_{{r}}=2,\gamma_{{p}}=1,K=
0.5,a_{{1}}=1,a_{{2}}=0,b_{{1}}= 0.01,b_{{2}}= 0.001,b_{{3}}=
0.001 \right\}\;.
\end{eqnarray}

\vspace{-10pt}
\begin{figure}[H]
\centering
  \includegraphics[width=11cm]{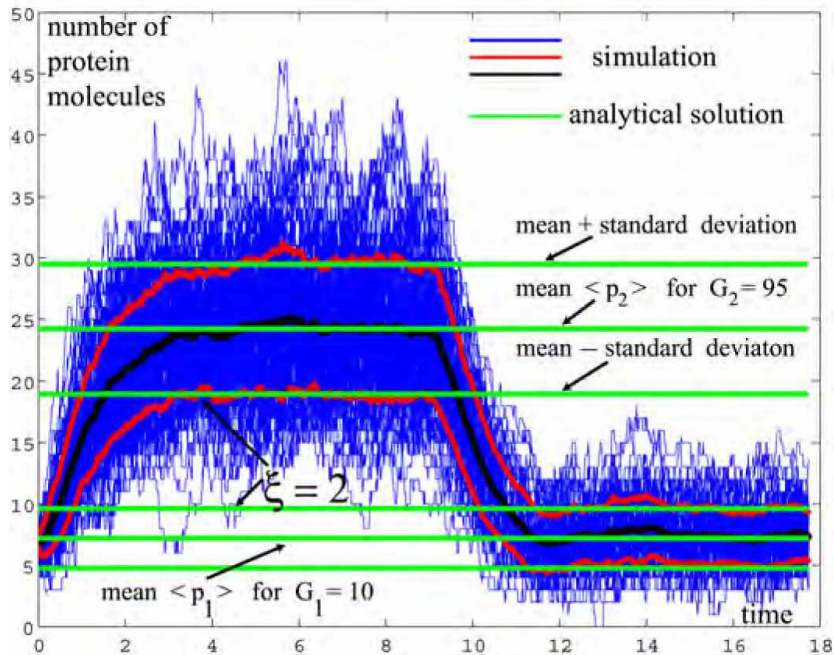}
  \caption{Pulse design.}
\end{figure}

 To design a shape for a pulse we first fix $G_1$ for the logic 0
and then choose a value for the ratio $\xi$. From Fig. 7 we read
the input $G_2$ for the logic 1. For example, if $G_1=10$ and
$\xi=2$ we obtain $G_2=95$, which is the point A on the graph. The
pulse for these values is shown in Fig. 8. This figure also shows
that the analytical values are confirmed by a Monte Carlo
simulation using the direct Gillespie algorithm, see Materials and
Methods. The simulated means and standard deviations are based on
500 independent stochastic processes. The pulse is separated from
the baseline level by a factor of $\xi=2$ and only a few of the
simulations drop down in the region $\langle p_1 \rangle
\pm\sigma_1$. The comparison of the analytic formulas with the
Monte Carlo results proves the power of the method outlined,
especially that the variables $X_m$ are well suited for numerical
computations. The next paragraph, and other examples that follow,
will show the effectiveness of the factorial cumulants.

\subsubsection{The Autoregulatory Gene Driven Only by its Protein
Level}

If $G=0$, the generator is closed, leaving only the feedback to
sustain the mRNA production. An equilibrium between mRNA and the
protein level will take place. Using the traditional method of
mass action (chemical equilibrium), we can compute this
equilibrium by equating the production rates with the degradation
rates:

\begin{eqnarray}\label{MassAction}
  \frac{a_1+a_2p}{b_1+b_2p+b_3p(p-1)} &=& \gamma_{r}r \;,\\
   k r &=& \gamma_{p}p\;.
\end{eqnarray}

 The mass action procedure assumes that the stochastic process is Poisson, so the size of the standard deviation from the mean
 equals the square root of the mean.
We found that the mass action procedure does not explain the data
obtained from Monte Carlo simulations, Fig. 9 and Table 3.
However, we match the simulations by solving the first $N$
equations of the infinite system of equations
(\ref{CumulantsForFeedback}). As $N$ increases, the solutions more
closely approach the simulated data; see Fig. 9 where $N=7$ and
$15$.

\begin{table}[H]
\tablenum{\large \bf {3}} \vspace{-2pt}
 \caption{${\mbox {\large \bf
\,\,\,Analytical Solutions Explain Monte Carlo Simulations}}$}
\begin{center}
\resizebox{!}{0.8cm}
 {\begin{tabular}{|c|c|c|c|c|}\hline
    &Monte Carlo & 15 equations & 7 equations & mass
    action\\\hline
   mean & 12.38 & 12.37 & 12.16 & 11.54  \\\hline
   mean + standard deviation & 17.55& 17.50 & 16.76 & 14.94 \\\hline
 \end{tabular}}
 \end{center}
 \end{table}
\vspace{-15pt}
\begin{figure}[H]
\centering
  \includegraphics[width=10cm]{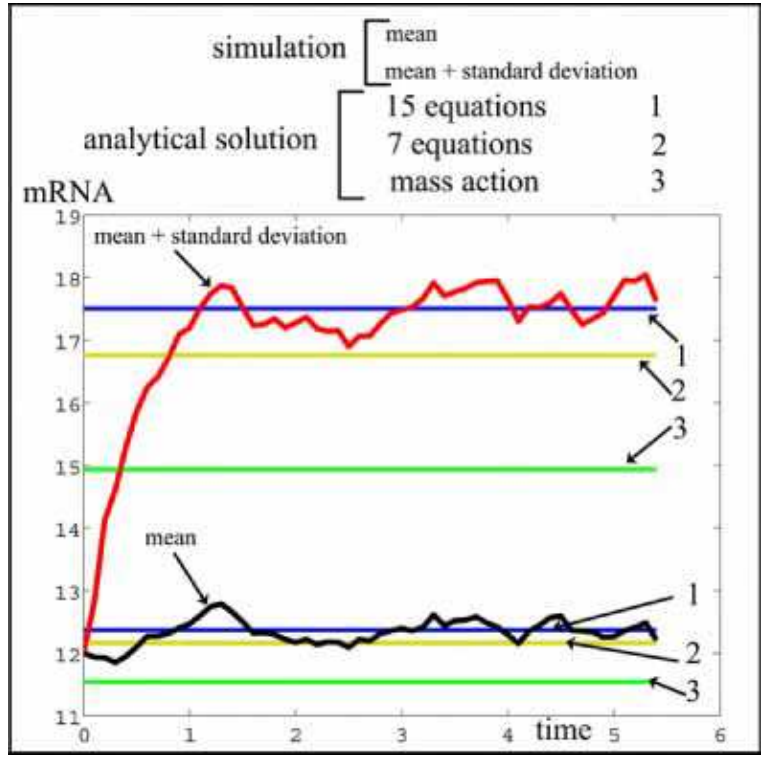}
  \caption{Analytical solution explaining the simulated data.}
\end{figure}

 As we take more equations we obtain better
results, so the  variables $X_m$ (factorial cumulants) are suited
for numerical and analytical approximations. The conclusion is
that the mean values depend on cumulants of higher orders and thus
equations that mix the means with cumulants of higher orders must
be solved simultaneously.

\subsubsection{Response of the Nonlinear Autoregulatory Gene to a
Time Variable Input Signal Generator}

The generator is now time dependent (\ref{GeneratorWithEta})

\begin{eqnarray}
  \nonumber G(t) &=& G+g(t)
\end{eqnarray}

The solution in the first order, (\ref{SolXVariational}), is
$X_m(t)=X_{m,0}+X_{m,1}(t)$. At this point we can move from tensor
notations to a matrix notation to present the solutions in the
usual form for input-output relations from control theory. We
construct thus column vectors $X_{(0)}$ and $X_{(1)}$ using the
lexicographic order for the tensor index $m$. For example
\begin{eqnarray}
  X_{(1)} &=&
  \left[X_{r,1},X_{p,1},X_{rr,1},X_{rp,1},X_{pp,1},\dots\right]^t\,.
\end{eqnarray}
which is written in a transposed form to save space.

Using (\ref{GeneratorWithEta}) and equating the terms which
contain $\eta $ from both sides of (\ref{CumulantsForFeedback}) we
obtain a linear time evolution equation for $X_{(1)}$

\begin{eqnarray}\label{EAB}
  E\dot X_{(1)}=A\,X_{(1)}+B\,g(t)\,,
\end{eqnarray}

where $E$ and $A$ are infinite square matrices and $B$ is an
infinite column vector. The entries of these matrices depend on
the parameters of the system as well as the stationary solution
vector $X_{(0)}$. From these infinite systems of equations we
construct a finite system with a dimension depending on how many
orders for the factorial cumulants $X_m$ we want to keep (that is
what is the maximum value for $\left | m\right |$ we need). To
obtain a nonsingular matrix $E$ the first equation in
(\ref{CumulantsForFeedback}) must be omitted. This equation is a
consequence of the transition probabilities being a rational
function. If the transition probabilities were only polynomials in
the state variables, then this first equation will become a
trivial $0=0$. A $2\times 2$ finite system will have the following
matrices:

\begin{eqnarray}
\nonumber  E=\left[ \begin {array}{cc}
b_1+b_2X_{p,0}+b_3X_{pp,0}+b_{{3}}{X_{{p,0}}}^{2}&2\,b_{{3}}X_{{rr,0}}\\\noalign{\medskip}0&b_{{1}}
+b_{{2}}+b_{{2}}X_{{p,0}}+4\,b_{{3}}X_{{p,0}}+3\,b_{{3}}X_{{pp,0}}+b_{{
3}}{X_{{p,0}}}^{2}\end {array} \right]
\end{eqnarray}

\begin{eqnarray}
 \nonumber A_{{1,1}}&=&-\gamma_{{r}} \left( b_{{1}}+b_{{2}}X_{{p,0}}+b_{{3}}
 \left( X_{{pp,0}}+{X_{{p,0}}}^{2} \right)  \right) +k \left( b_{{2}}+2
\,b_{{3}}X_{{p,0}} \right) +2\,kb_{{3}}X_{{rp,0}}\\
\nonumber
 A_{{1,2}}&=&G \left( b_{{2}}+2\,b_{{3}}X_{{p,0}} \right)
+a_{{2}}-\gamma _{{r}} \left( b_{{2}}X_{{r,0}}+b_{{3}} \left(
2\,X_{{rp,0}}+2\,X_{{r,0} }X_{{p,0}} \right)  \right)\\\nonumber &
&+2\,kb_{{3}}X_{{r,0}}+2\,kb_{{3}}X_{{rr,0}}
-4\,\gamma_{{p}}b_{{3}}X_{{rp,0}}
\\ \nonumber
A_{{2,1}}&=&k \left(
b_{{1}}+b_{{2}}+b_{{2}}X_{{p,0}}+4\,b_{{3}}X_{{p,0}
}+2\,b_{{3}}X_{{pp,0}}+b_{{3}} \left( X_{{pp,0}}+{X_{{p,0}}}^{2}
 \right)  \right)\\ \nonumber
 A_{{2,2}}&=&k \left( b_{{2}}X_{{r,0}}+4\,b_{{3}}X_{{r,0}}+2\,b_{{3}}X_{{
rp,0}}+b_{{3}} \left( 2\,X_{{rp,0}}+2\,X_{{r,0}}X_{{p,0}} \right)
 \right) -\\ \nonumber
 & &\gamma_{{p}} \left( b_{{1}}+b_{{2}}+2\,b_{{2}}X_{{p,0}}+8\,
b_{{3}}X_{{p,0}}+4\,b_{{3}}X_{{pp,0}}+b_{{3}} \left(
3\,X_{{pp,0}}+3\,{X _{{p,0}}}^{2} \right)  \right)
\end{eqnarray}

\begin{eqnarray}
 \nonumber  {\it B}= \left[ \begin {array}{c}    b_{{1}}+b_{{2}}X_{{p,0}}+b_{{3}} \left( X_{{pp,0}}+{X_
{{p,0}}}^{2} \right)   \\\noalign{\medskip}0\end {array}
 \right]\;.
\end{eqnarray}

However, finite systems of larger dimensions are needed to obtain
accurate solutions for the mean and standard deviations of the
molecule numbers. Large systems of equations are easily generated
with symbolic software like Maple or Mathematica; however, these
equations are too large to be displayed within the article.
However, using numerical values
(\ref{NumericalParametersFeedback}) and $G=30$ we can go further
and display the final results. The stationary point solution up to
the second order
\begin{eqnarray}
  X_{{r,0}}= 20.251,X_{{p,0}}= 10.125,X_{{rr,0}}= 0.909,X_{{rp,0}}=- 0.935,X_{{pp,0}
}=- 0.468
\end{eqnarray}
shows that the mean value for mRNA is about 20 molecules and the
protein number is about 10. The factorial cumulants of higher
order have absolute values smaller that the corresponding
factorial moments. For example $F_{rr,0}$ would be of order
$20(20-1)=380$ whereas $X_{rr,0}$ is about 1.

In the spirit of control theory, the solution to (\ref{EAB}) can
be written as an input-output relation using the Laplace transform
of the $X_{m,1}$ variables:

\begin{eqnarray}\label{FeddbackLaplace}
  X_{p,1}(s)&=&\,{\frac { 0.50\left( s+ 5.93 \right)  \left( s+ 2.50 \right)  \left( {s}^{2}+ 5.94\,s+ 11.9 \right) }{ \left( {s}^{2}+ 3.0\,s+ 2.62
 \right)  \left( {s}^{2}+ 4.14\,s+ 6.87 \right)  \left( {s}^{2}+ 10.2
\,s+ 29.0 \right) }}\, g(s)
\\\nonumber
  X_{pp,1}(s)&=&\,{\frac { 0.0075\left( s+ 65.9 \right)  \left( s+ 2.04 \right)  \left( {s}^{2}+ 10.2\,s+ 29.8 \right) }{ \left( {s}^{2}+ 3.0\,s+ 2.62
 \right)  \left( {s}^{2}+ 4.14\,s+ 6.87 \right)  \left( {s}^{2}+ 10.2
\,s+ 29.0 \right) }}
 \, g(s)
\end{eqnarray}

Thus the mean and the fluctuation of the protein number can be
directly related with the input signal $g(s)$ which is the Laplace
transform of $g(t)$.

\subsection{ Michaelis-Menten  Amplifier}

Catalytic enzymatic processes, like phosphorylation, are
fundamental for biological processes. The process requires a
substrate S reacting with an enzyme E to form a complex C which in
turn is converted into a product P and the enzyme E, Fig. 10. In a
test tube, the reaction proceeds in one direction, that is
$k_{-1}=0$ and $k_{-2}=0$ which is a special case of Fig. 10.
However, it is possible that in a cell a more general scheme where
$k_{-1}\neq0$ and $k_{-2}\neq0$ can take place \cite{Harvard};
thus we study the case in Fig. 10. The substrate S is usually
supplied in large quantities compared with the enzyme E, and the
goal of the process is to transform the substrate S into the
product P. We choose an input oscillatory signal generator to act
on the enzyme E. Then, we follow the signal through the complex C
to the output product P.
 It is possible to drive large oscillations in the product P using small oscillations in the enzyme E.
In this case the catalytic process behaves like a molecular
amplifier. This situation is analogous with how a transistor
amplifies the input signal on its base. A constant voltage source
is necessary to supply the energy for the electrical
amplification. Here the role of the source is played by the
substrate S, the signal in the transistor's base by the enzyme E
and the output signal from the transistor's collector by the
product P. The state is $q=(E,S,C,P)$ and the transition
probabilities are polynomials in the state variables, Table 4. The
molecular diagram, Fig. 11, depicts all possible transitions in
the system and which variables control these transitions.

\
\begin{figure}[H]
\centering
  \includegraphics[width=9cm]{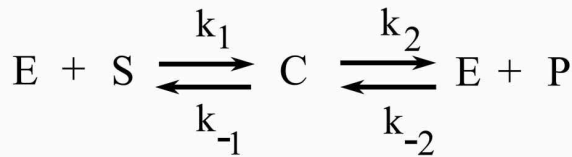}
  \caption{Catalytic reaction.}
\end{figure}
 \vspace{10pt}

\begin{table}[h]
\tablenum{\bf {\large 4}} \caption{${\mbox {\large \bf
\,\,\,Michaelis-Menten Process}}$}
\begin{center}
\resizebox{!}{1.6cm}
 {\begin{tabular}{|c|c|c|c|} \hline
  $\phantom{\frac{\sigma_D^Y}{\sigma_D^Y}}\epsilon_1 = (1, 0, 0, 0)\phantom{\frac{\sigma_D^Y}{\sigma_D^Y}}$ & $T_1 = G(t)$ & $\phantom{\frac{\sigma_D^Y}{\sigma_D^Y}}\epsilon_3 = (-1,-1,1,0)\phantom{\frac{\sigma_D^Y}{\sigma_D^Y}}$ & $T_3 = k_1 E\cdot S$ \\\hline
  $\phantom{\frac{\sigma_D^Y}{\sigma_D^Y}}\epsilon_{-1} = (-1, 0, 0, 0)\phantom{\frac{\sigma_D^Y}{\sigma_D^Y}}$ & $\phantom{\frac{\sigma_D^Y}{\sigma_D^Y}}\hspace{3pt}T_{-1} = \gamma_E E\phantom{\frac{\sigma_D^Y}{\sigma_D^Y}}$ & $\phantom{\frac{\sigma_D^Y}{\sigma_D^Y}}\epsilon_{-3} = (1, 1, -1,0)\phantom{\frac{\sigma_D^Y}{\sigma_D^Y}}$ & $T_{-3} = k_{-1} C$ \\\hline
  $\phantom{\frac{\sigma_D^Y}{\sigma_D^Y}}\epsilon_2 = (0, 1, 0, 0)\phantom{\frac{\sigma_D^Y}{\sigma_D^Y}}$  & $T_2 =K_S $& $\phantom{\frac{\sigma_D^Y}{\sigma_D^Y}}\epsilon_{4}= (1, 0, -1, 1)\phantom{\frac{\sigma_D^Y}{\sigma_D^Y}} $ & $T_4 = k_2 C $\\\hline
  $\phantom{\frac{\sigma_D^Y}{\sigma_D^Y}}\epsilon_{-2} = (0, -1, 0, 0)\phantom{\frac{\sigma_D^Y}{\sigma_D^Y}}$ & $T_{-2} = \gamma_S S $& $\phantom{\frac{\sigma_D^Y}{\sigma_D^Y}}\epsilon_{-4} = (-1, 0, 1, -1)\phantom{\frac{\sigma_D^Y}{\sigma_D^Y}} $& $\phantom{\frac{\sigma_D^Y}{\sigma_D^Y}}\hspace{3pt}T_{-4}=k_{-2}E\cdot P \phantom{\frac{\sigma_D^Y}{\sigma_D^Y}}$\\\hline
  $\phantom{\frac{\sigma_D^Y}{\sigma_D^Y}}\epsilon_{-5} = (0, 0, 0, -1)\phantom{\frac{\sigma_D^Y}{\sigma_D^Y}}$ & $T_{-5} = \gamma_P P$& &\\\hline
 \end{tabular}}
\end{center}
\end{table}

\begin{figure}[t]
\centering
  \includegraphics[height=13cm]{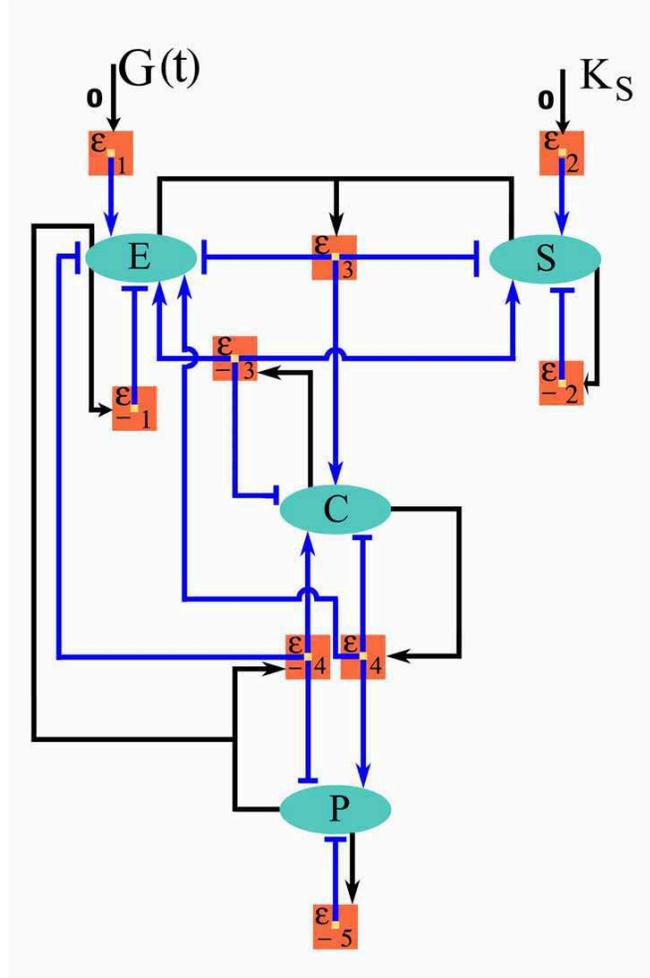}
  \caption{Molecular diagram for Michaelis-Menten process.}
\end{figure}

The transition probabilities will generate a stochastic process
described by the time evolution of its factorial cumulants:
\begin{eqnarray}
  \nonumber \partial_tX &=& k_1(z_C-z_Ez_S)\left(\partial_{z_Ez_S} X +\partial_{z_E}X\partial_{z_S} X \right
)+k_{-1}(z_Ez_S-z_C)\partial_{z_C}X+k_2(z_Ez_P-z_C)\partial_{z_C}X \\
  \nonumber & & +G(t)(z_E-1)+\eta_E(1-z_E)\partial_{z_E}X+\eta_S(1-z_S)\partial_{z_S}X +\eta_P(1-z_P)\partial_{z_P}X\\
  \nonumber & & +k_{-2}(z_C-z_Ez_P)\left( \partial_{z_Ez_P} X +\partial_{z_E}X\partial_{z_P} X
\right)+ K_S(z_S-1)\,.
\end{eqnarray}

The term $G(t)(z_E-1)$ represents the time variable input
generator which acts on the enzyme $E$, whereas  $K_S(z_S-1)$
comes from the constant production of the source $S$. An advantage
of the variable $X(z,t)$ is that the generator terms do not
contain the variable $X(z,t)$. This will translate into equations
with constant coefficients for the $X_m$ variables:
\begin{eqnarray}\label{MMDynamicalEquations}
  \dot {X}_E  &=&-k_1 \left( X_{ES}+X_E  X_S
      \right) +k_{-1}X_C  +k_2X_C +G(t) -\gamma_{{E}}X_E
      \\ \nonumber & &-k_{-2} \left( X_{EP}  +X_EX_P   \right)\;, \\
  \nonumber \dot{ X_{S}}  &=&-k_1 \left( X_{ES} +X_E  X_S \right)
      +k_{-1}X_C -\gamma_{{S}}X_S  +K_{{S}}\;,
\end{eqnarray}
\begin{eqnarray}\nonumber
  \nonumber \dot{X_C}
      &=& k_1 \left( X_{ES} +X_E  X_S  \right) -k_{-1}X_C
      -k_2X_C
      +k_ {{-2}} \left( X_{EP}  +X_E  X_P  \right) ,\\
\nonumber \dot{X}_P  &=& k_2X_C -\gamma_{{P}}X_P  -
      k_{-2} \left( X_{EP}  +X_E  X_P  \right) ,
\\
\nonumber\dot{X}_{EE}&=& -2\,k_1 \left( X_{EE}  X_S +X_E X_{ES}
      \right) + 2\,k_{-1}X_{EC}  +2\,k_2X_{EC}-2\,\gamma_{{E}}X_{EE}\\
      \nonumber & & -2\,k_{-2} \left( X_{EE} X_P
      +X_E  X_{EP}   \right) ,\\
\nonumber \dot{X}_{ES} &=& -k_1 \left( X_{ES}  +X_E
      X_S   \right) -k_1 \left( X_{ES}  X_S  +X_E X_{SS}   \right) -k_1 \left( X_{EE}
      X_S  +X_E
      X_{ES}   \right)\\ \nonumber & & +k_{-1}X_C  +K_
      {{-1}}X_{SC}  +k_{-1}X_{EC}  \\
      \nonumber & &+k_2X_{SC}  -\gamma_{{E}}X_{ES}  - \gamma_{{S}}X_{ES}
      -k_{-2} \left( X_{ES}X_P  +X_E  X_{SP}\right)\;, \\
\nonumber \dot{X}_{EC}  &=& -
      k_1 \left( X_{EC}  X_S  +X_E  X_{SC} \right) +k_1
      \left( X_{EE}  X_S  +X_E
      X_{ES}   \right) \\ \nonumber & &+k_{-1}X_{CC}
      -k_{-1}X_{EC}  +k_2X_{CC}
      -k_2X_{EC}  \\ \nonumber & &-\gamma_{{E}}X_{EC}
      -k_{-2} \left( X_{EC}  X_P
      +X_E  X_{CP}
      \right) +k_{-2} \left( X_{EE}  X_P +X_E  X_{EP}   \right)
      ,\\\nonumber
\dot{X}_{EP}  &=& -k_1 \left( X_{EP}X_S  +X_E  X_{SP} \right)
+k_{-1}X_{CP}  + k_2X_C +k_2X_{CP}
      +k_2X_{EC}  -\gamma_{{E}}X_{EP}  \\
      \nonumber & &-\gamma _{{P}}X_{EP}  -k_{-2} \left( X_{EP}+X_E X_P
      \right) - k_{-2} \left( X_{EP}  X_P  +X_E X_{PP} \right)
      \\ \nonumber & &-k_{-2}\left( X_{EE}  X_P  +X_E
      X_{EP}   \right) , \\
\nonumber \dot{X}_{SS}  &=&-2\,k_1 \left( X_{ES}  X_S  +X_E
      X_{SS} \right) +2\,k_{-1}X_{SC}  -2\,\gamma_{{S}} X_{SS}  , \\
\nonumber \dot{X}_{SC}  &=&-K _{{1}} \left( X_{EC} X_S +X_E
      X_{SC} \right) +k_1\left( X_{ES}  X_S  +X_E
      X_{SS}   \right) +k_{-1}X_{CC}\\ \nonumber & & -k_{-1}X_{SC}  -k_2X_{SC}
      -\gamma_{{S}}X_{SC}  +k_{-2}\left( X_{ES}  X_P  +X_E
      X_{SP}   \right) , \\
\nonumber \dot{X}_{SP}  &=&-k_1 \left( X_{EP}  X_S  +X_E
      X_{SP} \right) +k_{-1}X_{CP}  +k_2X_{SC}
      -\gamma_{{S}}X_{SP}  -\gamma_{{P}}X_{SP}  \\ \nonumber & &-k_{-2} \left( X_{ES} X_P  +X_E  X_{SP}
      \right) , \\
\nonumber \dot{X}_{CC}  &=&2\,k_{{1} } \left( X_{EC}
      X_S  +X_E X_{SC}   \right) -2\,k_{-1}X_{CC}  \\ \nonumber &
      &-2\,k_2X_{CC}  +2\,k_{-2} \left( X_{EC}X_P  +X_EX_{CP}   \right) , \\
\nonumber \dot{X}_{CP}  &=&k_1 \left( X_{EP}  X_S  +X_E
      X_{SP} \right) -k_{-1}X_{CP}  -k_2X_{CP}
      +k_2X_{CC}  \\ \nonumber & &-\gamma_{{P}}X_{CP}  +k_{-2} \left( X_{EP}  X_P
      +X_E  X_{PP}\right) -k_{-2} \left( X_{EC}  X_P+X_E  X_{CP}   \right) , \\
\nonumber \dot{X}_{PP}
      &=&2\,k_2X_{CP}-2\,\gamma_{{P}}X_{PP} -2\,k_{-2} \left(
      X_{EP} X_P  +X_E X_{PP} \right)\;.
\end{eqnarray}

 The generator
$G(t)=G+g(t)$ will determine a stationary state by $G$ and a time
variation by $g(t)$. For an oscillatory input of the form
\begin{eqnarray}
  G(t) &=& G+G \cos(\omega t)
\end{eqnarray}
and for the following numerical coefficients

$\{\gamma_{{P}}=1,K_{{-2}}=1,K_{{1}}=
  0.1,K_{{2}}=6,K_{{S}}=50,
  G=300,\gamma_{{S}}= 0.003,K_{{-1}}= 0.1,\gamma_{{E}}=50\}\;,$

\noindent the stationary state is

$\{X_{E,0} = 6.00, X_{S,0} = 92.71,X_{C,0} = 58.14,X_{P,0} =
49.72, X_{EE,0} = -0.04,  X_{ES,0} = -.88, X_{EC,0} = 0.04,
X_{EP,0} = 0.76, X_{SS,0} = 12.28, X_{SC,0} = -7.82, X_{SP,0} =
-1.23, X_{CS,0} = 3.10, X_{CP,0} = 3.60, X_{PP,0} = -2.33\}$.

The stationary state was solved up to the second order cumulants
using the equations (\ref{MMDynamicalEquations}). In the first
order, the time variation of the $X_m$ variables is
$X_m(t)=X_{m,0}+X_{m,1}(t)$, (\ref{SolXVariational}). Each
variable $X_{m,1}(t)$ will contain an oscillatory component
$A_m(\omega) e^{i\omega t}$. As a function of $\omega$, and for
the parameters considered above, the ratio between the amplitude
of the protein oscillations and the amplitude of the enzyme
oscillations is presented in Fig. 12. The maximum of this ratio is
$3.71$ at $\omega=3$.
\begin{figure}[H]
\centering
  \includegraphics[height=7cm]{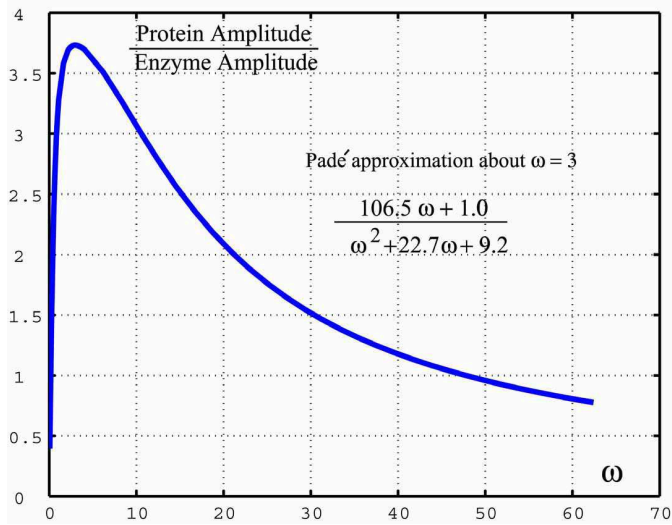}
  \caption{Protein amplification, $\frac{\left |A_P(\omega)\right|}{\left |A_E(\omega)\right|}$.}
\end{figure}

 A Pad\'{e} approximation about $\omega=3$ for the amplitudes' ratio
 is

\begin{eqnarray}\label{Amplification}
  \frac{\left |A_P(\omega)\right|}{\left |A_E(\omega)\right|}\approx{\frac { 106.5\,\omega+ 1.0}{{\omega}^{2}+ 22.7\,\omega+
  9.2}}\;.
\end{eqnarray}

The frequency $\omega $ is measured in $[time]^{-1}$, see section
{\bf {C}}. Thus, the enzyme oscillations are amplified as they
pass to the product P, Fig. 12.
  So far, we analyzed stationary solutions and stationary periodic
  regimes. Another question to address is how the system behaves
  in a transitory regime. We did 500  Monte Carlo simulations for the system that starts at the time $t=0$ from the
  zero initial conditions (all molecule numbers are zero). As the
  energy is pumped into the system by the signal generator
  $G(t)=G+G \cos(\omega t)$, the molecule numbers will grow
  towards a stable periodic state. The transitory process is an oscillatory
  variation superimposed on a growing exponential trend, Fig. 13.
  The numerical solution of the equations
  (\ref{MMDynamicalEquations}),
  show that the simulated data are explained by the dynamical
  equations. The error measured by the $L_2$ norm of
  the difference between the simulated data and the computed data
  is $1\% $ of the norm of the simulated data. Thus we can work with numerical solutions to (\ref{MMDynamicalEquations})
instead of using Monte Carlo simulations. Moreover, on
(\ref{MMDynamicalEquations}) we can apply a different analytical
approximation (i.e. harmonic balance, expansion in a small
parameter) which can capture the behavior of the system as a
function of its parameters. We observe, Fig. 13, that the product
oscillates in antiphase with respect to the enzyme, a phenomenon
also present also in a basic transistor amplifier. \vspace{10pt}
\begin{figure}[h]
\centering
  \includegraphics[height=13cm]{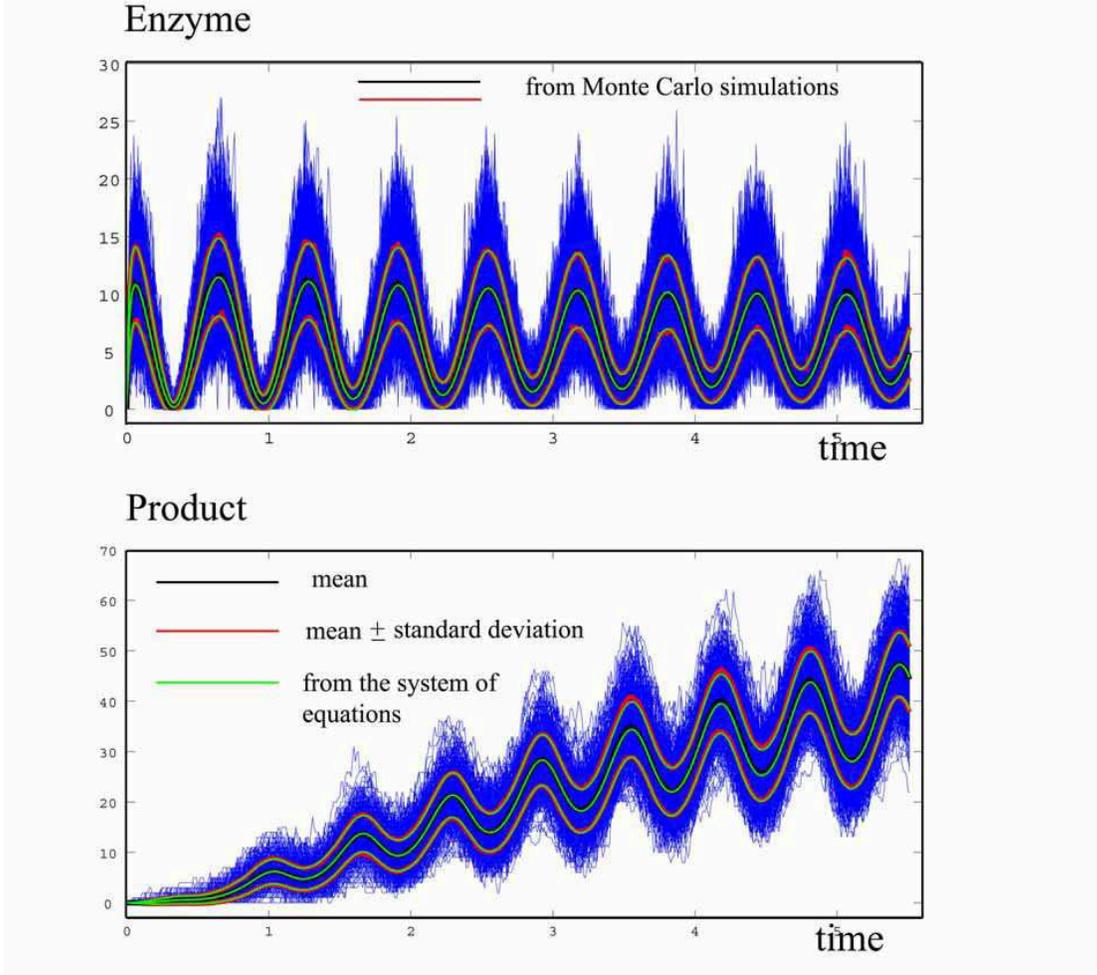}
  \caption{Transitory regime: Numerical solutions of the equations
  (\ref{MMDynamicalEquations}) agree with Monte Carlo simulations.}
\end{figure}

\subsection{E2F1 Regulatory Element}

The systems studied in the preceding paragraphs were excited by
one signal generator. There are many cases in living organisms
where a gene is regulated by more than one signal. In what follows
we will study a regulatory element inspired by the E2F1, a member
of the E2F family of transcription factors \cite{E2F1Paper1}. In
addition to its established proliferative effect, E2F1 has also
been implicated in the induction of apoptosis through
p53-dependent and p53-independent pathways \cite{E2F1Paper2}. The
components of the E2F1 system were described in \cite{Kohn}. The
mRNA level, Fig.14, is regulated by 3 transcription factors E2F1,
pRb and DP1. The regulation is done by the dimer E2F1:DP1 which,
for short, is denoted by the letter $a$. This dimer binds to the
DNA and its effect is to increase the rate of transcription. The
second control of transcription is from the complex a:pRB, that
binds to the DNA and repress the transcription. The state of the
system is thus an 8 component vector $q=(E2F1,
DP1,pRB,a,b,c,d,mRNA)$. This example shows that a state is not
just a list of different species of molecules. The same dimer
E2F1:DP1 is present in the state vector as a component $a$ unbound
to DNA and a state component $c$ when the dimer is bound to the
DNA. These are 2 different situations of the same molecular
species and we treat them as different components of the state of
the system. In the diagram of Fig. 14, we use an oval glued to a
rectangle to graphically depict the DNA-bound form of a
transcription factor. The transition probability that controls the
mRNA production is $T_{\epsilon_8}=k_8(l_1-l_2d)$, Table 5. The
transition $\epsilon_8$ is upregulated by $c$ so $T_{\epsilon_8}$
is proportional to $c$. The repression by $d$ is described by the
term $l_1-l_2 d$ chosen to obey two criteria: (1) it  decreases as
$d$ increases,and (2) is positive so the transition probability
will be a positive number. A specified set of parameters together
with a solution to the equation of motions are realistic if the
transition probabilities stay positive or zero for any instant of
time. One control line starting from $c$ ends with an arrow on the
transition $\epsilon_8$ which signifies that $c$ upregulates the
transcription. This control line corresponds to the term $k_8 l_1
c$ from $T_{\epsilon_8}$. Another control line starts from $c$ and
$d$ and ends in a short bar, thus repressing the mRNA
transcription. This line represents the term $-k_8 l_2 c d$ from
$T_{\epsilon_8}$. Variations in the number of $c$ and $d$
molecules will depend on variations in E2F1, DP1 and pRB. Thus we
will insert 3 signal generators $G_1(t)$, $G_2(t)$ and $G_3(t)$ to
modulate the levels of E2F1, DP1 and pRB respectively. The
transitions $\epsilon_i$, with $i=1,2,3$, represent these input
generators. The degradation transitions $\epsilon_{-i},$
$\,i=1,2,3$, are taken to be proportional with the number of
molecules being degraded. The formation of the dimer $a$ is
described by $\epsilon_4$ which is controlled by a nonlinear
transition probability (the product of E2F1 with DP1). The
transition $\epsilon_5$ represents the binding of the dimer $a$ to
DNA. That is the creation of one $c$ from one $a$, as is readable
from the components of the transition $\epsilon_5$. The transition
probability for this binding event is proportional to the number
of $a$ dimers and with the free space available on the DNA. This
free space available for DNA binding should be of the form $n-c-d$
where $n$ is the maximum number of proteins that can bind to DNA
to regulate the transcription. We subtract from $n$ the space
already occupied which is $c+d$. In order to cover the situation
when the binding properties of $c$ and $d$ are different we use
$T_{\epsilon_5}=k_5a(n_1-n_2c-n_3d)$, with $n_1,n_2$ and $n_3$
some constant coefficients. Then, the transition  $\epsilon_6$ is
like $\epsilon_4$ and $\epsilon_7$ like $\epsilon_5$. The
transitions $\epsilon_{-i}$, $\, i=4,5,6,7,$ represent reverse
processes. The equations are more compactly written if we index
the state by integer numbers:
$q_1=E2F1,q_2=DP1,q_3=pRb,q_4=a,q_5=b,q_6=c,q_7=d,q_8=mRNA$. The
time evolution for $X(z,t)$ is given by:
\begin{table}
\tablenum{\large \bf {5}}
 \caption{${\mbox {\large \bf \,\,\,E2F1
Regulatory Element}}$}\vspace{-10pt}
\begin{center}
\resizebox{!}{2.4cm}{\begin{tabular}{|c|c|c|c|}
  \hline
  $\phantom{\frac{\sigma_D^Y}{\sigma_D^Y}}\epsilon_{{1}}=(1,0,0,0,0,0,0,0)\phantom{\frac{\sigma_D^Y}{\sigma_D^Y}}$ & $T_{\epsilon_{1}}=G_{{1}}(t) $ & $\phantom{\frac{\sigma_D^Y}{\sigma_D^Y}}\epsilon_{{-1}}=(-1,0,0,0,0,0,0,0)\phantom{\frac{\sigma_D^Y}{\sigma_D^Y}}$ & $\phantom{\frac{\sigma_D^Y}{\sigma_D^Y}}T_{{\epsilon_{-1}}}=k_{{-1}}\,{ E2F1}\phantom{\frac{\sigma_D^Y}{\sigma_D^Y}}$
  \\\hline
  $\phantom{\frac{\sigma_D^Y}{\sigma_D^Y}}\epsilon_{{2}}=(0,1,0,0,0,0,0,0)\phantom{\frac{\sigma_D^Y}{\sigma_D^Y}}$ & $T_{\epsilon_{2}}=G_2(t) $ & $\phantom{\frac{\sigma_D^Y}{\sigma_D^Y}}\epsilon_{{-2}}=(0,-1,0,0,0,0,0,0)\phantom{\frac{\sigma_D^Y}{\sigma_D^Y}}$ & $T_{{\epsilon_{-2}}}=k_{{-2}}\,DP1$ \\\hline
  $\phantom{\frac{\sigma_D^Y}{\sigma_D^Y}}\epsilon_{{3}}=(0,0,1,0,0,0,0,0)\phantom{\frac{\sigma_D^Y}{\sigma_D^Y}}$ & $T_{\epsilon_{3}}=G_3(t)$ & $\phantom{\frac{\sigma_D^Y}{\sigma_D^Y}}\epsilon_{{-3}}=(0,0,-1,0,0,0,0,0)\phantom{\frac{\sigma_D^Y}{\sigma_D^Y}}$ & $T_{{\epsilon_{-3}}}=k_{{-3}}\,pRb$ \\\hline
  $\phantom{\frac{\sigma_D^Y}{\sigma_D^Y}}\epsilon_{{4}}=(-1,-1,0,1,0,0,0,0)\phantom{\frac{\sigma_D^Y}{\sigma_D^Y}}$ & $T_{\epsilon_4}=k_{{4}}{ E2F1}\cdot{DP1}$ & $\phantom{\frac{\sigma_D^Y}{\sigma_D^Y}}\epsilon_{{-4}}=((1,1,0,-1,0,0,0,0)\phantom{\frac{\sigma_D^Y}{\sigma_D^Y}}$ & $T_{\epsilon_{-4}}=k_{{-4}}a$ \\\hline
  $\phantom{\frac{\sigma_D^Y}{\sigma_D^Y}}\epsilon_{{5}}=(0,0,0,-1,0,1,0,0)\phantom{\frac{\sigma_D^Y}{\sigma_D^Y}}$ & $T_{\epsilon_{5}}=k_{{5}}a \left( n_1-n_2 c- n_3 d\right)$ & $\phantom{\frac{\sigma_D^Y}{\sigma_D^Y}}\epsilon_{{-5}}=(0,0,0,1,0,-1,0,0)\phantom{\frac{\sigma_D^Y}{\sigma_D^Y}}$ & $T_{\epsilon_{-5}}=k_{-5}c$ \\\hline
  $\phantom{\frac{\sigma_D^Y}{\sigma_D^Y}}\epsilon_{{6}}=(0,0,-1,-1,1,0,0,0)\phantom{\frac{\sigma_D^Y}{\sigma_D^Y}}$ & $T_{\epsilon_6}=k_6\,a\cdot pRb$ & $\phantom{\frac{\sigma_D^Y}{\sigma_D^Y}}\epsilon_{{-6}}=(0,0,1,1,-1,0,0,0)\phantom{\frac{\sigma_D^Y}{\sigma_D^Y}}$ & $T_{\epsilon_{-6}}=k_{-6}b$ \\\hline
  $\phantom{\frac{\sigma_D^Y}{\sigma_D^Y}}\epsilon_{{7}}=(0,0,0,0,-1,0,1,0)\phantom{\frac{\sigma_D^Y}{\sigma_D^Y}}$ & $\phantom{\frac{\sigma_D^Y}{\sigma_D^Y}}T_{\epsilon_7}=k_7\,b\,( m_1-m_2 c- m_3 d)\phantom{\frac{\sigma_D^Y}{\sigma_D^Y}}$ & $\phantom{\frac{\sigma_D^Y}{\sigma_D^Y}}\epsilon_{{-7}}=(0,0,0,0,1,0,-1,0)\phantom{\frac{\sigma_D^Y}{\sigma_D^Y}}$ & $T{\epsilon_{-7}}=k_{-7}\,d$ \\\hline
  $\phantom{\frac{\sigma_D^Y}{\sigma_D^Y}}\epsilon_{{8}}=(0,0,0,0,0,0,0,1)\phantom{\frac{\sigma_D^Y}{\sigma_D^Y}}$ & $T_{\epsilon_8}=k_8 c(l_1-l_2\,d)$ & $\phantom{\frac{\sigma_D^Y}{\sigma_D^Y}}\epsilon_{{-8}}=(0,0,0,0,0,0,0,-1)\phantom{\frac{\sigma_D^Y}{\sigma_D^Y}}$ & $T_{\epsilon_{-8}}=k_{-8}r$\\
  \hline
\end{tabular}}
\end{center}
\end{table}
\begin{figure}[H]
\centering
  \includegraphics[width=18cm]{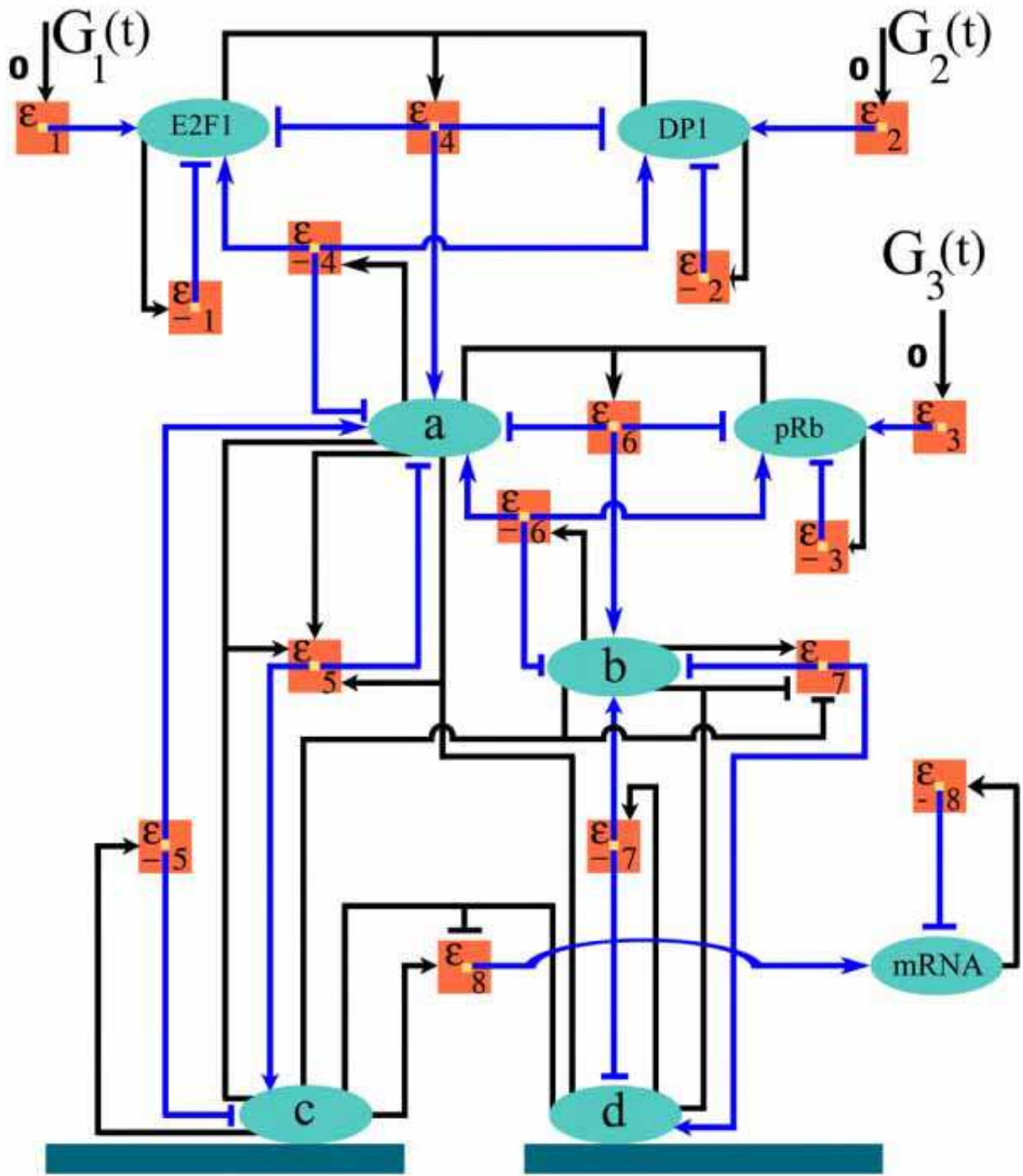}
 \vspace{-100pt} \caption{Molecular diagram for E2F1 regulatory element.}
\end{figure}

\begin{eqnarray}
  \partial_t X  &=&  {\it G_1} \left( t \right)   \left( z_{{1}}-1 \right) +
  k_{{-1}} \left( 1-z_{{1}} \right)\partial_1 X
  \\ \nonumber
&+&  {\it G_2} \left( t
  \right)    \left( z_{{2}}-1 \right)
  +k_{{-2}} \left( 1-z_{{2}}
  \right) \partial_2 X \\ \nonumber
&+&  {\it G_3} \left( t \right)    \left(
  z_{{3}}-1 \right)+k_{{-3}} \left( 1-z_{{3}} \right) \partial_3 X
  \\ \nonumber
&+&k_{{4}} \left( z_{{4}}-z_{{1}}z_{{2}} \right)  \left(
  \partial_{12}X+
  \partial_1 X
  \partial_2 X \right)+k_{{-4}} \left( z_{{1}}z_{{2}}-z_{{4}} \right) \partial_4 X
  \\ \nonumber
&+&k_{{5}} \left( z_{{6}}-z_{{4}} \right)  \left( n_{{1}}
  \partial_4 X -n_{{2}}z_{{6}} \left(
  \partial_{46} X +
  \partial_4 X
  \partial_6 X  \right) -n_{{3}}z_{{ 7}}
  \left( \partial_{47}X
  + \partial_4 X
  \partial_7 X
  \right)  \right)\\ \nonumber
&+&k_{{-5}} \left( z_{{4}}-z_{{6}} \right) \partial_6 X
  \\ \nonumber
&+&k_{{6}} \left(
  z_{{5}}-z_{{3}}z_{{4}} \right)  \left(
  \partial_{34}X +
  \partial_3 X
  \partial_4 X \right)+k_{{-6}} \left( z_{{3}}z_{{4}}-z_{{5}} \right)
  \partial_5X
  \\ \nonumber
&+&k_{{7}} \left( z_{{7}}-z_{{5}} \right)  \left( m_{{1}}
  \partial_5 X  -m_{{2}}z_{{6}} \left( {
  \partial_{56}X} +
  \partial_5 X
  \partial_6 X  \right) -m_{{3}}z_{{ 7}}
  \left( \partial_{57}X
  +  \partial_5 X
  \partial_7 X
  \right)  \right)\\ \nonumber
&+&k_{{-7}} \left( z_{{5}}-z_{{7}} \right)\partial_7X
  \\ \nonumber
&+&k_{{8}} \left( z_{{6}}z_{{8}}-z_{{6}} \right)  \left(
   l_{{1}}
   \partial_6 X -l_{{2}}z_{{7}} \left(
   \partial_{67}X +
   \partial_6 X
   \partial_7 X   \right)  \right)\\ \nonumber
 &+& k_{{-8}} \left( 1-z_{{8}} \right) \partial_8 X\,.
\end{eqnarray}

 The equations for
the mean of the state components are:
\begin{eqnarray}
  \dot{X}_{{1}}&=&{\it G_1} \left( t \right) -k_{{-1}}X_{{1}}-k_{{4}}
     \left( X_{{12}}+X_{{1}}X_{{2}} \right) +k_{{-4}}X_{{4}}\;,\\ \nonumber
 \dot{X}_{{2}}&=&{\it G_2} \left( t \right) -k_{{-2}}X_{{2}}-k_{{4}}
    \left( X_{{12}}+X_{{1}}X_{{2}} \right) +k_{{-4}}X_{{4}}\;,\\\nonumber
 \dot{X}_{{3}}&=&{\it G_3} \left( t \right) -k_{{-3}}X_{{3}}-k_{{6}}
   \left( X_{{34}}+X_{{3}}X_{{4}} \right) +k_{{-6}}X_{{5}}\;,\\ \nonumber
 \dot{X}_{{4}}&=&k_{{4}} \left( X_{{12}}+X_{{1}}X_{{2}} \right) -k_{{-4}}X_{{4}
   }-k_{{5}} \left( n_{{1}}X_{{4}}-n_{{2}} \left(
   X_{{46}}+X_{{4}}X_{{6}}
   \right) -n_{{3}} \left( X_{{47}}+X_{{4}}X_{{7}} \right)  \right)\\ \nonumber
 &+&k_{
   {-5}}X_{{6}}-k_{{6}} \left( X_{{34}}+X_{{3}}X_{{4}} \right)
   +k_{{-6}}X _{{5}}\;,\\ \nonumber
\dot{X}_{{5}}&=&k_{{6}} \left( X_{{34}}+X_{{3}}X_{{4}} \right)
   -k_{{-6}}X_{{5} }-k_{{7}} \left( m_{{1}}X_{{5}}-m_{{2}} \left(
   X_{{56}}+X_{{5}}X_{{6}}
   \right) -M_{{3}} \left( X_{{57}}+X_{{5}}X_{{7}} \right)  \right) +k_{
   {-7}}X_{{7}}\;,\\ \nonumber
\dot{X}_{{6}}&=&k_{{5}} \left( n_{{1}}X_{{4}}-n_{{2}} \left(
   X_{{46}}+X_{{4}}X _{{6}} \right) -n_{{3}} \left(
   X_{{47}}+X_{{4}}X_{{7}} \right)
    \right) -k_{{-5}}X_{{6}}\;,\\ \nonumber
\dot{X}_{{7}}&=&k_{{7}} \left( m_{{1}}X_{{5}}-m_{{2}} \left(
   X_{{56}}+X_{{5}}X _{{6}} \right) -m_{{3}} \left(
   X_{{57}}+X_{{5}}X_{{7}} \right)
   \right) -k_{{-7}}X_{{7}}\;,\\ \nonumber
\dot{X}_{{8}}&=&k_{{8}} \left( l_{{1}}X_{{6}}-l_{{2}} \left(
   X_{{67}}+X_{{6}}X _{{7}} \right)  \right)
  -k_{{-8}}X_{{8}}\;.
\end{eqnarray}

As in the previous examples, the system of equations is infinite
because in the equation for an $X_m$, variables $X_{m'}$ with
higher order indices $m'>m$ are present. For example, to compute
the standard deviation of the the mRNA number, we need to include
the following equation in the system

\begin{eqnarray}
  \dot{X}_{{88}}&=&2\,k_{{8}} \left( l_{{1}}X_{{68}}-l_{{2}}
\left( X_{{678}}+X_ {{6}}X_{{78}}+X_{{68}}X_{{7}} \right) \right)
-2\,k_{{-8}}X_{{88}}\;,
\end{eqnarray}

 which contains the third order variable $X_{{678}}$. Because the state has 8 components, there will be a total of 164
equations for the $X_m(t)$ variables, up to third order in $m$
($\left|m\right|\leq3$). We will disregard any fourth order
variables in the following, so that we will work with a set of 164
equations. Again, the constant part of the generators $G_1,G_2$
and $G_3$ will fix a stationary state. The equations for the
stationary point are polynomial in $X_{m,0}$, and thus can be
solved by using one of the existing algorithms for polynomial
systems of equations \cite{Polynomial}. The system is large and
many unphysical solutions will be generated by solving it
directly. A strategy to obtain the desired solution is to use at
the beginning only the equations for the first order cumulants
where we set all higher order factorial cumulants equal to zero.
This partial solution will be used as a starting point for finding
a solution for the entire 164 equations. For the following set of
parameters: $\{G_1=340,G_2=275,G_3=2.86,k_{{4}}=1,k_{{-3}}= 0.13,
k_{{-6}}= 3.025,k_{{-4}}=10,k_{{5}}=14, l_{{1}}=60, k_{{-7}}=
7.773,k_{{-1}}=17,k_{{-2}}=11,k_{{8}}=1, k_{{6}}= 0.11 ,k_{{7}}=
0.011,k_{{-8}}= 22.800,k_{{-5}}=9275,
n_1=60,n_2=1,n_3=1,m_1=60,m_2=1,m_3=1,l_1=60,l_2=1\}$, the
stationary point for the mean values of the state is $\{X_{1,0}=
20,X_{2,0}= 25,X_{3,0}=
 22,X_{4,0}= 50,
 X_{5,0}=40,X_{6,0}=4
,X_{7,0}= 3,X_{8,0}= 10\}$. We notice that the constraints imposed
upon the transition probabilities were effective in the sense that
the values for the molecules $c$ and $d$ that bind to the DNA are
much less than the values for the unbound molecules $a$ and $b$.
Out of the 164 values for the $X_{m,0}$ at the stationary point,
the maximum value is $X_{48,0}=6.88$ which expresses the
correlation between the mRNA and the dimer $a$. The minimum
negative number is $X_{78,0}=-0.73$, for the correlation between
mRNA and the DNA-bound molecule $d$.
 The generators will modulate the mRNA level through their time
 dependant term $g_i(t)$,$\,i=1,2,3$. The effect of these input
 variations can be computed like we did in the previous examples,
 using an expansion in the small parameter $\eta$. Then the mRNA
 level will vary according to

 \begin{eqnarray}\label{E2F1TransferFunctions}
   X_{8,1}(s)
   &=&H_8^1(s)g_1(s)+H_8^2(s)g_2(s)+H_8^3(s)g_3(s)\;,\\\nonumber
   X_{88,1}(s)&=&H_{88}^1(s)g_1(s)+H_{88}^2(s)g_2(s)+H_{88}^3(s)g_3(s)\;,
 \end{eqnarray}

were the transfer functions are

\bigskip


$\begin{array}{cc}
  \displaystyle{H_{8}^1(s) =\frac{0.06\,{s}^{2}+ 0.65\,s+
  0.87}{\left( s+ 1.57 \right)  \left( s+ 1.48 \right)  \left( s+ 1.38 \right)}}\;, & \displaystyle{ \hspace{20pt}H_{88}^1(s)=\frac{0.0002\,{s}^{2}+ 0.0228\,s+ 0.0340}{\left( s+ 80.9 \right)  \left( s+ 1.57 \right)  \left( s+
  1.48\right)}}\;,
  \\[10pt]
  \displaystyle{H_8^2(s)=\frac{0.01\,{s}^{2}+ 0.18\,s+
 0.26}{\left( s+ 1.57 \right)  \left( s+ 1.48 \right)  \left( s+ 3.79 \right)}}\;, &  \displaystyle{\hspace{20pt}H_{88}^2(s)=\frac{0.01\,{s}^{2}+ 0.14\,s+ 0.29}{\left( s+ 1.57
\right) \left( s+ 11.30 \right)  \left( s+ 3.01 \right)}}\;, \\[10pt]
  \displaystyle{H_8^3(s)=\frac{- 0.16\,{s}^{2}- 2.16\,s-
  0.52}{\left( s+ 0.043 \right)  \left( s+ 1.48 \right)  \left( s+ 22.8 \right)}}\;, & \displaystyle{\hspace{20pt}H_{88}^2(s)=\frac{0.006\,{s}^{2}- 0.066\,s-
  0.023}{\left( s+ 1.57 \right)  \left( s+ 13.08 \right)  \left( s+ 0.043
  \right)}}\;.
\end{array}$

\bigskip

 The transfer functions obtained from all 164
equations are actually rational functions of much higher powers
than those above. The inverse Laplace transform of these solutions
consists of a sum of many exponential decaying components. Some
components are very small, so they can be neglected. We did an
exhaustive search through all groups of 3 components to find the
best approximation. The error was computed as the ratio of the
$L_2$ norm of the difference between the solution and its
approximation, over the $L_2$ norm of the solution. The average
error for the formulas above is $6\%$. Using more than 3 decaying
components to approximate the transfer function will decrease the
error, but will increase the complexity of the rational function.
  From the expression of the transfer function $H_8^3$, we see that the
  pRb acts with a negative sign, as was expected.
  However, the relative strength of the action of each signal
  generator on the mRNA level can not be assessed unless we
  solve the dynamical equations.

\section{Time Evolution Equation for a Gene Regulatory Network Excited by Signal Generators}

The systems studied in the previous sections proved the usefulness
of the equations for the variables $X_m(t)$ and showed how to use
these equations to solve practical problems. Now we aim to write
the time evolution equation for $X_m(t)$ for a general stochastic
nonlinear regulatory network. Nonlinear refers here to transition
probabilities that are rational functions in the state variables.
 A genetic regulatory network is represented by a vector
 $q=(q_1,\dots,q_n)$. Each $q_i$ represent the number of molecules
 for the component $i$ of the state $q$. The components
 $i=1\dots n$ can represent different proteins, mRNAs, or the
 same protein but in different configurations or localizations (in
 nucleus, on the membrane, in Golgi apparatus etc.). The state
 components $q_i$ change in time due to a set of possible
 transitions $\epsilon$. If at time $t$ the state is $q$, then at
 time $t+dt$ the state will be $q+\epsilon$, with $\epsilon$ one
 of the possible transitions. Each transition is governed by its
 probability of transition $T_{\epsilon}(q,t)$ which depends on
 the state variables $q$ and time $t$. To express this state dependance we need some notations.

A vector $\overline m=(m_1,m_2,...,m_n)$ will index a set of
polynomials which will constitute a basis for the set of all
polynomials in $q$:
\begin{eqnarray}
  e_{\overline m}(q)=e_{m_1}(q_1)...e_{m_n}(q_n)
\end{eqnarray}
 with
\begin{eqnarray}
 e_{m_k}(q_k)=q_k(q_k-1)...(q_k-m_k+1)\,.
\end{eqnarray}
The $e_{m_k}$ is known as the falling factorial and is related
with the Pochhammer symbol. We will refer to $e_{\overline m}(q)$
as the factorial basis. Each vector $\overline m$  can be written
in a tensor notation
\begin{eqnarray}
  m=\underbrace{11...1}_{m_1}\underbrace{22...2}_{m_2}...\underbrace{nn...n}_{m_n}
\end{eqnarray}
and each tensor index $m$ can be transferred into a vector
notation $\overline m$. The modulus of $m$ or $\overline m$ is the
degree of the polynomial: $\left|m\right| \equiv\left|\overline
m\right|=\sum_{k=1}^n m_k$.

To find the standard deviations or moments of the state $q$, the
transformation between the factorial basis and the basis
$\{1,x,x^2,\dots,x^l\dots\}$ is helpful

\begin{eqnarray}
  x^l &=& \sum_{k=0}^l S(l,k)e_k(x)\;,
\end{eqnarray}

where $S(l,k)$ are the Stirling numbers of the second kind. The
inverse transformation depends on the Stirling numbers of the
first kind $s(l,k)$

\begin{eqnarray}
  e_l(x) &=& \sum_{k=0}^l s(l,k)x^k\;.
\end{eqnarray}

In general,

\begin{equation}\label{stirling}
    e_{\overline m}(q)=\prod_{i=1}^ne_{\overline
m_i}(q_i)=\prod_{i=1}^n\sum_{k_i=0}^{m_i}s(m_i,k_i)q_i^{k_i}=\sum_{\overline
k=0}^{\overline m}\prod_{i=1}^ns(m_i,k_i)q_i^{k_i}\;;
\end{equation}

therefore

\begin{equation}\label{stir}
    <e_{\overline m}(q)>=\sum_{\overline k=0}^{\overline
m}s(m_1,k_1)s(m_2,k_2)...s(m_n,k_n)<q_1^{k_1}q_2^{k_2}...q_n^{k_n}>
\end{equation}

and

\begin{equation}\label{stir2}
    <q_1^{m_1}q_2^{m_2}...q_n^{m_n}>=\sum_{\overline k=0}^{\overline
m}S(m_1,k_1)S(m_2,k_2)...S(m_n,k_n)<e_{\overline k}(q)>\;.
\end{equation}

With the help of the factorial basis we can write 3 types of
transition probabilities,

\noindent linear transition:
 \begin{eqnarray}
   T_\epsilon(q,t)=\sum_kM_\epsilon^k(t)q_k\;,
 \end{eqnarray}
polynomial transition:
\begin{eqnarray}
  T_\epsilon(q,t)=\sum_{m}M_\epsilon^{
m}(t)e_{\overline m}(q)\;,
\end{eqnarray}
and rational transition:
\begin{eqnarray}\label{RationalTransition}
  T_\epsilon(q,t)=\frac {\sum_{m}M_\epsilon^{
m}(t)e_{\overline m}(q)}{\sum_{ n}M_\epsilon^{ n}(t)e_{\overline
n}(q)}\,.
\end{eqnarray}

To keep the components of the state $q$ positive or zero the
transition probabilities should obey some boundary conditions. The
state with at least one component set on zero is on the boundary
of the set of all possible states. On the boundary, some of the
$\epsilon$'s will point outside, and there is a danger that the
system will jump into a state with at least one state component
with negative molecule numbers. To avoid the unphysical negative
states, we will impose boundary conditions on the transition
probabilities. Consider a system that is in a positive state at
$t=0$. This system will never jump in a negative region if
$T_{\epsilon}(q,t)=0$ for $q_i$ such that $q_i+\epsilon_i\leq -1$,
$i=1,\dots,n$. The last inequality can be expressed in terms of
$\overline m$ if we look at the structure of the transition
probability $T_{\epsilon}(q,t)=\sum_{m}
M_{\epsilon}^{m}(t)e_{\overline m}(q)$ and the roots of
$e_{\overline m}(q)$. Namely, the condition $q_i+\epsilon_i\leq
-1$, $i=1,\dots,n$ is fulfilled if

\begin{eqnarray}\label{BoundaryCondition}
{\overline m}\geq -\epsilon
\end{eqnarray}
for every ${ m}$ in $T_{\epsilon}(q,t)$ with $ M_{\epsilon}^{
m}(t)\neq 0$. The boundary condition (\ref{BoundaryCondition}) for
rational transition probabilities refers to the numerator of
(\ref{RationalTransition}).

We will first deduce the time variation equation for the
polynomial transitions:

\begin{eqnarray}
  T_\epsilon(q,t)=\sum_m M^m_\epsilon(t) {\bf{e}}_{\overline
m}(q)\;.
\end{eqnarray}

 To excite a gene regulatory network, an experimental scientist
 must act on it through a set of signal generators. The signal generators are present in the coefficients
$M^m_\epsilon(t)$. The most obvious way to introduce a signal
generator is by adding it to a transition probability
\begin{eqnarray}
  T_\epsilon(q,t)=\sum_mM^m_\epsilon e_{\overline
m}(q)+G(t)\,.
\end{eqnarray}

Written in the factorial basis the generator is
$G(t)=M_{\epsilon}^0(t)e_0(q)$.
 This type of signal generator can be implemented using a light switch to control the promoter
 of a gene, Fig. 2.
A different type of signal generator modulates protein
degradation. An experimental implementation of such a generator
will have a tremendous impact on the protein function prediction.
The influence of such a generator can
 be written as
 \begin{eqnarray}
   T_\epsilon (q,t)=\sum_mM^m_\epsilon e_{\overline m}(q)+G(t)e_{\overline n}(q)
 \end{eqnarray}
where $ G(t)=M_\epsilon^n(t)$, with $n\geq -\epsilon$ to satisfy
the boundary conditions.

The equation for the probability of the gene regulatory network
$P(q_1,...q_n,t)$ to be in the state $q$ at time $t$ is
\begin{eqnarray}
{\frac {\partial P(q,t)}{\partial t}}=\sum_\epsilon T_\epsilon
(q-\epsilon,t)P(q-\epsilon,t)-\sum_\epsilon
T_\epsilon(q,t)P(q,t)\,.
\end{eqnarray}

We are interested in the mean values for different molecules, as
well as their correlations. The generating function for such
variables is $F(z,t)$ which is the $\cal Z$-transform of $P(q,t)$:

\begin{eqnarray}
  F(z,t)\equiv{\cal Z}(P(q,t)) = \sum_q z^{q}P(q,t)\,.
\end{eqnarray}

Here $z=(z_1,\dots,z_n)$ and $z^q=z_1^{q_1}\dots z_n^{q_n}$.

Taking care of the boundary condition (\ref{BoundaryCondition}),
the equation for the variable $F(z,t)$ is

\begin{equation}\label{ZtransformofF}
{\frac {\partial F}{\partial t}}=\sum_{\epsilon ,m}(z^{\epsilon
+{\overline m}}-z^{\overline m})M_\epsilon^m(t)\partial_m F
\end{equation}

where we used the property that
\begin{eqnarray}
{\cal Z}(e_{\overline m}(q)P(q,t)))=z^{\overline
m}\partial_mF(z,t)\,.
\end{eqnarray}

The variables that describe the dynamic of the system are
generated by taking partial derivatives of $F(z,t)$ with respect
to $z$. For this process the following relation is useful:
\begin{eqnarray}\label{QProp}
\partial_\alpha(z^\epsilon)=Q_\alpha (\epsilon)z^{\epsilon -{\overline
\alpha}}\;,
\end{eqnarray}
with
\begin{eqnarray}
Q_\alpha (\epsilon)=\epsilon_\alpha \;.
\end{eqnarray}
 In what follows, the Greek letters will refer to a one dimensional index that runs
from $1$ to $n$. The Latin letters will refer to a tensor index
$m=\alpha_1\alpha_2\alpha_3...\, .$ Then
$\partial_{\alpha\beta}(z_1^{\epsilon _1}...z_n^{ \epsilon
_n})=\partial_\beta(Q_\alpha(\epsilon)z^{\epsilon-{\overline
\alpha}})=Q_{\alpha\beta}(\epsilon)z^{\epsilon-{\overline \alpha
}-\overline \beta} $, with
$Q_{\alpha\beta}(\epsilon)=Q_\alpha(\epsilon)Q_\beta(\epsilon-\overline
\alpha ) $.

In general
$$ \partial _mz^\epsilon=Q_m(\epsilon)z^{\epsilon-{\overline
m}}\;,$$ with
$$Q_{m\alpha}(\epsilon)=Q_m(\epsilon)Q_\alpha(\epsilon-{\overline
m}).
$$
From (\ref{ZtransformofF}) we obtain

\begin{eqnarray}
  \nonumber \partial_\alpha {\dot{F}} &=& \sum_{m,\epsilon}(Q_\alpha(\epsilon+{\overline m})z^{\epsilon +{\overline m}-{\overline \alpha}}-
  Q_\alpha({\overline m})z^{{\overline m}-{\overline \alpha}})M_\epsilon^m(t)\partial_mF
  +\sum_{m,\epsilon }(z^{\epsilon +{\overline m}}-z^{\overline   m})M_\epsilon^m(t)\partial_{m\alpha}F
\end{eqnarray}

and for $z=1$

\begin{equation}\label{ecuatia1}
   \dot{F}_\alpha =\sum_{m,\epsilon}(Q_\alpha(\epsilon+{\overline m})-Q_\alpha({\overline
   m}))M_\epsilon^m(t)F_m\,.
\end{equation}

With the compact notation
\begin{eqnarray}
R_\alpha^m(t)=\sum_\epsilon(Q_\alpha(\epsilon +{\overline
m})-Q_\alpha({\overline  m}))M_\epsilon^m(t)\;,
\end{eqnarray}

the equation (\ref{ecuatia1}) becomes
\begin{eqnarray}
\dot{F}_\alpha =R_\alpha ^m(t)F_m\;,
\end{eqnarray}
where summation over the dummy index $m$ is implied.

Similarly:
\begin{eqnarray}\label{Falphabeta}
\dot{F}_{\alpha \beta }=R_{\alpha \beta }^m(t)F_m+R_\alpha
^m(t)F_{m\beta}+R_\beta ^m(t)F_{m\alpha}\;,
\end{eqnarray}

with

\begin{eqnarray}\label{R2}
R_{\alpha \beta }^m(t)=\sum_\epsilon(Q_{\alpha\beta}(\epsilon
+{\overline  m})-Q_{\alpha\beta}({\overline m}))M_\epsilon^m(t)\;.
\end{eqnarray}

The general equations for the $F_m$ variables are obtained by
applying the operator $\partial_{\alpha_1\dots\alpha_n}$ to
(\ref{ZtransformofF}). We write the action of this operator on a
product of two functions as:

\begin{eqnarray}
  \partial_{\alpha_1\dots\alpha_n}(f g) =\left\{
  \partial_{\alpha_1\dots\alpha_{k}}f\,\partial_{\alpha_{k+1}\dots\alpha_{n}}g\right\}_{\alpha}\;,
\end{eqnarray}

where the braces indicate the summation for all pairs of disjoint
sets $\left(\alpha_1\dots\alpha_{k}
\right)$,$\left(\alpha_{k+1}\dots\alpha_{n}\right)$  that form a
partition of the tensor index $\alpha_1\dots\alpha_n$. When
listing all possible partitions, we take care that a permutation
of the elements of a set does not change said set. Also $R^m$ with
an empty index set is zero, because $Q(\epsilon)=1$, which comes
from $z^{\epsilon}=Q(\epsilon)z^{\epsilon}$, see (\ref{QProp}).

Then the equation for the time variation of $F_m(t)$ is
\begin{eqnarray}\label{EqInF}
\dot F_{\alpha_1 \dots \alpha_n}=\left\{R_{\alpha_1 \dots
\alpha_k}^m(t)F_{m\alpha_{k+1} \dots \alpha_n}\right\}_{\alpha}\;,
\end{eqnarray}
where summation over the dummy index $m$ is implied.  The tensor
$R_{\alpha_1 \dots \alpha_k}^m(t)$ is given by (\ref{R2}) with
$\alpha_1 \dots \alpha_k$ instead of the index $\alpha \beta$. To
see the structure of the equation (\ref{EqInF}) we specialize it
for for $n=3$.
\begin{eqnarray}
  {\dot F}_ {\alpha_1\alpha_2\alpha_3}&=& R_{\alpha_1\alpha_2\alpha_3}^m(t)F_m
  \\\nonumber
   & & +R_{\alpha_1\alpha_2}^m(t)F_{m\alpha_3}+R_{\alpha_1\alpha_3}^m(t)F_{m\alpha_2}+R_{\alpha_2\alpha_3}^m(t)F_{m\alpha_1}
   \\\nonumber
  & &
  +R_{\alpha_1}^m(t)F_{m\alpha_2\alpha_3}+R_{\alpha_2}^m(t)F_{m\alpha_1\alpha_3}+R_{\alpha_3}^m(t)F_{m\alpha_1\alpha_2}\;.
\end{eqnarray}

\subsection{Factorial Cumulants and Filled Young Tableaux}

 The equation (\ref{EqInF}) has a similar structure with what is
 called a bilinear system in Nonlinear Control Theory. A bilinear
 system is represented by a time evolution equation that is linear
 in state, linear in control, but not jointly linear in both

 \begin{eqnarray}\label{BiLin}
   \frac{dx}{dt} &=& Ax + \sum_{k=1}^n N_k u_k x +B u\;,
 \end{eqnarray}

where $x\in \mathbb{R}^n$ and $A, N_k,\,k=1\dots n,\,B$ are
appropriate matrices \cite{Mohler}. The controls are $u_k,k=1\dots
n$ and the state is described by $x$. However, the system
(\ref{EqInF}) is not finite like (\ref{BiLin}), and as we
explained in the first section, discarding $F_m$ with higher order
$m$ will produce an unstable finite system of equations. Our goal
is twofold: (1) to change the variables $F_m$ so that the
discarding process for obtaining a finite number of equations
become meaningful and (2) to keep the bilinear structure of the
equations in the new variable.

The change of variable:

\begin{eqnarray}
  F(z,t) &=& e^{X(z,t)}\,
\end{eqnarray}

is similar with the change from moments to cumulants
\cite{McCullagh}. Because $F(z,t)$ generates the factorial
moments, $X(z,t)$ will generate the factorial cumulants.

The time dependent variables, $F_m$, will be replaced by
$X_m=\partial_m X(z,t)\mid_{ z=1}$. The transformation relations
between $F_m$ and $X_m$ follow from the Fa\`{a} Di Bruno's formula
for the derivative of the composition of functions
\cite{FaaDiBruno}. To keep the bilinear structure, we need to
construct an appropriate index notation for the terms of the
Fa\`{a} Di Bruno's formula. We introduce the index construction by
way of an example. The fourth order derivative of $F$ at $z=1$ is
\begin{eqnarray}\label{FdB4}
  F_{\alpha\beta\gamma\delta} &=&  X_{\alpha\beta\gamma\delta}+ \\
  \nonumber & & X_{\alpha\beta\gamma}X_{\delta}+X_{\alpha\beta\delta}X_\gamma+X_{\alpha\gamma\delta}X_\beta+X_{\beta\gamma\delta}X_\alpha+ \\
  \nonumber & & X_{\alpha\beta}X_{\gamma\delta}+X_{\alpha\gamma}X_{\beta\delta}+X_{\alpha\delta}X_{\beta\gamma}+ \\
  \nonumber & & X_{\alpha\beta}X_\gamma X_\delta+X_{\alpha\gamma}X_\beta X_\delta+X_{\alpha\delta}X_\beta X_\gamma+X_{\beta\gamma}X_\alpha X_\delta+X_{\beta\delta}X_\alpha X_\gamma+X_{\gamma\delta}X_\alpha X_\beta +\\
  \nonumber & & X_\alpha X_\beta X_\gamma X_\delta
\end{eqnarray}

 Given the index $\alpha\beta\gamma\delta$ from the
 left side of (\ref{FdB4}) we need to generate all the
 indices that appear in the right  side of (\ref{FdB4}).
 If the term $X_{\alpha\gamma}X_\beta X_\delta $ is present in the
 sum, then any symmetric version of it, like $X_\beta X_{\gamma\alpha}
 X_\delta$, cannot be present. If we classify all possible
 symmetries of a term, then we will find an index notation that will eliminate all the equivalent terms. The symmetries come from the
 commutativity of the product and the commutativity of the partial
 derivatives. Young tableaux and filled Young tableaux will help to classify the symmetries
 and also to construct an index notation which will keep the
 bilinear structure. A Young tableau is associated with a partition
 of an integer N. For a fixed positive integer $N$, to each partition
$N=\mu_1\lambda_1+\mu_2\lambda_2+...+\mu_k\lambda_k$,
$\lambda_1>\lambda_2>...>\lambda_k>0$, we associate an empty Young
tableau consisting of $\mu_i$ rows of $\lambda_i$ empty boxes,
$i=1,..,k$. For example consider the partition
$8=3+2+2+1=3+2\cdot2+1$, that is
$\mu_1=1,\lambda_1=3,\mu_2=2,\lambda_2=2,\mu_3=1,\lambda_3=1$. The
Young tableau corresponding to this partition is

\begin{center}
 Y=$ \unitlength 4pt
\begin{picture}(9,3)(-1,-2.65)
\put(0,0){\line(1,0){6.9}} \put(0,2.3){\line(1,0){6.9}}
\put(0,-2.3){\line(1,0){4.6}} \put(0,-4.6){\line(1,0){4.6}}
\put(0,-6.9){\line(1,0){2.3}} \put(0,2.3){\line(0,-1){9.2}}
\put(2.3,2.3){\line(0,-1){9.2}}\put(2.3,2.3){\line(0,-1){6.9}}
\put(4.6,2.3){\line(0,-1){6.9}}\put(6.9,2.3){\line(0,-1){2.3}}
\end{picture}$
\end{center}

\bigskip \bigskip

 We will use these Young tableaux filled with indices
$\alpha,\beta,\gamma,...$, to represent products of $X_m$
variables:
\begin{center}
 X\,\begin{picture}(9,3)(1.8,8)\unitlength 4pt
            \put(0,0){\line(1,0){6.9}} \put
            (0,-2.3){\line(1,0){2.3}} \put (0,2.3){\line(1,0){6.9}}
            \put(0,2.3){\line(0,-1){4.6}} \put(2.3,2.3){\line(0,-1){4.6}}
            \put(4.6,0){\line(0,1){2.3}} \put(6.9,0){\line(0,1){2.3}}
            \put(0.6,1.1){${}_\alpha $}\put(2.9,0.9){${}_\beta$}
            \put(5.2,1.1){${}_\gamma$}\put(0.6,-1.4){${}_\delta$}
         \end{picture}\,\,\,\,\,\,\,\,\,\,\,=$X_{\alpha\beta\gamma}X_\delta$
\end{center}
\bigskip \bigskip

Then ({\ref{FdB4}}) is written using filled Young tableaux as
indices as follows:

\begin{picture}(100,100)(0,-80)

  \begin{picture}(10,10)

   F\,\,\,\unitlength 4pt
      \begin{picture}(9,3)(1.8,2)
        \put(0.6,1.1){${}_\alpha $} \put(2.9,0.9){${}_\beta$} \put(5.2,1.1){${}_\gamma$} \put(7.8,0.9){${}_\delta$}
        \put(0,0){\line(1,0){9.2}} \put(0,2.3){\line(1,0){9.2}}  \put(0,0){\line(0,1){2.3}} \put(2.3,0){\line(0,1){2.3}}
        \put(4.6,0){\line(0,1){2.3}} \put(6.9,0){\line(0,1){2.3}} \put(9.2,0){\line(0,1){2.3}}
      \end{picture}
  = X\,\,\,\,\unitlength 4pt
      \begin{picture}(9,3)(1.8,2)
          \put(0.6,1.1){${}_\alpha $} \put(2.9,0.9){${}_\beta$} \put(5.2,1.1){${}_\gamma$} \put(7.8,0.9){${}_\delta$}
          \put(0,0){\line(1,0){9.2}} \put(0,2.3){\line(1,0){9.2}}  \put(0,0){\line(0,1){2.3}} \put(2.3,0){\line(0,1){2.3}}
          \put(4.6,0){\line(0,1){2.3}} \put(6.9,0){\line(0,1){2.3}}\put(9.2,0){\line(0,1){2.3}}
      \end{picture}+

  \end{picture}

\begin{picture}(10,10)(-52,25)

     X\,\,\,\,\unitlength 4pt
        \begin{picture}(9,3)(1.8,2)
             \put(0,0){\line(1,0){6.9}} \put (0,-2.3){\line(1,0){2.3}} \put (0,2.3){\line(1,0){6.9}}
             \put(0,2.3){\line(0,-1){4.6}} \put(2.3,2.3){\line(0,-1){4.6}} \put(4.6,0){\line(0,1){2.3}}
             \put(6.9,0){\line(0,1){2.3}}
             \put(0.6,1.1){${}_\alpha $}\put(2.9,0.9){${}_\beta$} \put(5.2,1.1){${}_\gamma$}\put(0.6,-1.4){${}_\delta$}
        \end{picture}
       $\!\!\!\!\!\!\!\!$+
   X\,\,\,\,\unitlength 4pt
         \begin{picture}(9,3)(1.8,2)
            \put(0,0){\line(1,0){6.9}} \put
            (0,-2.3){\line(1,0){2.3}} \put (0,2.3){\line(1,0){6.9}}
            \put(0,2.3){\line(0,-1){4.6}} \put(2.3,2.3){\line(0,-1){4.6}}
            \put(4.6,0){\line(0,1){2.3}} \put(6.9,0){\line(0,1){2.3}}
            \put(0.6,1.1){${}_\alpha $}\put(2.9,0.9){${}_\beta$}
            \put(5.2,0.9){${}_\delta$}\put(0.6,-1.2){${}_\gamma$}
         \end{picture}
       $\!\!\!\!\!\!\!$+
   X\,\,\,\,\unitlength 4pt
         \begin{picture}(9,3)(1.8,2)
             \put(0,0){\line(1,0){6.9}}
             \put (0,-2.3){\line(1,0){2.3}} \put (0,2.3){\line(1,0){6.9}}
             \put(0,2.3){\line(0,-1){4.6}} \put(2.3,2.3){\line(0,-1){4.6}}
             \put(4.6,0){\line(0,1){2.3}} \put(6.9,0){\line(0,1){2.3}}
             \put(0.6,1.1){${}_\alpha $}\put(2.9,1.1){${}_\gamma$}
             \put(5.2,0.9){${}_\delta$}\put(0.6,-1.4){${}_\beta$}
         \end{picture}
       $\!\!\!\!\!\!\!$+
   X\,\,\,\,\unitlength 4pt
         \begin{picture}(9,3)(1.8,2) \put(0,0){\line(1,0){6.9}}
             \put(0,-2.3){\line(1,0){2.3}} \put (0,2.3){\line(1,0){6.9}}
             \put(0,2.3){\line(0,-1){4.6}} \put(2.3,2.3){\line(0,-1){4.6}}
             \put(4.6,0){\line(0,1){2.3}} \put(6.9,0){\line(0,1){2.3}}
             \put(0.6,0.9){${}_\beta $}\put(2.9,1.1){${}_\gamma$}
             \put(5.2,0.9){${}_\delta$}\put(0.6,-1.2){${}_\alpha$}
         \end{picture}
       $\!\!\!\!\!\!\!$+\
\end{picture}

\begin{picture}(10,10)(-38,58)
     \unitlength 4pt
  X\,\,\,
   \begin{picture}(9,3)(1.8,2)
      \put(0,0){\line(1,0){4.6}}\put(0,2.3){\line(1,0){4.6}}
      \put(0,-2.3){\line(1,0){4.6}} \put(0,-2.3){\line(0,1){4.6}}
      \put(2.3,-2.3){\line(0,1){4.6}}\put(4.6,-2.3){\line(0,1){4.6}}
      \put(0.6,1.1){${}_\alpha $} \put(2.9,0.9){${}_\beta$}
      \put(0.6,-1.2){${}_\gamma $}\put(2.9,-1.4){${}_\delta$}
   \end{picture}
  $\!\!\!\!\!\!\!\!\!\!\!\!\!\!$+
     \unitlength 4pt
  X$\,\,\,$
   \begin{picture}(9,3)(1.8,2)
       \put(0,0){\line(1,0){4.6}}\put(0,2.3){\line(1,0){4.6}}
       \put(0,-2.3){\line(1,0){4.6}} \put(0,-2.3){\line(0,1){4.6}}
       \put(2.3,-2.3){\line(0,1){4.6}}\put(4.6,-2.3){\line(0,1){4.6}}
       \put(0.6,1.1){${}_\alpha $} \put(2.9,1.1){${}_\gamma$}
       \put(0.6,-1.4){${}_\beta $}\put(2.9,-1.4){${}_\delta$}
   \end{picture}
  $\!\!\!\!\!\!\!\!\!\!\!\!\!\!+
  \unitlength 4pt
  X\,\,$
   \begin{picture}(9,3)(1.8,2)
       \put(0,0){\line(1,0){4.6}}\put(0,2.3){\line(1,0){4.6}}
       \put(0,-2.3){\line(1,0){4.6}} \put(0,-2.3){\line(0,1){4.6}}
       \put(2.3,-2.3){\line(0,1){4.6}}\put(4.6,-2.3){\line(0,1){4.6}}
       \put(0.6,1.1){${}_\alpha $} \put(2.9,0.9){${}_\delta$}
       \put(0.6,-1.4){${}_\beta $} \put(2.9,-1.2){${}_\gamma$}
   \end{picture}$\!\!\!\!\!\!\!\!\!\!\!\!\!$+
\end{picture}

  \begin{picture}(10,10)(-25,92)\unitlength 4pt
     X\,\,\,
        \begin{picture}(9,3)(1.8,2)
        \put(0,0){\line(1,0){4.6}}\put(0,2.3){\line(1,0){4.6}}
        \put(0,2.3){\line(0,-1){6.9}}\put(2.3,2.3){\line(0,-1){6.9}}
        \put(4.6,2.3){\line(0,-1){2.3}}
        \put(0,-2.3){\line(1,0){2.3}}\put(0,-4.6){\line(1,0){2.3}}
        \put(0.6,1.1){${}_\alpha $} \put(2.9,0.9){${}_\beta$}
        \put(0.6,-1.2){${}_\gamma $}\put(0.6,-3.7){${}_\delta $}
        \end{picture}
    $\!\!\!\!\!\!\!\!\!\!\!\!\!$+

X\,\,\,
        \begin{picture}(9,3)(1.8,2)
        \put(0,0){\line(1,0){4.6}}\put(0,2.3){\line(1,0){4.6}}
        \put(0,2.3){\line(0,-1){6.9}}\put(2.3,2.3){\line(0,-1){6.9}}
        \put(4.6,2.3){\line(0,-1){2.3}}
        \put(0,-2.3){\line(1,0){2.3}}\put(0,-4.6){\line(1,0){2.3}}
        \put(0.6,1.1){${}_\alpha $} \put(2.9,1.1){${}_\gamma$}
        \put(0.6,-1.4){${}_\beta $}\put(0.6,-3.7){${}_\delta $}
        \end{picture}
    $\!\!\!\!\!\!\!\!\!\!\!\!\!$+

X\,\,\,
        \begin{picture}(9,3)(1.8,2)
        \put(0,0){\line(1,0){4.6}}\put(0,2.3){\line(1,0){4.6}}
        \put(0,2.3){\line(0,-1){6.9}}\put(2.3,2.3){\line(0,-1){6.9}}
        \put(4.6,2.3){\line(0,-1){2.3}}
        \put(0,-2.3){\line(1,0){2.3}}\put(0,-4.6){\line(1,0){2.3}}
        \put(0.6,1.1){${}_\alpha $} \put(2.9,0.9){${}_\delta$}
        \put(0.6,-1.4){${}_\beta $}\put(0.6,-3.5){${}_\gamma $}
        \end{picture}
    $\!\!\!\!\!\!\!\!\!\!\!\!\!$+

X\,\,\,
        \begin{picture}(9,3)(1.8,2)
        \put(0,0){\line(1,0){4.6}}\put(0,2.3){\line(1,0){4.6}}
        \put(0,2.3){\line(0,-1){6.9}}\put(2.3,2.3){\line(0,-1){6.9}}
        \put(4.6,2.3){\line(0,-1){2.3}}
        \put(0,-2.3){\line(1,0){2.3}}\put(0,-4.6){\line(1,0){2.3}}
        \put(0.6,0.9){${}_\beta $} \put(2.9,1.1){${}_\gamma$}
        \put(0.6,-1.2){${}_\alpha $}\put(0.6,-3.7){${}_\delta $}
        \end{picture}
    $\!\!\!\!\!\!\!\!\!\!\!\!\!$+

X\,\,\,
        \begin{picture}(9,3)(1.8,2)
        \put(0,0){\line(1,0){4.6}}\put(0,2.3){\line(1,0){4.6}}
        \put(0,2.3){\line(0,-1){6.9}}\put(2.3,2.3){\line(0,-1){6.9}}
        \put(4.6,2.3){\line(0,-1){2.3}}
        \put(0,-2.3){\line(1,0){2.3}}\put(0,-4.6){\line(1,0){2.3}}
        \put(0.6,0.9){${}_\beta$} \put(2.7,0.9){${}_\delta$}
        \put(0.6,-1.4){${}_\alpha $}\put(0.6,-3.5){${}_\gamma $}
        \end{picture}
    $\!\!\!\!\!\!\!\!\!\!\!\!\!$+

X\,\,\,
        \begin{picture}(9,3)(1.8,2)
        \put(0,0){\line(1,0){4.6}}\put(0,2.3){\line(1,0){4.6}}
        \put(0,2.3){\line(0,-1){6.9}}\put(2.3,2.3){\line(0,-1){6.9}}
        \put(4.6,2.3){\line(0,-1){2.3}}
        \put(0,-2.3){\line(1,0){2.3}}\put(0,-4.6){\line(1,0){2.3}}
        \put(0.6,1.1){${}_\gamma $} \put(2.9,0.9){${}_\delta$}
        \put(0.6,-1.2){${}_\alpha $}\put(0.6,-3.7){${}_\beta $}
        \end{picture}
    $\!\!\!\!\!\!\!\!\!\!\!\!\!$+

\end{picture}



  \begin{picture}(10,10)(-12,135)\unitlength 4pt
     X\,\,\,
         \begin{picture}(9,3)(1.8,2)
         \put(0,2.3){\line(0,-1){9.2}}\put(2.3,2.3){\line(0,-1){9.2}}
         \put(0,0){\line(1,0){2.3}}\put(0,2.3){\line(1,0){2.3}}
         \put(0,-4.6){\line(1,0){2.3}}\put(0,-2.3){\line(1,0){2.3}}
         \put(0,-6.9){\line(1,0){2.3}}
         \put(0.6,1.1){${}_\alpha $}
         \put(0.6,-1.2){${}_\beta $}
         \put(0.6,-3.4){${}_\gamma $}
         \put(0.6,-5.9){${}_\delta $}
         \end{picture}
   \end{picture}
\end{picture}


\vspace{100pt}

The rows of a Young tableau are listed in decreasing order of
their length, which will enforce an order in the product of the
variables $X_m$. Thus the symmetry due to the commutativity of the
product is lifted. However, in a block of rows of equal length
there is still an ambiguity in ordering the terms in a product. To
lift the ambiguity, we will order rows of equal length in
decreasing lexicographic order of the words that are placed in
rows. For example, in the 6th term of the above formula
$\alpha\beta>\gamma \delta$ and thus $\alpha\beta$ is placed above
$\gamma \delta$. The lexicographic order between the tensor
indices is induced by the order of the components of the state:
$q_i\prec q_j$ if $i<j$. There is one more symmetry left to be
lifted. This symmetry is generated by the commutativity of the
partial derivatives and is lifted by ordering the letters in a row
from left to right. For example, the 4th term in the formula above
has $\alpha \gamma \delta $ on its first row and not $\gamma
\alpha \delta$. Mathematically, the symmetries are described with
the help of a set of permutations that act on the tensor index
that fill a Young tableau. The components of a tensor index $m$
will be denoted using superscripts not to be confused with the
components of the vector $\overline m$; thus $m=m^1m^2m^3\dots\;$.
A Young tableau $Y$ filled with a tensor index $m$ from $F_m$ is
denoted by $Y[m]$. The filling process starts from the upper left
box of $Y$ where $m^1$ is inserted, and moves from left to right
and top to bottom. The action of a permutation $\sigma $ on the
elements of $Y[m]$ is denoted by $Y[m^\sigma]$ and is exemplified
below:
\bigskip \bigskip
\begin{center}\hspace{-30pt}
  Y[m]=\,\,\begin{picture}(9,3)(1.8,3)\unitlength 12.8pt
            \put(0,0){\line(1,0){6.9}} \put
            (0,-2.3){\line(1,0){2.3}} \put (0,2.3){\line(1,0){6.9}}
            \put(0,2.3){\line(0,-1){4.6}} \put(2.3,2.3){\line(0,-1){4.6}}
            \put(4.6,0){\line(0,1){2.3}} \put(6.9,0){\line(0,1){2.3}}
            \put(0.6,0.9){${m^1} $}\put(2.9,0.9){${m^2}$}
            \put(5.2,0.9){${m^3}$}\put(0.6,-1.2){${m^4}$}
         \end{picture}\hspace{130pt} $Y[m^\sigma]$=\,\,
         \begin{picture}(9,3)(1.8,3)\unitlength 12.8pt
            \put(0,0){\line(1,0){6.9}} \put
            (0,-2.3){\line(1,0){2.3}} \put (0,2.3){\line(1,0){6.9}}
            \put(0,2.3){\line(0,-1){4.6}} \put(2.3,2.3){\line(0,-1){4.6}}
            \put(4.6,0){\line(0,1){2.3}} \put(6.9,0){\line(0,1){2.3}}
            \put(0.4,1.1){${}_{m^{\sigma(1)}} $}\put(2.7,1.1){${}_{m^{\sigma(2)}}$}
            \put(5,1.1){${}_{m^{\sigma(3)}}$}\put(0.4,-1.2){${}_{m^{\sigma(4)}}$}
         \end{picture}
\end{center}

\bigskip \bigskip \bigskip \bigskip

For example, if $Y$=\,\,\begin{picture}(9,3)(1.8,-3)\unitlength
3.8pt
            \put(0,0){\line(1,0){6.9}} \put
            (0,-2.3){\line(1,0){2.3}} \put (0,2.3){\line(1,0){6.9}}
            \put(0,2.3){\line(0,-1){4.6}} \put(2.3,2.3){\line(0,-1){4.6}}
            \put(4.6,0){\line(0,1){2.3}} \put(6.9,0){\line(0,1){2.3}} \end{picture}\hspace{20pt},\hspace{8pt}$m=\alpha \beta \gamma \delta$\;, $\sigma(1)=1, \sigma(2)=2, \sigma(3)=4$ and $\sigma(4)=3$ we have:

\medskip

Y[m]=\,\,\begin{picture}(9,3)(1.8,-3)\unitlength 3.8pt
            \put(0,0){\line(1,0){6.9}} \put
            (0,-2.3){\line(1,0){2.3}} \put (0,2.3){\line(1,0){6.9}}
            \put(0,2.3){\line(0,-1){4.6}} \put(2.3,2.3){\line(0,-1){4.6}}
            \put(4.6,0){\line(0,1){2.3}} \put(6.9,0){\line(0,1){2.3}}
            \put(0.6,1){${}_{\alpha} $}\put(2.9,1.1){${}_{\beta}$}
            \put(5.2,1.1){${}_{\gamma}$}\put(0.6,-1.2){${}_{\delta}$}
         \end{picture}\hspace{50pt}  $Y[m^\sigma]$=\,\,
         \begin{picture}(9,3)(1.8,-3)\unitlength 3.8pt
            \put(0,0){\line(1,0){6.9}} \put
            (0,-2.3){\line(1,0){2.3}} \put (0,2.3){\line(1,0){6.9}}
            \put(0,2.3){\line(0,-1){4.6}} \put(2.3,2.3){\line(0,-1){4.6}}
            \put(4.6,0){\line(0,1){2.3}} \put(6.9,0){\line(0,1){2.3}}
            \put(0.6,1.1){${}_{\alpha} $}\put(2.9,1){${}_{\beta}$}
            \put(5.2,1.1){${}_{\delta}$}\put(0.6,-1.2){${}_{\gamma}$}
         \end{picture}

\bigskip
and
\begin{eqnarray}\label{Exrpp}
    X_{Y[m]}&=&X_{\alpha \beta \gamma}X_{\delta}\\
    X_{Y[m^\sigma]}&=&X_{\alpha \beta \delta}X_{\gamma}\;.
\end{eqnarray}

 To lift the last symmetry, we need to find the permutations $\sigma$ which leave the term $X_{Y[m]}$ invariant,
that is $X_{Y[m^\sigma]}=X_{Y[m]}$. In general, the components
$m^i$ of the tensor index $m$ need not be distinct. For example we
have m=rpp in (\ref{CumulantsForFeedback}). However, to obtain the
Fa\`{a} Di Bruno formula, when we solve for $\sigma$ in
$X_{Y[m^\sigma]}=X_{Y[m]}$, the index $m$ must have distinct
components ($m^i\neq m^j$ for all $i\neq j$). The set of
permutations $\sigma $ thus found form a subgroup of the
permutation group $S_{\left |Y\right|}$. Here $\left|Y\right|$ is
the dimension of the Young tableau $Y$ which equals the total
number of its boxes. This subgroup is denoted as $H^{Y}$.

The terms in the Fa\`{a} Di Bruno formula will be generated using
filled Young tableaux  and a set of representative permutations
$\sigma_i,i=1\dots J $ chosen form  each set of the coset space
\begin{eqnarray}
  S_{\left|Y\right|}/H^{Y}=\{\sigma_1H^{Y},\dots,\sigma_JH^{Y}\}\;.
\end{eqnarray}

Here $J$ is $\left| Y\right|!/{\hbox{cardinal}}(H^{Y})$.  The
lexicographic order is a practical method to select a set of
representative permutations $\sigma_1,\dots,\sigma_J$, without
computing the invariant subgroup $H^{Y}$ and the coset space; we
will use the lexicographic order for simple cases. However, for
general results we will use the coset space. Some examples of
invariant subgroups and coset spaces are presented in Table 4.
\bigskip \bigskip
\begin{table}[!h]\begin{center}
\tablenum{\large \bf {6}} \caption{${\mbox {\large \bf
\,\,\,Subgroups and Coset Spaces}}$}
 {\renewcommand{\arraystretch}{1.4}
\hbox{\hspace{25pt}}\begin{tabular}{|c|c|c|}
  \hline
   \hspace{30pt}$Y[\alpha \beta \gamma \delta]$\hspace{35pt} &$\phantom{\frac{\sigma^{\frac{\frac{H}{U}}{U}}}{\sigma_{\frac{T}{U}}}} {H^{Y}}$ & $S_{|Y|}/H^{Y}$ \\ \hline
   &  &  \\\begin{picture}(9,3)(1.8,3)\unitlength 4pt
             \put(0.6,1.1){${}_\alpha $} \put(2.9,0.9){${}_\beta$} \put(5.2,1.1){${}_\gamma$} \put(7.8,0.9){${}_\delta$} \put(0,0){\line(1,0){9.2}} \put(0,2.3){\line(1,0){9.2}}  \put(0,0){\line(0,1){2.3}} \put(2.3,0){\line(0,1){2.3}} \put(4.6,0){\line(0,1){2.3}} \put(6.9,0){\line(0,1){2.3}} \put(9.2,0){\line(0,1){2.3}}
        \end{picture}& $S_4$ & $\{ (1)\}$ \\
        &  &  \\
  \begin{picture}(9,3)(1.8,3)\unitlength 4pt
            \put(0,0){\line(1,0){6.9}} \put
            (0,-2.3){\line(1,0){2.3}} \put (0,2.3){\line(1,0){6.9}}
            \put(0,2.3){\line(0,-1){4.6}} \put(2.3,2.3){\line(0,-1){4.6}}
            \put(4.6,0){\line(0,1){2.3}} \put(6.9,0){\line(0,1){2.3}}
            \put(0.6,1.1){${}_\alpha $}\put(2.9,0.9){${}_\beta$}
            \put(5.2,1.1){${}_\gamma$}\put(0.6,-1.4){${}_\delta$}
         \end{picture} & $\{(123),(12)\}$  & $\{(1),(34),(24),(14) \}$ \\
   &  &  \\
  \begin{picture}(9,3)(1.8,3)\unitlength 4pt
               \put(0,0){\line(1,0){4.6}}\put(0,2.3){\line(1,0){4.6}}
      \put(0,-2.3){\line(1,0){4.6}} \put(0,-2.3){\line(0,1){4.6}}
      \put(2.3,-2.3){\line(0,1){4.6}}\put(4.6,-2.3){\line(0,1){4.6}}
      \put(0.6,1.1){${}_\alpha $} \put(2.9,0.9){${}_\beta$}
      \put(0.6,-1.2){${}_\gamma $}\put(2.9,-1.4){${}_\delta$}
         \end{picture} & $\{(12),(34),(13)(24)\}$ & $\{(1),(23), (243)\}$ \\
  &  &  \\
  \begin{picture}(9,3)(1.8,3)\unitlength 4pt
            \put(0,0){\line(1,0){4.6}}\put(0,2.3){\line(1,0){4.6}}
        \put(0,2.3){\line(0,-1){6.9}}\put(2.3,2.3){\line(0,-1){6.9}}
        \put(4.6,2.3){\line(0,-1){2.3}}
        \put(0,-2.3){\line(1,0){2.3}}\put(0,-4.6){\line(1,0){2.3}}
        \put(0.6,1.1){${}_\alpha $} \put(2.9,0.9){${}_\beta$}
        \put(0.6,-1.2){${}_\gamma $}\put(0.6,-3.7){${}_\delta $}
        \end{picture} & $\{(12),(34),(12)(34)\}$ & $\{ (1),
(23),(24),(13),(14),(13)(24)\}$ \\
   &  &  \\[5pt]
  \begin{picture}(9,3)(1.8,3)\unitlength 4pt
         \put(0,2.3){\line(0,-1){9.2}}\put(2.3,2.3){\line(0,-1){9.2}}
         \put(0,0){\line(1,0){2.3}}\put(0,2.3){\line(1,0){2.3}}
         \put(0,-4.6){\line(1,0){2.3}}\put(0,-2.3){\line(1,0){2.3}}
         \put(0,-6.9){\line(1,0){2.3}}
         \put(0.6,1.1){${}_\alpha $}
         \put(0.6,-1.4){${}_\beta $}
         \put(0.6,-3.4){${}_\gamma $}
         \put(0.6,-6){${}_\delta $}
        \end{picture} & &  \\
   & $S_4$  & $\{ (1)\}$ \\&  &  \\
\hline
\end{tabular}}
\end{center}
\end{table}

\vspace{15pt}

In the above table we used the cycle notation for permutations:
$(123)$ means $\sigma(1)=2,\, \sigma(2)=3,\, \sigma(3)=1$.

Finally we can write the Fa\`{a} Di Bruno formula in Young
tableaux notation:

\begin{equation}\label{FaaDiBruno1}
    F_m= \sum_
    {\mid Y \mid=\mid m \mid }\, \, \,  \sum _{\sigma \in S_{\mid Y\mid }/H^{Y}}X_{Y[m^\sigma]}\,.
\end{equation}

Here the tensor index $m$ can have any form; there is no need for
the components $m^i$ to be distinct (as it was when we defined
$H^Y$).

 For partial derivatives with $z$ not fixed to $1$ the
formula is similar
\begin{equation}\label{FaaDiBruno2}
    \partial_mF(z,t)= \sum_
    {\mid Y \mid=\mid m \mid }\, \, \,  \sum _{\sigma \in S_{\mid Y\mid
    }/H^{Y}}\partial_{Y[m^\sigma]}X(z,t)e^{X(z,t)}\;,
\end{equation}

with the convention that the derivation, with respect to a filled
Young tableau, is the product of the derivatives along each line
of the tableau. Thus, using the example (\ref{Exrpp}) we have:

\begin{equation}
    \partial_{Y[m]}X(z,t)=\partial_{\alpha}\partial_{\beta}\partial_{\gamma}
X(z,t)\partial_{\delta} X(z,t)\,.
\end{equation}

\subsection{Equation of Motion for Polynomial Transition Probabilities}

The equation (\ref{ZtransformofF}) in the variable $X(z,t)$ is

\begin{equation}\label{ecXdez}
    \partial_t X(z,t)=\sum_{m,\epsilon} (z^{\epsilon +\overline {m}}-z^{\overline {m}}\,
    )M_\epsilon^m(t)\left [ \sum_
    {\mid Y \mid=\mid m \mid }\, \, \,  \sum _{\sigma \in S_{\mid Y\mid }/H^{Y}}\partial_{Y[m^\sigma]}X(z,t)\right
    ]\;.
\end{equation}

The time dependent variables will now be
$\partial_mX(z,t)\mid_{z=1}$, so that we must take partial
derivatives with respect to $z$ of (\ref{ecXdez}). The
concatenation notation $\partial_{\alpha
\beta}=\partial_{\alpha}\partial_{\beta}$ must be generalized for
filled Young tableaux

\begin{eqnarray}\label{Concatenation}
  \partial_{\alpha |
  Y[m]}X(z,t):=\partial_\alpha(\partial_{Y[m]}X(z,t))\;.
\end{eqnarray}

 From the definition (\ref{Concatenation}), the concatenation $\alpha | Y[m]$ means that a box containing
 $\alpha$ must be glued to each row of $Y[m]$ and the object thus
 obtained must be rearranged into a lexicographical order filled
 Young tableau. Here is an example of concatenation with a box
 filled with the index 2:

$\partial_{\bf 2}(\,\partial\,\,\,\,\unitlength 4pt
\begin{picture}(9,3)(1.8,2) \put(0,0){\line(1,0){6.9}}
\put(0,2.3){\line(1,0){6.9}} \put(0,-2.3){\line(1,0){4.6}}
\put(0,-4.6){\line(1,0){4.6}} \put(0,-6.9){\line(1,0){2.3}}
\put(0,2.3){\line(0,-1){9.2}}
\put(2.3,2.3){\line(0,-1){9.2}}\put(2.3,2.3){\line(0,-1){6.9}}
\put(4.6,2.3){\line(0,-1){6.9}}\put(6.9,2.3){\line(0,-1){2.3}}
\put(0.6,1.1){${}_1 $}\put(2.9,1.1){${}_2
$}\put(5.2,1.1){${}_3$}\put(0.6,-1.2){${}_4$}\put(2.9,-1.2){${}_5$}
        \put(0.6,-3.5){${}_6$}\put(2.9,-3.5){${}_7$}
       \put(0.6,-5.7){${}_8 $}
\end{picture}\!\!\!\!\!\!\!X(z,t))=\partial\,\,\,\,\unitlength 4pt \begin{picture}(9,3)(1.8,2)
\put(0,0){\line(1,0){9.2}} \put(0,2.3){\line(1,0){9.2}}
\put(0,-2.3){\line(1,0){4.6}} \put(0,-4.6){\line(1,0){4.6}}
\put(0,-6.9){\line(1,0){2.3}} \put(0,2.3){\line(0,-1){9.2}}
\put(2.3,2.3){\line(0,-1){9.2}}\put(2.3,2.3){\line(0,-1){6.9}}
\put(4.6,2.3){\line(0,-1){6.9}}\put(6.9,2.3){\line(0,-1){2.3}}\put(9.2,2.3){\line(0,-1){2.3}}
\put(0.6,1.1){${}_1 $}\put(2.9,1.1){${}_2 $}\put(5.2,1.1){${}_{\bf
2}$}\put(0.6,-1.2){${}_4$}\put(2.9,-1.2){${}_5$}\put(7.5,1.1){${}_3$}
        \put(0.6,-3.5){${}_6$}\put(2.9,-3.5){${}_7$}
       \put(0.6,-5.7){${}_8 $}
\end{picture}\hspace{-5pt}X(z,t)+\partial\,\,\,\,\unitlength 4pt \begin{picture}(9,3)(1.8,2)
\put(0,0){\line(1,0){6.9}} \put(0,2.3){\line(1,0){6.9}}
\put(0,-2.3){\line(1,0){6.9}} \put(0,-4.6){\line(1,0){4.6}}
\put(0,-6.9){\line(1,0){2.3}} \put(0,2.3){\line(0,-1){9.2}}
\put(2.3,2.3){\line(0,-1){9.2}}\put(2.3,2.3){\line(0,-1){6.9}}
\put(4.6,2.3){\line(0,-1){6.9}}\put(6.9,2.3){\line(0,-1){4.6}}
\put(0.6,1.1){${}_1 $}\put(2.9,1.1){${}_2
$}\put(5.2,1.1){${}_3$}\put(0.6,-1.2){${}_{\bf 2}$}
\put(2.9,-1.2){${}_4$}\put(0.6,-3.5){${}_6$}\put(2.9,-3.5){${}_7$}
        \put(0.6,-5.7){${}_8 $}\put(5.2,-1.2){${}_5$}
\end{picture}\hspace{-12pt}X(z,t)+\partial\,\,\,\,\unitlength 4pt \begin{picture}(9,3)(1.8,2)
\put(0,0){\line(1,0){6.9}} \put(0,2.3){\line(1,0){6.9}}
\put(0,-2.3){\line(1,0){6.9}} \put(0,-4.6){\line(1,0){4.6}}
\put(0,-6.9){\line(1,0){2.3}} \put(0,2.3){\line(0,-1){9.2}}
\put(2.3,2.3){\line(0,-1){9.2}}\put(2.3,2.3){\line(0,-1){6.9}}
\put(4.6,2.3){\line(0,-1){6.9}}\put(6.9,2.3){\line(0,-1){4.6}}
\put(0.6,1.1){${}_1 $}\put(2.9,1.1){${}_2
$}\put(5.2,1.1){${}_3$}\put(0.6,-1.2){${}_{\bf 2}$}
\put(2.9,-1.2){${}_6$}\put(0.6,-3.5){${}_4$}\put(2.9,-3.5){${}_5$}
         \put(0.6,-5.7){${}_8 $}\put(5.2,-1.2){${}_7$}
\end{picture}\hspace{-12pt}X(z,t)+\partial\,\,\,\,\unitlength 4pt \begin{picture}(9,3)(1.8,2)
\put(0,0){\line(1,0){6.9}} \put(0,2.3){\line(1,0){6.9}}
\put(0,-2.3){\line(1,0){4.6}} \put(0,-4.6){\line(1,0){4.6}}
\put(0,-6.9){\line(1,0){4.6}} \put(0,2.3){\line(0,-1){9.2}}
\put(2.3,2.3){\line(0,-1){9.2}}\put(2.3,2.3){\line(0,-1){9.2}}
\put(4.6,2.3){\line(0,-1){9.2}}\put(6.9,2.3){\line(0,-1){2.3}}
\put(0.6,1.1){${}_1 $}\put(2.9,1.1){${}_2
$}\put(5.2,1.1){${}_3$}\put(0.6,-1.2){${}_4$}\put(2.9,-1.2){${}_5$}
       \put(0.6,-3.5){${}_6$}\put(2.9,-3.5){${}_7$}
       \put(0.6,-5.7){${}_{\bf 2} $}\put(2.9,-5.7){${}_8 $}
\end{picture}$\hspace{-10pt}$X(z,t)$
\bigskip \bigskip \bigskip

Inductively we define
\begin{eqnarray}
\partial_{\alpha |\beta |...\gamma |Y[m]}=\partial_\alpha
(\partial_{\beta |...|\gamma |Y[m]})\;.
\end{eqnarray}

 The
concatenation notation will be also applied to the $X_m$ variable:
$$X_{\alpha |Y[m]}:=\partial_{\alpha | Y[m]}X(z,t)\mid _{z=1}\;.$$

The equations for the factorial cumulants are now a consequence of
(\ref{ecXdez})

\begin{equation}\label{ecX1}
    \dot
    {X}_\alpha=R_\alpha^m(t)\sum_{Y,\sigma}X_{Y[m^\sigma]}\;,
\end{equation}

\begin{equation}\label{ecX2}
    \dot {X}_{\alpha \beta}=R_{\alpha
    \beta}^m(t)\sum_{Y,\sigma}X_{Y[m^\sigma]}+R_{\alpha }^m(t)\sum_{Y,\sigma}X_{\beta \mid Y[m^\sigma]}+R_{\beta
}^m(t)\sum_{Y,\sigma}X_{\alpha \mid Y[m^\sigma]}\;,
\end{equation}

with $\sum_{Y,\sigma}$ being a short notation for the the sums
over $Y$ and $\sigma$ in (\ref{ecXdez}).

In general

\begin{equation}\label{eclinie}
\dot {X}_{\alpha_1\alpha_2...\alpha_n}=\left \{
R_{\alpha_1\alpha_2...\alpha_k
}^m(t)\sum_{Y,\sigma}X_{\alpha_{k+1}\mid...\mid\alpha_n\mid
Y[m^\sigma]}\right \}_{\alpha}\;.
\end{equation}

On the right side of (\ref{eclinie}) there are more types of Young
tableaux than the one row tableau of the indices from the left
side. This is a consequence of the nonlinearity of the system. To
obtain a closed system of equations, though infinite, we must
obtain the time evolution equation for $X_{Y[m]}$ for any type of
filled Young tableau $Y[m]$, not only for one row, as in
(\ref{eclinie}). The equations written in the variables $X_{Y[m]}$
will be bilinear, as desired; this procedure is known as Carleman
bilinearization \cite{Rough}. To obtain these equations, we need
to introduce the sum of two filled Young tableaux. Let $Y_1[m]$
and $Y_2[\tilde m]$ be two filled Young tableaux with
corresponding partitions
$N=\mu_1\lambda_1+\mu_2\lambda_2+...+\mu_k\lambda_k$ and $\tilde
N=\tilde \mu_1\tilde \lambda_1+\tilde \mu_2\tilde
\lambda_2+...+\tilde \mu_k\tilde \lambda_k$, respectively. We
define their sum, denoted $Y_1[m]\oplus Y_2[\tilde m]$, by
interlacing and ordering the rows of $Y_1[m]$ and $Y_2[\tilde m]$
. The sum corresponds to the partition $N+\tilde
N=\mu_1\lambda_1+\mu_2\lambda_2+...+\mu_k\lambda_k+\tilde
\mu_1\tilde \lambda_1+\tilde \mu_2\tilde \lambda_2+...+\tilde
\mu_k\tilde \lambda_k$. For example:
\begin{center}
\unitlength 4pt \begin{picture}(9,3)(1.8,1)
\put(0,0){\line(1,0){9.2}} \put(0,2.3){\line(1,0){9.2}}
\put(0,-2.3){\line(1,0){4.6}} \put(0,-4.6){\line(1,0){4.6}}
\put(0,-6.9){\line(1,0){2.3}} \put(0,2.3){\line(0,-1){9.2}}
\put(2.3,2.3){\line(0,-1){9.2}}\put(2.3,2.3){\line(0,-1){6.9}}
\put(4.6,2.3){\line(0,-1){6.9}}\put(6.9,2.3){\line(0,-1){2.3}}\put(9.2,2.3){\line(0,-1){2.3}}
\put(0.6,0.9){${}_{\bf 2} $}\put(2.9,0.9){${}_{\bf 3}
$}\put(5.2,0.9){${}_{\bf 4}$}\put(0.6,-1.4){${}_{\bf 5}$}
\put(2.9,-1.4){${}_{\bf 7}$}\put(7.5,0.9){${}_{\bf 6}$}
         \put(0.6,-3.7){${}_{\bf 8}$}\put(2.9,-3.7){${}_{\bf 9}$}
         \put(0.6,-6.1){${}_{\bf 1} $}
\end{picture}$\oplus$ \hspace{6pt} \begin{picture}(9,3)(1.8,1)
\put(0,0){\line(1,0){11.5}} \put(0,2.3){\line(1,0){11.5}}
\put(0,-2.3){\line(1,0){6.9}} \put(0,-4.6){\line(1,0){4.6}}
 \put(0,2.3){\line(0,-1){6.9}}\put(11.5,2.3){\line(0,-1){2.3}}
\put(2.3,2.3){\line(0,-1){6.9}}\put(2.3,2.3){\line(0,-1){6.9}}
\put(4.6,2.3){\line(0,-1){6.9}}\put(6.9,2.3){\line(0,-1){4.6}}\put(9.2,2.3){\line(0,-1){2.3}}
\put(0.6,0.9){${}_3 $}\put(2.9,0.9){${}_4
$}\put(5.2,0.9){${}_5$}\put(0.6,-1.4){${}_1$}\put(2.9,-1.4){${}_2$}\put(4.7,-1.4){${}_{10}$}\put(7.5,0.9){${}_7$}
         \put(0.6,-3.7){${}_6$}\put(2.9,-3.7){${}_9$}\put(9.8,0.9){${}_8$}
\end{picture}\hspace{6pt}:=\hspace{10pt}\begin{picture}(9,3)(1.8,1)
\put(0,0){\line(1,0){11.5}}
\put(0,2.3){\line(1,0){11.5}}\put(9.8,0.9){${}_8$}
\put(0,-2.3){\line(1,0){9.2}} \put(0,-4.6){\line(1,0){6.9}}
\put(0,-6.9){\line(1,0){4.6}}\put(0,-9.2){\line(1,0){4.6}}\put(0,-11.5){\line(1,0){4.6}}
 \put(0,2.3){\line(0,-1){16.1}}\put(11.5,2.3){\line(0,-1){2.3}}
\put(2.3,2.3){\line(0,-1){16.1}}\put(0,-13.8){\line(1,0){2.3}}
\put(4.6,2.3){\line(0,-1){13.8}}\put(6.9,2.3){\line(0,-1){6.9}}\put(9.2,2.3){\line(0,-1){4.6}}
\put(0.6,0.9){${}_3 $}\put(2.9,0.9){${}_4
$}\put(5.2,0.9){${}_5$}\put(0.6,-1.4){${}_{\bf
2}$}\put(2.9,-1.4){${}_{\bf 3}$}\put(5.2,-1.4){${}_{\bf
4}$}\put(7.3,-1.4){${}_{\bf 6}$} \put(7.5,0.9){${}_7$}
         \put(0.6,-3.7){${}_1$}\put(2.9,-3.7){${}_2$} \put(4.7,-3.7){${}_{10}$}
         \put(0.6,-6.1){${}_{\bf 5}$}\put(2.9,-6.1){${}_{\bf 7}$}
         \put(0.6,-8.2){${}_6$}\put(2.9,-8.2){${}_9$}
         \put(0.6,-10.5){${}_{\bf 8}$}\put(2.9,-10.5){${}_{\bf 9}$}
         \put(0.6,-12.8){${}_{\bf 1}$}
\end{picture}
\end{center}

\bigskip \bigskip \bigskip \bigskip

In other words,
\begin{center}
X\,\,\,\,\unitlength 4pt \begin{picture}(9,3)(1.8,2)
\put(0,0){\line(1,0){9.2}} \put(0,2.3){\line(1,0){9.2}}
\put(0,-2.3){\line(1,0){4.6}} \put(0,-4.6){\line(1,0){4.6}}
\put(0,-6.9){\line(1,0){2.3}} \put(0,2.3){\line(0,-1){9.2}}
\put(2.3,2.3){\line(0,-1){9.2}}\put(2.3,2.3){\line(0,-1){6.9}}
\put(4.6,2.3){\line(0,-1){6.9}}\put(6.9,2.3){\line(0,-1){2.3}}\put(9.2,2.3){\line(0,-1){2.3}}
\put(0.6,0.9){${}_{\bf 2} $}\put(2.9,0.9){${}_{\bf 3}
$}\put(5.2,0.9){${}_{\bf 4}$}\put(0.6,-1.4){${}_{\bf 5}$}
\put(2.9,-1.4){${}_{\bf 7}$}\put(7.5,0.9){${}_{\bf 6}$}
         \put(0.6,-3.7){${}_{\bf 8}$}\put(2.9,-3.7){${}_{\bf 9}$}
         \put(0.6,-6.1){${}_{\bf 1} $}
\end{picture}$\cdot\,\,\,$X\,\,\,\,\unitlength 4pt \begin{picture}(9,3)(1.8,2)
\put(0,0){\line(1,0){11.5}} \put(0,2.3){\line(1,0){11.5}}
\put(0,-2.3){\line(1,0){6.9}} \put(0,-4.6){\line(1,0){4.6}}
 \put(0,2.3){\line(0,-1){6.9}}\put(11.5,2.3){\line(0,-1){2.3}}
\put(2.3,2.3){\line(0,-1){6.9}}\put(2.3,2.3){\line(0,-1){6.9}}
\put(4.6,2.3){\line(0,-1){6.9}}\put(6.9,2.3){\line(0,-1){4.6}}\put(9.2,2.3){\line(0,-1){2.3}}
\put(0.6,0.9){${}_3 $}\put(2.9,0.9){${}_4
$}\put(5.2,0.9){${}_5$}\put(0.6,-1.4){${}_1$}\put(2.9,-1.4){${}_2$}\put(4.7,-1.4){${}_{10}$}\put(7.5,0.9){${}_7$}
         \put(0.6,-3.7){${}_6$}\put(2.9,-3.7){${}_9$}\put(9.8,0.9){${}_8$}
\end{picture}\,\,\,=\,X\,\,\,\,\unitlength 4pt \begin{picture}(9,3)(1.8,2)
\put(0,0){\line(1,0){11.5}}
\put(0,2.3){\line(1,0){11.5}}\put(9.8,0.9){${}_8$}
\put(0,-2.3){\line(1,0){9.2}} \put(0,-4.6){\line(1,0){6.9}}
\put(0,-6.9){\line(1,0){4.6}}\put(0,-9.2){\line(1,0){4.6}}\put(0,-11.5){\line(1,0){4.6}}
 \put(0,2.3){\line(0,-1){16.1}}\put(11.5,2.3){\line(0,-1){2.3}}
\put(2.3,2.3){\line(0,-1){16.1}}\put(0,-13.8){\line(1,0){2.3}}
\put(4.6,2.3){\line(0,-1){13.8}}\put(6.9,2.3){\line(0,-1){6.9}}\put(9.2,2.3){\line(0,-1){4.6}}
\put(0.6,0.9){${}_3 $}\put(2.9,0.9){${}_4
$}\put(5.2,0.9){${}_5$}\put(0.6,-1.4){${}_{\bf
2}$}\put(2.9,-1.4){${}_{\bf 3}$}\put(5.2,-1.4){${}_{\bf
4}$}\put(7.3,-1.4){${}_{\bf 6}$} \put(7.5,0.9){${}_7$}
         \put(0.6,-3.7){${}_1$}\put(2.9,-3.7){${}_2$} \put(4.7,-3.7){${}_{10}$}
         \put(0.6,-6.1){${}_{\bf 5}$}\put(2.9,-6.1){${}_{\bf 7}$}
         \put(0.6,-8.2){${}_6$}\put(2.9,-8.2){${}_9$}
         \put(0.6,-10.5){${}_{\bf 8}$}\put(2.9,-10.5){${}_{\bf 9}$}
         \put(0.6,-12.8){${}_{\bf 1}$}
\end{picture}\hspace{6pt}=\,\,X\hspace{7pt}\begin{picture}(9,3)(1.8,2)
\put(0,0){\line(1,0){9.2}} \put(0,2.3){\line(1,0){9.2}}
\put(0,-2.3){\line(1,0){4.6}} \put(0,-4.6){\line(1,0){4.6}}
\put(0,-6.9){\line(1,0){2.3}} \put(0,2.3){\line(0,-1){9.2}}
\put(2.3,2.3){\line(0,-1){9.2}}\put(2.3,2.3){\line(0,-1){6.9}}
\put(4.6,2.3){\line(0,-1){6.9}}\put(6.9,2.3){\line(0,-1){2.3}}\put(9.2,2.3){\line(0,-1){2.3}}
\put(0.6,0.9){${}_{\bf 2} $}\put(2.9,0.9){${}_{\bf 3}
$}\put(5.2,0.9){${}_{\bf 4}$}\put(0.6,-1.4){${}_{\bf 5}$}
\put(2.9,-1.4){${}_{\bf 7}$}\put(7.5,0.9){${}_{\bf 6}$}
         \put(0.6,-3.7){${}_{\bf 8}$}\put(2.9,-3.7){${}_{\bf 9}$}
         \put(0.6,-6.1){${}_{\bf 1} $} \put(10.5,0.9){${}_{\bf \oplus}$}
\end{picture} \hspace{6pt} \begin{picture}(9,3)(1.8,2)
\put(0,0){\line(1,0){11.5}} \put(0,2.3){\line(1,0){11.5}}
\put(0,-2.3){\line(1,0){6.9}} \put(0,-4.6){\line(1,0){4.6}}
 \put(0,2.3){\line(0,-1){6.9}}\put(11.5,2.3){\line(0,-1){2.3}}
\put(2.3,2.3){\line(0,-1){6.9}}\put(2.3,2.3){\line(0,-1){6.9}}
\put(4.6,2.3){\line(0,-1){6.9}}\put(6.9,2.3){\line(0,-1){4.6}}\put(9.2,2.3){\line(0,-1){2.3}}
\put(0.6,0.9){${}_3 $}\put(2.9,0.9){${}_4
$}\put(5.2,0.9){${}_5$}\put(0.6,-1.4){${}_1$}\put(2.9,-1.4){${}_2$}\put(4.7,-1.4){${}_{10}$}\put(7.5,0.9){${}_7$}
         \put(0.6,-3.7){${}_6$}\put(2.9,-3.7){${}_9$}\put(9.8,0.9){${}_8$}
\end{picture}
\end{center}

\bigskip \bigskip \bigskip \bigskip\bigskip

 Using (\ref{eclinie}) and the definition of $X_{\Theta}$ where
$\Theta$ is a filled Young tableau we obtain the equation of
motion:

\begin{equation}\label{ecY}
    \dot X _{\Theta}=\sum_{i \in Rows(\Theta)}\left (\left \{R_{\theta_1\theta_2...\theta_k }^{m}\sum_
    {\mid Y \mid=\mid m \mid }\, \, \,  \sum _{\sigma \in S_{\mid Y\mid }/H^{Y}}X_{{\theta_{k+1}\mid...\mid\theta_{n_i}} \mid Y[m^\sigma]\oplus
\widehat {\Theta^i}}\right \}_{\theta_1,...,\theta_{n_i}} \right)
\end{equation}

The rows of $\Theta$ are indexed by $i$ and the indices filling
the row $i$ are denoted by $\theta_1\dots\theta_{n_i}$. The length
of the row $i$ of $\Theta$ is $n_i$. Here, the summation by the
tensor index $m$ is understood. The hat above $\Theta ^i$ means
that the row $i$ was removed from $\Theta$. The brace indicates
that we have to sum over all possible subsets
$(\theta_1\theta_2...\theta_k )$ of the set
$(\theta_1,...,\theta_{n_i}).$ The order of operations in the
index of $X$ in (\ref{ecY}) is first concatenation $|$ and then
$\bigoplus$.
  An example of an equation of type (\ref{ecY}) for
  the Michaelis-Menten  system analyzed in a preeceding section
  follows:

$\dot{X}\,\,\,$\unitlength 4pt
\begin{picture}(9,3)(1.8,2)
 \put(0,0){\line(1,0){4.6}} \put(0,2.3){\line(0,-1){4.6}}\put(4.6,2.3){\line(0,-1){2.3}}
 \put(2.3,2.3){\line(0,-1){4.6}}\put(0,2.3){\line(1,0){4.6}}\put(0,-2.3){\line(1,0){2.3}}
 \put(0.45,0.9){${}_E$}\put(0.45,-1.4){${}_S$}\put(2.75,0.9){${}_E$}
\end{picture}$\hspace{-22pt}=-2k_1$X\,\,
\begin{picture}(9,3)(1.8,2)
 \put(0,0){\line(1,0){6.9}} \put(0,2.3){\line(0,-1){4.6}}\put(4.6,2.3){\line(0,-1){2.3}}
 \put(2.3,2.3){\line(0,-1){4.6}}\put(0,2.3){\line(1,0){6.9}}\put(0,-2.3){\line(1,0){2.3}}
 \put(0.45,0.9){${}_E$}\put(0.45,-1.4){${}_S$}\put(2.75,0.9){${}_E$}\put(6.9,2.3){\line(0,-1){2.3}}\put(5.05,0.9){${}_E$}
\end{picture}$\hspace{-12pt}-2k_1$X\,\,
\begin{picture}(9,3)(1.8,2)
 \put(0,0){\line(1,0){4.6}} \put(0,2.3){\line(0,-1){6.9}}\put(4.6,2.3){\line(0,-1){2.3}}\put(0,-4.6){\line(1,0){2.3}}
 \put(2.3,2.3){\line(0,-1){6.9}}\put(0,2.3){\line(1,0){4.6}}\put(0,-2.3){\line(1,0){2.3}}
 \put(0.45,0.9){${}_E$}\put(0.45,-1.4){${}_S$}\put(2.75,0.9){${}_E$}\put(0.45,-3.7){${}_S$}
\end{picture}$\hspace{-20pt}-2k_1$X\,\,
\begin{picture}(9,3)(1.8,2)
 \put(0,0){\line(1,0){4.6}} \put(0,2.3){\line(0,-1){6.9}}\put(4.6,2.3){\line(0,-1){2.3}}\put(0,-4.6){\line(1,0){2.3}}
 \put(2.3,2.3){\line(0,-1){6.9}}\put(0,2.3){\line(1,0){4.6}}\put(0,-2.3){\line(1,0){2.3}}
 \put(0.45,0.9){${}_E$}\put(0.45,-1.4){${}_E$}\put(2.75,0.9){${}_S$}\put(0.45,-3.7){${}_S$}
\end{picture}$\hspace{-22pt}-2k_{-2}$X\,\,
\begin{picture}(9,3)(1.8,2)
 \put(0,0){\line(1,0){6.9}} \put(0,2.3){\line(0,-1){4.6}}\put(4.6,2.3){\line(0,-1){2.3}}
 \put(2.3,2.3){\line(0,-1){4.6}}\put(0,2.3){\line(1,0){6.9}}\put(0,-2.3){\line(1,0){2.3}}
 \put(0.45,0.9){${}_E$}\put(0.45,-1.4){${}_S$}\put(2.75,0.9){${}_E$}\put(6.9,2.3){\line(0,-1){2.3}}\put(5.05,0.9){${}_P$}
\end{picture}$\!\!\!\!\!\!\!\!\!$ $-2k_{-2}$X\,\,
\begin{picture}(9,3)(1.8,2)
 \put(0,0){\line(1,0){4.6}} \put(0,2.3){\line(0,-1){6.9}}\put(4.6,2.3){\line(0,-1){2.3}}\put(0,-4.6){\line(1,0){2.3}}
 \put(2.3,2.3){\line(0,-1){6.9}}\put(0,2.3){\line(1,0){4.6}}\put(0,-2.3){\line(1,0){2.3}}
 \put(0.45,0.9){${}_E$}\put(0.45,-1.4){${}_S$}\put(2.75,0.9){${}_E$}\put(0.45,-3.7){${}_P$}
\end{picture}$\hspace{-22pt}-2k_{-2}$X\,\,
\begin{picture}(9,3)(1.8,2)
 \put(0,0){\line(1,0){4.6}} \put(0,2.3){\line(0,-1){6.9}}\put(4.6,2.3){\line(0,-1){2.3}}\put(0,-4.6){\line(1,0){2.3}}
 \put(2.3,2.3){\line(0,-1){6.9}}\put(0,2.3){\line(1,0){4.6}}\put(0,-2.3){\line(1,0){2.3}}
 \put(0.45,0.9){${}_E$}\put(0.45,-1.4){${}_E$}\put(2.75,0.9){${}_P$}\put(0.45,-3.7){${}_S$}
\end{picture}

\bigskip \bigskip

\hspace{40pt}$-k_1$X\,\,
\begin{picture}(9,3)(1.8,2)
 \put(0,0){\line(1,0){4.6}} \put(0,2.3){\line(0,-1){6.9}}\put(4.6,2.3){\line(0,-1){2.3}}\put(0,-4.6){\line(1,0){2.3}}
 \put(2.3,2.3){\line(0,-1){6.9}}\put(0,2.3){\line(1,0){4.6}}\put(0,-2.3){\line(1,0){2.3}}
 \put(0.45,0.9){${}_E$}\put(0.45,-1.4){${}_E$}\put(2.75,0.9){${}_E$}\put(0.45,-3.7){${}_S$}
\end{picture}$\hspace{-22pt}-k_1$X\,\,
\begin{picture}(9,3)(1.8,2)
 \put(0,0){\line(1,0){4.6}} \put(0,2.3){\line(0,-1){4.6}}\put(4.6,2.3){\line(0,-1){4.6}}
 \put(2.3,2.3){\line(0,-1){4.6}}\put(0,2.3){\line(1,0){4.6}}\put(0,-2.3){\line(1,0){4.6}}
 \put(0.45,0.9){${}_E$}\put(0.45,-1.4){${}_E$}\put(2.75,0.9){${}_E$}\put(2.75,-1.4){${}_S$}
\end{picture}$\hspace{-23pt}+2k_{-1}$X\,\,
\begin{picture}(9,3)(1.8,2)
 \put(0,0){\line(1,0){4.6}} \put(0,2.3){\line(0,-1){4.6}}\put(4.6,2.3){\line(0,-1){2.3}}
 \put(2.3,2.3){\line(0,-1){4.6}}\put(0,2.3){\line(1,0){4.6}}\put(0,-2.3){\line(1,0){2.3}}
 \put(0.45,0.9){${}_E$}\put(0.45,-1.4){${}_S$}\put(2.75,0.9){${}_C$}
\end{picture}\hspace{-23pt}$+2k_2$X\,\,
\begin{picture}(9,3)(1.8,2)
 \put(0,0){\line(1,0){4.6}} \put(0,2.3){\line(0,-1){4.6}}\put(4.6,2.3){\line(0,-1){2.3}}
 \put(2.3,2.3){\line(0,-1){4.6}}\put(0,2.3){\line(1,0){4.6}}\put(0,-2.3){\line(1,0){2.3}}
 \put(0.45,0.9){${}_E$}\put(0.45,-1.4){${}_S$}\put(2.75,0.9){${}_C$}
\end{picture}$\hspace{-22pt}-2\gamma_E$X\,\,
\begin{picture}(9,3)(1.8,2)
 \put(0,0){\line(1,0){4.6}} \put(0,2.3){\line(0,-1){4.6}}\put(4.6,2.3){\line(0,-1){2.3}}
 \put(2.3,2.3){\line(0,-1){4.6}}\put(0,2.3){\line(1,0){4.6}}\put(0,-2.3){\line(1,0){2.3}}
 \put(0.45,0.9){${}_E$}\put(0.45,-1.4){${}_S$}\put(2.75,0.9){${}_E$}
\end{picture}$\hspace{-22pt}-\gamma_S$X\,\,
\begin{picture}(9,3)(1.8,2)
 \put(0,0){\line(1,0){4.6}} \put(0,2.3){\line(0,-1){4.6}}\put(4.6,2.3){\line(0,-1){2.3}}
 \put(2.3,2.3){\line(0,-1){4.6}}\put(0,2.3){\line(1,0){4.6}}\put(0,-2.3){\line(1,0){2.3}}
 \put(0.45,0.9){${}_E$}\put(0.45,-1.4){${}_S$}\put(2.75,0.9){${}_E$}
\end{picture}$\hspace{-22pt}+k_{-1}$X\,\,
\begin{picture}(9,3)(1.8,2)
 \put(0,0){\line(1,0){4.6}} \put(0,2.3){\line(0,-1){4.6}}\put(4.6,2.3){\line(0,-1){2.3}}
 \put(2.3,2.3){\line(0,-1){4.6}}\put(0,2.3){\line(1,0){4.6}}\put(0,-2.3){\line(1,0){2.3}}
 \put(0.45,0.9){${}_E$}\put(0.45,-1.4){${}_C$}\put(2.75,0.9){${}_E$}
\end{picture}

\bigskip \bigskip

\hspace{40pt}$+k_S$X\,\,
\begin{picture}(9,3)(1.8,2)
 \put(0,0){\line(1,0){4.6}} \put(0,2.3){\line(0,-1){2.3}}\put(4.6,2.3){\line(0,-1){2.3}}
 \put(2.3,2.3){\line(0,-1){2.3}}\put(0,2.3){\line(1,0){4.6}}
 \put(0.45,0.9){${}_E$}\put(2.75,0.9){${}_E$}
\end{picture}

\bigskip \bigskip

The terms on the right side of the above formula were obtained
from (\ref{ecY}). For example, for
$i=$\begin{picture}(9,3)(-1.4,0.5)
 \put(0,0){\line(1,0){4.6}} \put(0,2.3){\line(0,-1){2.3}}\put(4.6,2.3){\line(0,-1){2.3}}
 \put(2.3,2.3){\line(0,-1){2.3}}\put(0,2.3){\line(1,0){4.6}}
 \put(0.45,0.9){${}_E$}\put(2.75,0.9){${}_E$}
\end{picture}\hspace{-10pt}, $m=EP$, $|Y|=2$, we have

$-2k_{-2}$X\,\,
\begin{picture}(9,3)(1.8,2)
 \put(0,0){\line(1,0){6.9}} \put(0,2.3){\line(0,-1){4.6}}\put(4.6,2.3){\line(0,-1){2.3}}
 \put(2.3,2.3){\line(0,-1){4.6}}\put(0,2.3){\line(1,0){6.9}}\put(0,-2.3){\line(1,0){2.3}}
 \put(0.45,0.9){${}_E$}\put(0.45,-1.4){${}_S$}\put(2.75,0.9){${}_E$}\put(6.9,2.3){\line(0,-1){2.3}}\put(5.05,0.9){${}_P$}
\end{picture}\hspace{-13pt}$=$$R_E^{EP}X\,\,$
\begin{picture}(9,3)(-1.6,2)
 \put(0,0){\line(1,0){4.6}} \put(0,2.3){\line(0,-1){2.3}}\put(-3.55,0.9){${}_E$}\put(-1.2,0.9){${}_\mid$}
 \put(2.3,2.3){\line(0,-1){2.3}}\put(0,2.3){\line(1,0){4.6}}\put(4.6,2.3){\line(0,-1){2.3}}
 \put(0.45,0.9){${}_E$}\put(2.75,0.9){${}_P$}\put(5.05,0.9){$\oplus$}
\end{picture}\hspace{-15pt}
\begin{picture}(9,3)(-2.8,2)
 \put(0,0){\line(1,0){2.3}} \put(0,2.3){\line(0,-1){2.3}}
 \put(2.3,2.3){\line(0,-1){2.3}}\put(0,2.3){\line(1,0){2.3}}
 \put(0.45,0.9){${}_S$}
\end{picture}\hspace{-22pt}

\bigskip \bigskip

$-2k_{-2}$X\,\,
\begin{picture}(9,3)(1.8,2)
 \put(0,0){\line(1,0){4.6}} \put(0,2.3){\line(0,-1){6.9}}\put(4.6,2.3){\line(0,-1){2.3}}\put(0,-4.6){\line(1,0){2.3}}
 \put(2.3,2.3){\line(0,-1){6.9}}\put(0,2.3){\line(1,0){4.6}}\put(0,-2.3){\line(1,0){2.3}}
 \put(0.45,0.9){${}_E$}\put(0.45,-1.4){${}_S$}\put(2.75,0.9){${}_E$}\put(0.45,-3.7){${}_P$}
\end{picture}$\hspace{-22pt}-2k_{-2}$X\,\,
\begin{picture}(9,3)(1.8,2)
 \put(0,0){\line(1,0){4.6}} \put(0,2.3){\line(0,-1){6.9}}\put(4.6,2.3){\line(0,-1){2.3}}\put(0,-4.6){\line(1,0){2.3}}
 \put(2.3,2.3){\line(0,-1){6.9}}\put(0,2.3){\line(1,0){4.6}}\put(0,-2.3){\line(1,0){2.3}}
 \put(0.45,0.9){${}_E$}\put(0.45,-1.4){${}_E$}\put(2.75,0.9){${}_P$}\put(0.45,-3.7){${}_S$}
\end{picture}\hspace{-22pt}$=$$R_E^{EP}X\,\,$
\begin{picture}(9,3)(-1.6,2)
 \put(0,0){\line(1,0){2.3}} \put(0,2.3){\line(0,-1){4.6}}\put(-3.55,0.9){${}_E$}\put(-1.2,0.9){${}_\mid$}
 \put(2.3,2.3){\line(0,-1){4.6}}\put(0,2.3){\line(1,0){2.3}}\put(0,-2.3){\line(1,0){2.3}}
 \put(0.45,0.9){${}_E$}\put(0.45,-1.4){${}_P$}\put(2.75,0.9){$\oplus$}
\end{picture}\hspace{-23pt}
\begin{picture}(9,3)(-2.8,2)
 \put(0,0){\line(1,0){2.3}} \put(0,2.3){\line(0,-1){2.3}}
 \put(2.3,2.3){\line(0,-1){2.3}}\put(0,2.3){\line(1,0){2.3}}
 \put(0.45,0.9){${}_S$}
\end{picture}\hspace{-22pt}

\vspace{30pt}

\subsection{ Equation of Motion for Rational Transition
Probabilities}

The Master Equation is now

\begin{eqnarray}\label{MasterRational}
  \frac{\partial P(q,t)}{\partial t} &=& \sum_{\epsilon}\frac{f_{\epsilon}(q-\epsilon,t)}{{\tilde
  f}_{\epsilon}(q-\epsilon,t)}P(q-\epsilon,t)-\sum_{\epsilon}\frac{f_{\epsilon}(q,t)}{{\tilde
  f}_{\epsilon}(q,t)}P(q,t)\,,
\end{eqnarray}

where $f(q,t)$ and$\tilde f(q,t)$ are polynomial functions in the
state variable $q$.

Multiplying both sides of (\ref{MasterRational}) with
$h(q,t)=\prod_{\epsilon} {\tilde f}_{\epsilon}(q,t) {\tilde
f}_{\epsilon}(q-\epsilon,t)$ produces a Master Equation with
polynomial coefficients:

\begin{eqnarray}
  h(q,t)\frac{\partial P(q,t)}{\partial t} &=&
  \sum_{\epsilon}T_{\epsilon}^{(1)}(q-\epsilon,t)P(q-\epsilon,t)-\sum_{\epsilon}T_{\epsilon}^{(2)}(q,t)P(q,t)\,.
\end{eqnarray}

The decomposition in the factorial base will be

\begin{eqnarray}
  h(q,t) &=&\sum_{m} M^m(t) {\mbox {\large \bf e}}_{\overline m}(q) \\
  T_{\epsilon}^1(q,t)&=& \sum_{m_1}M_{\epsilon}^{m_1}(t) {\mbox {\large \bf e}}_{{\overline m}_1}(q)  \\
  T_{\epsilon}^2(q,t)&=& \sum_{m_2}M_{\epsilon}^{m_2}(t) {\mbox {\large \bf e}}_{{\overline m}_2}(q)
\end{eqnarray}

The boundary condition $m\geq -\epsilon$ that applies to
$f_{\epsilon}(q,t)$ also applies to $ T_{\epsilon}^1(q,t)$ and $
T_{\epsilon}^2(q,t)$ and thus the equation for $F(z,t)$ is

\begin{eqnarray}\label{EqFRational}
  \sum_{\epsilon}M_{\epsilon}^m(t) z^{\overline m}\partial_m\partial_t F &=&\sum_{\epsilon}M_{\epsilon}^{m_1}(t) z^{\epsilon+{\overline
  m}_1}\partial_{m_1}F-\sum_{\epsilon}M_{\epsilon}^{m_2}(t) z^{{\overline
  m}_2}\partial_{m_2}F
\end{eqnarray}

Change the variable to $F(z,t)=e^{X(z,t)}$.

\begin{eqnarray}\label{EqXztRational}
\sum_{\epsilon }M_{\epsilon}^m(t)z^{\overline
m}\left(\sum_{Y,\sigma}\partial_{Y[m^\sigma]}{\dot X}(z,t)+{\dot
X}(z,t)\sum_{Y,\sigma}\partial_{Y[m^\sigma]}{ X}(z,t)
\right)&=&\\\nonumber
\sum_{\epsilon}M_{\epsilon}^{m_1}(t)z^{\epsilon+{\overline
m}_1}\sum_{Y_1,\sigma_1}\partial_{Y_1[{m_1}^{\sigma_1}]}{
X}(z,t)-\sum_{\epsilon}M_{\epsilon}^{m_2}(t)z^{{\overline
m}_2}\sum_{Y_2,\sigma_2}\partial_{Y_2[{m_2}^{\sigma_2}]}{ X}(z,t)
\end{eqnarray}

Take the partial derivative of (\ref{EqXztRational}) with respect
to the tensor index $\alpha_1\dots\alpha_n$  using the general
formula

\begin{eqnarray}
  \partial_{\alpha_1\dots\alpha_n}(f g h) =\left\{
  \partial_{\alpha_1\dots\alpha_{k_1}}f\,\partial_{\alpha_{k_1+1}\dots\alpha_{k_2}}g\,\partial_{\alpha_{k_2+1}\dots\alpha_{n}}h\right\}_{\alpha}\;.
\end{eqnarray}

The braces indicates the summation for all triplets of disjoint
sets $\left(\alpha_1\dots\alpha_{k_1} \right)$,
$\left(\alpha_{k_1+1}\dots\alpha_{k_2}\right)$ and
$\left(\alpha_{k_2+1}\dots\alpha_{n}\right)$ that form a partition
of the tensor index $\alpha_1\dots\alpha_n$.

The time evolution equation for the factorial cumulants is then
\vspace{-10pt}
\begin{eqnarray}\label{EqRational}
&&\\\nonumber
 &&\left\{Q_{\alpha_1 \dots \alpha_k} (\overline m){\dot
X}_{\alpha_{k+1}|\dots\alpha_n |Y[m^\sigma]}+Q_{\alpha_1 \dots
\alpha_{k_1}} (\overline m){\dot
X}_{\alpha_{k_1+1}|\dots\alpha_{k_2}
|Y[m^\sigma]}X_{\alpha_{k_2+1}|\dots\alpha_n|Y[m^\sigma]}\right\}_{\alpha}=\\\nonumber
&&\left\{M_{\epsilon}^{m_1}(t)Q_{\alpha_1\dots\alpha_k}(\epsilon+{\overline
m}_1)X_{\alpha_{k+1}|\dots\alpha_n |Y_1[{m_1}^{\sigma_1}]}-
M_{\epsilon}^{m_2}(t)Q_{\alpha_1\dots\alpha_k}({\overline
m}_2)X_{\alpha_{k+1}|\dots\alpha_n
|Y_2[{m_2}^{\sigma_2}]}\right\}_{\alpha}
\end{eqnarray}
$$ $$
 where summation over the dummy indices $m$, $m_1$ and $m_2$ is implied.
 This equation was used to solve the Hill feedback control
(\ref{CumulantsForFeedback}). Instead of the Carleman
bilinearization, which is difficult to apply for this case, we can
use the variational approach \cite{Rough} and
(\ref{GeneratorWithEta}, \ref{SolXVariational}).

\section{Discussion}

  In this work, we have extended the analysis carried out in \cite{Osc}
from linear to nonlinear stochastic networks. The genetic
regulatory networks are stimulated by a set of signal generators .
In an experimental setting, a specific set of molecules (mRNAs,
proteins) are selected and their time variation is controlled by
input signal generators. As a consequence, the number of the
different molecules that comprise the genetic regulatory network
will vary in time. The time variation of these molecular numbers
is subject to a system of equations. We deduce this system of
equations for a stochastic genetic regulatory network described by
a {\it state}, a set of {\it transitions}  and their {\it
transition probabilities}. The nonlinear effects are due to the
transition probabilities being polynomial or rational functions in
the state components.
 The system being stochastic, the variables of interest are the mean and the correlations
for the molecular species which comprise the genetic network.
 The time dependance of these means and correlations are expressed in terms of a set of factorial cumulants. The network's dynamic is described
 by the time variation of these factorial cumulants. The time
 evolution equations take the form
 (\ref{eclinie}) for polynomial transition probabilities and
 (\ref{EqRational}) for rational transition probabilities.

We solved the equation of motion for four genetic regulatory
networks.

The first example aims to further generalize the results of
\cite{Osc}. There, a linear stochastic genetic network was
analyzed and the equations for the factorial cumulants up to order
two were solved. In the present paper, we study a nonlinear
connection of two linear systems. We arrived at the conclusion
that the solution to the nonlinear coupled systems implies
factorial cumulants of order more than two. The equations for the
cumulants of a linear network can be organized in an hierarchical
structure with respect to the order of the cumulants, see Fig. 3
System 1. The cumulants up to a given order form a closed system
of equations, which is not the case for a typical nonlinear
network. However, we also have shown that for the special case of
linear systems, that nonlinear coupling does not require an
infinite system of equations. Indeed, if we need to solve for the
second order factorial cumulants for $\hbox{System 2}$, then we
need up to forth order cumulants for $\hbox{System 1}$.

 The second example uses the equations (\ref{EqRational}) to study an
 autoregulatory system that is frequently used to explain
 experimental results. The system is composed of one
 gene which regulates its own transcription. The protein acts on
 mRNA production through a term that is a rational function in
 protein number, see Table 2. We studied this system from a synthetic biology
 perspective, aiming to design a logic gate. The biomolecular
 device being intrinsically probabilistic, a logic 1 will be characterized by a mean value and a standard deviation
 from the mean; similar for the logic 0. The distance between the mean values of the
 logical levels should be sufficiently large to include the standard deviations of both logic levels.
  We presented a scheme to design a logic gate from an
 autoregulatory gene, see Fig. 7. From another point of view, the
 autoregulatory system is useful for checking the effectiveness of
 the factorial cumulants. Namely, the analytical solutions must match the
 results from  a Monte Carlo simulation. Such a comparison was
 done for the case when the signal generator is
 closed, causing the gene transcription to be completely under the control of
 its protein product. Because inside a living cell
 many regulatory proteins appear in small number, we choose the
 network parameters so that the mRNA number fluctuates around 12
 molecules, Table 3. We found that the traditional mass action
 equations (\ref{MassAction}) do not explain the Monte Carlo
 simulations; to explain the simulated results we had to use
 equations that involve higher order factorial cumulants. Next, we
 study the response of the system to a time variable signal
 generator. The input-output relations for protein are presented in terms of
 the Laplace transforms of the time dependent variables (\ref{FeddbackLaplace}).

 The stochastic version of the classical Michaelis-Menten mechanism for enzymic catalysis is
 the subject of the third example. An input signal generator acting on
 the enzyme will drive the source, complex and product. The
 time variation of these molecules is described by the system of
 equations (\ref{MMDynamicalEquations}). For an oscillatory signal
 generator, the Michaelis-Menten process behaves as a molecular
 amplifier. Namely, it is possible to drive large oscillations in the product P using
small oscillations in the enzyme E, Fig. 12. Another aspect that
we investigated for this process is its transitory regime, Fig.
13. We showed that the dynamical equations
(\ref{MMDynamicalEquations}) and the Monte Carlo simulations agree
with each other. Thus, we can use the dynamical equations to
produce the statistical variables, rather than generating many
instances of the stochastic process.

The last example is based on E2F1 regulatory element. The
transcription is controlled by three transcription factors: E2F1,
DP1 and pRb. Here we studied the interference of three signal
generators. Each generator will modulate the mRNA production as is
specified by the transfer functions from
(\ref{E2F1TransferFunctions}).
 The E2F1 regulatory element, as was studied here, is a part of a
 complex system of many transcription factors. If we see the E2F1
 as a module in a complex system, than we have to interconnect
 many modules to obtain the whole complex system. How to decompose
 a complex system into its modules and how to interconnect many
 modules to obtain a complex system is left for a future study.
 The dynamical equations for all these examples of
gene regulatory networks are special cases of a general system of
equations that were obtained in the last two sections of the
article. For transition probabilities that are polynomial in the
state components, the system of equations is (\ref{eclinie}). This
system of equations is polynomial in the $X_m$ variables. With the
Carleman bilinearization method, the system is transformed into
(\ref{ecY}). Filled Young tableaux that appear in (\ref{ecY}) help
to construct new variables from the products of the $X_m$
variables. For rational transition probabilities, the equation is
(\ref{EqRational}) and was used to solve the Hill feedback control
(\ref{CumulantsForFeedback}).

 The procedure outline above can be applied to many other genetic
 networks such as networks with multistable steady states \cite{Alex}. The
 simplest network from this class is bistable; it toggles between
 two stable steady states. Biological examples of bistable systems
 include the lambda lysis-lysogeny switch and the hysteretic lac
 repressor system.
 Another avenue of research is to understand the feedback theory in terms of factorial cumulants.
 Negative feedback stabilizes the system whereas positive feedback is responsible
 for oscillations and multistable states. The feedback design
 principles are important in developing biomolecular devices.
  The question is thus: How to translate the feedback design theory from the classical
   control theory into the language of nonlinear stochastic
   genetic networks? The control theory for nonlinear stochastic genetic
   networks should also contain studies about observability and
   reachibility.
From another perspective, studies on the factorial cumulants
discard will be important. What is the minimum cummulat order we
need to retain to reach a predefined precision for the output
variables?

 From an experimental point of view, practical
implementation of signal generators, on both the mRNA and protein
level, will boost research on cell signaling. Experimentally it
was proven that a source of oscillations propagates into a genetic
network. Namely, in \cite{Florian} a group of mice were exposed
for 2 weeks to an external source of 12 hours light followed by 12
hours of darkness. This input external oscillator entrained the
internal clock of the cell. The output signals (mRNA expression
levels) were measured by sacrificing a mouse from the entrained
group every 4 hours for 2 days. The mice were kept in compete
darkness during the 2 days measurement period, so only the
internal clock will affect the mRNA levels. The data, collected
with an Affymetrix (Santa Clara, CA) platform, showed that form
$\approx$ 6000 expressed genes, $\approx$ 500 oscillated with a
24-h period.
 The next experiment would be to implement the light switch from Fig.
 2, and drive one of the core component of the clock mechanism
 directly from its promoter. Then use a microarray experiment to measure the mRNA
 levels and find the set of genes that follow the frequency of signal
 generator. Besides a microarray design,  which screen large sets
 of genes at few time points, an experimental design based on a phototube, \cite{Yamazaki}, can
 record the expression of few genes but in real time. Thus detailed information about
 the time variation of specific genes can be recorded. Such detailed information is crucial for developing a proper
 mathematical description of gene interaction. Models
 developed in the field of system identification \cite{SystemId},
 in conjunction with the approach presented in this article, will help
 to better interpretate the measured data.
  With an input signal generator that acts on a target gene or
  protein, we can also measure the speed of propagation of the signal
  through the gene network. The speed of propagation can be very fast; for example the G protein-coupled receptor switches in
  milliseconds, \cite{Vilardaga}.
 To conclude, understanding how the behavior of
living cells emerges from a genetic network, experimental designs
should be correlated with mathematical theories. We hope that the
methods presented in this article will help to create new
experimental designs for systems biology.

\section{Materials and Methods}
\subsection{Monte Carlo Simulations for Time Dependent Transition Probabilities }

The time dependent Direct Gillespie algorithm was used to generate
the stochastic simulations \cite{GillespieBook}. We present here
the Gillespie algorithm using the notations introduced before.

We denote by $p_\epsilon(\tau\mid q,t)d\tau$ the probability  that
the system will jump in direction $\epsilon$ in the time interval
$[t+\tau,t+\tau+d\tau]$ if it stayed in the state $q$ in the time
interval $[t,t+\tau]$. In terms of transition probabilities:

\begin{eqnarray}\label{PP}
  p_\epsilon(\tau\mid q,t)d\tau=
e^{-\sum_\eta\int_0^{\tau}T_\eta(q,\,t+\tau')d\tau'}T_\epsilon(q,t+\tau)d\tau
\end{eqnarray}

where $T_\epsilon(q,t+\tau)d\tau$ is the probability that the
system will jump from the state $q$ in direction $\epsilon$ in the
time interval $[t+\tau,t+\tau+d\tau]$, regardless of the system's
history before $t+\tau$. The other term of (\ref{PP})
\begin{eqnarray}\label{Equation52}
e^{-\sum_\eta\int_0^{\tau}T_\eta(q,\,t+\tau')d\tau'}
\end{eqnarray}
is the probability that the system will stay in the state $q$ in
the time interval $(t,t+\tau)$.

To obtain (\ref{Equation52}) divide the interval $(t,t+\tau)$ in
small pieces $(t+k\delta,t+(k+1)\delta)$ for $k=0,...,N-1$, with
$N\delta=\tau$. Then the probability that the system will stay in
the state $q$ in the time interval $(t,t+\tau)$ is the product of
the probabilities that the system will stay in the state $q$ in
the intervals $(t+k\delta,t+(k+1)\delta)$. Because  $\delta$ is
small we can use the definition of the transition probability to
find that $1-\sum_\eta T_\eta (q,t+k\delta)\delta$ is the
probability that the system will stay in the state $q$ in the time
interval $(t+k\delta,t+(k+1)\delta)$. The fact that $\delta $ is
small makes also possible to bring this probability into an
exponential form

\begin{eqnarray}\label{ExpT}
e^{-\int_{k\delta}^{(k+1)\delta}\sum_\eta T_\eta
(q,t+\tau')d\tau'}\;.
\end{eqnarray}

 The exponential form will
help to transform the product of the probabilities into a sum.
Multiplying (\ref{ExpT}) for $k=1,...,N-1$ we obtain
$e^{-\sum_\eta\int_0^{\tau}T_\eta(q,\,t+\tau')d\tau'}$, which
appear in the right side of (\ref{PP}).

The cumulative distribution function of $p_\epsilon(\tau\mid
q,t)d\tau$ is

$$F(\tau\mid q,t)=\int_0^\tau \sum_\epsilon  p_\epsilon(\tau''\mid q,t)d\tau''$$

$$=\int_0^\tau \sum_\epsilon e^{-\sum_\eta\int_0^{\tau''}T_\eta(q,\,t+\tau')d\tau'}T_\epsilon(q,t+\tau'')d\tau''$$

$$=1-e^{{-\int_0^\tau}\sum_\epsilon T_\epsilon
(q,t+\tau')d\tau'}\;.$$

After the transition took place at time $t$, the next transition
will take place at $t+\tau$, with $\tau $ a solution of the
equation $F(\tau\mid q,t)=U_1$. Here $U_1$ is a uniform random
number from $[0,1]$ and $q$ and $t$ are known. We will find $\tau$
using the bisection method \cite{Num}. The root is bracketed in
the interval $[-ln(1-U_1)M(q)^{-1}, -ln(1-U_1)m(q)^{-1}]$, where

\begin{equation}\label{}
    m(q)=inf_{x\in {\bf R}}\sum_\epsilon T_\epsilon (q,x)\;,
\end{equation}

\begin{equation}\label{}
    M(q)=sup_{x\in {\bf R}}\sum_\epsilon T_\epsilon (q,x)\;,
\end{equation}

and the procedure is stopped when an accuracy of $10^{-\alpha}$ is
reached. We used $\alpha =5.$

After $\tau$ was found, a second random number $U_2$ is necessary
to find which transition will take place. In other words, we have
to find one $\epsilon_\mu$ from all possible transitions  $\mu
=1\dots \Upsilon$. The unknown $\mu$ from the set of indices
$1,...,\Upsilon$ is obtained from
$$ \sum_{k=1} ^{\mu-1} T_{\epsilon_{k}} (q,t+\tau)\leq U_2\sum_{k=1} ^\Upsilon T_{\epsilon_{k}} (q,t+\tau)\leq \sum_{k=1} ^\mu T_{\epsilon_{k}} (q,t+\tau).$$

Analytical computations and numerical analysis were done with
Maple (Waterloo Maple Inc., Waterloo, Ontario, Canada) and Matlab
(Mathworks, Natick, Massachusetts, United States).

\section{Acknowledgments}

O. Lipan is grateful to Wing H. Wong for his initial impulse for
writing this article, continuous encouragements and critical
inputs. We thank our colleagues at the Center for Biotechnology
and Genomic Medicine for their support. The paper was supported by
O. Lipan startup package for which we thank the Medical College of
Georgia  and the Center for Biotechnology and Genomic Medicine.


\end{document}